\documentclass[12pt,letterpaper,twoside,openright]{report} 

\usepackage[english]{babel}
\usepackage[dvips]{graphicx} 
\usepackage{epsfig}
\usepackage{psfrag}
\usepackage{makeidx} 
\usepackage{amsmath} 
\usepackage{bbm,amsfonts}
\usepackage{amsthm}
\usepackage{mathdots}

\input{Qcircuit}

\theoremstyle{definition}
\newtheorem{theorem}{Theorem}[section]

\newtheorem{lemma}[theorem]{Lemma}
\newtheorem{df}[theorem]{Definition}
\newtheorem{fact}[theorem]{Fact}
\newtheorem{cor}[theorem]{Corollary}
\newtheorem{corollary}[theorem]{Corollary}
\newtheorem{conj}[theorem]{Conjecture}
\DeclareMathOperator{\tr}{tr}
\DeclareMathOperator{\thr}{tr}
\DeclareMathOperator{\supp}{supp}
\DeclareMathOperator{\Sum}{Sum}
\DeclareMathOperator{\Int}{Int}
\newcommand{\Ref}[1]{(\ref{#1})}

\def\qdots{\push{\hspace{1ex} \cdots \hspace{1ex}}}
\def\qvdots{\push{\vspace{1em} \vdots \vspace{1em}}}
\def\qddots{\push{\hspace{1ex} \vspace{1em} \ddots \vspace{1em} \hspace{1ex}}}

\def\C{{\mathbbm C}}
\def\R{{\mathbbm R}}
\def\N{{\mathbbm N}}
\def\Z{{\mathbbm Z}}
\def\F{{\mathbbm F}}
\def\Id{{\mathbbm 1}}
\def\T{{\mathbbm T}}
\def\E{{\mathcal E}}
\def\Hilbert{{\mathcal H}}
\def\Borel{{\mathcal B}}
\def\B{{\mathcal B}}
\def\U{{\mathcal U}}
\def\Pauli{{\mathcal P}}
\def\Clifford{{\mathcal C}}
\def\SL{{\mathcal SL}}
\def\Integrable{{\mathcal I}}
\def\Tau{{\mathcal T}}

\def\littlel{l}

\def\>{\rangle}
\def\<{\langle}
\renewcommand{\bra}[1]{\langle {#1}|}
\renewcommand{\ket}[1]{|{#1}\rangle}
\newcommand{\bracket}[2]{\langle {#1}|{#2} \rangle}

\def\ahalf{\frac{1}{2}}
\def\OneOverSqrtTwo{\frac{1}{\sqrt{2}}}
\def\bfact{\begin{fact}}
\def\efact{\end{fact}}

\def\f{\frac}
\def\bv{\left( \begin{matrix}}
\def\ev{\end{matrix} \right)}

\renewcommand{\v}[1]{{\mathbf #1}}
\def\dag{^\dagger}
\def\what{\widehat}
\def\wtilde{\widetilde}

\def\be{\begin{equation}}
\def\ee{\end{equation}}
\def\bes{\begin{eqnarray}}
\def\ees{\end{eqnarray}}
\def\bess{\begin{eqnarray*}}
\def\eess{\end{eqnarray*}}
\def\nn{\nonumber}

\begin{document}

\pagestyle{empty} 

\begin{center}

\vspace*{1.0cm}
\Huge
{Efficient Simulation of Random Quantum States and Operators}

\vspace*{1.0cm}
\normalsize
by \\
\vspace*{1.0cm}
\Large
Christoph Dankert\\
\vspace*{2.0cm}
\normalsize
A thesis \\
presented to the University of Waterloo \\ 
in fulfilment of the \\
thesis requirement for the degree of \\
Master of Mathematics\\
in \\
Computer Science\\

\vspace*{2.0cm}
Waterloo, Ontario, Canada, 2005\\

\vspace*{1.0cm}
\copyright \hspace{.1ex} Christoph Dankert 2005\\

\end{center}

\cleardoublepage


\pagestyle{plain} 
\pagenumbering{roman} 
\setcounter{page}{3}

\noindent
I hereby declare that I am the sole author of this thesis.

\noindent
I authorize the University of Waterloo to lend this thesis to other
institutions or individuals for the purpose of scholarly research.
\vspace{4cm}

\noindent
Christoph Dankert

\vspace{4cm}

\noindent
I further authorize the University of Waterloo to reproduce this thesis by
photocopying or other means, in total or in part, at the request of other
institutions or individuals for the purpose of scholarly research.
\vspace{4cm}

\noindent
Christoph Dankert
\newpage

\begin{center}
\Large
\textbf{Abstract}
\end{center}

We investigate the generation of quantum states and unitary operations that are ``random'' in certain respects. We show how to use such states to estimate the average fidelity, an important measure in the study of implementations of quantum algorithms. We re-discover the result that the states of a maximal set of mutually-unbiased bases serve this purpose. An efficient circuit is presented that generates an arbitrary state out of such a set.

Later on, we consider unitary operations that can be used to turn any quantum channel into a depolarizing channel. It was known before that the Clifford group serves this and a related purpose, and we show that these are actually the same. We also show that a small subset of the Clifford group is already sufficient to accomplish this. We conclude with an efficient construction of the elements of that subset.
\newpage

\begin{center}\textbf{Acknowledgements}\end{center}
I want to thank my supervisor Richard Cleve for uncountably many inspiring discussions and his helpful ideas and comments throughout my studies at the university of Waterloo. This thesis would not have been possible without his careful guidance. I also want to thank our collaborators Joseph Emerson and Etera Livine. In the research group with Richard, Joseph, and Etera, the results underlying this thesis emerged. The inspiring atmosphere of the Perimeter Institute and the fruitful environment of the Institute for Quantum Computing definitely contributed to the success of that collaboration. 

Thanks also go to Daniel Gottesman, who contributed the significant idea, and to Debbie Leung, who helped solving some of the problems that came up in the final steps of some of the proofs. Special thanks go to my parents for the moral support during difficult phases in writing this thesis. In particular, I want to thank my dad for proof-reading this thesis and finding many typos that would not have been discovered otherwise. Also, I would like to thank my readers Andris Ambainis and Raymond Laflamme. 

Furthermore, I would like to thank J.~Niel de Beaudrap, David Cory, Michele Mosca, Ashin Nayak, and Pranab Sen for inspiring discussions. Thanks go to Connie Slaughter, Lorna Schmalz, and Wendy Reibel for their support in so many administrative tasks. Finally, I want to thank Arvid Bessen, Peter Richter, and Franziska Steiger for their support during the preparation of the presentation of this work.

\newpage

\setcounter{page}{6} 

\clearpage
\thispagestyle{empty}
\cleardoublepage

\tableofcontents

\clearpage
\thispagestyle{empty}
\cleardoublepage

\listoftables

\clearpage
\thispagestyle{empty}
\cleardoublepage

\listoffigures
\newpage

\pagenumbering{arabic}
 

\pagestyle{myheadings}
\markboth{Random Quantum States and Operators}
\normalsize

\thispagestyle{empty}
\cleardoublepage

\chapter{Introduction}
\markright{Introduction}

\section{Preface}


Quantum Computing is a multi-disciplinary subject that tries to make use of the laws of quantum mechanics that govern our physical reality. Since Richard Feynman illustrated how to simulate quantum mechanical systems in the 1980's \cite{Feynman1982}, quantum computing gained a lot of attention. This is particularly due to Peter Shor's factoring algorithm \cite{Shor1994}, that, provided a quantum computers can be built efficiently, would break most of the public-key cryptosystems in use these days. Besides this drawback, quantum computers would enable us to efficiently simulate molecular dynamics and thus would help developing new materials, and would dramatically improve our understanding of molecular biology, for example. 

These widespread applications of quantum computers lead to an enormous effort that has been put into their physical realization. However, it has not been possible to control more than a dozen qubits---far from the applications outlined above, which will need many dozens, hundreds, or even thousands of qubits. Out of the many obstacles, noise is the most prominent one that hinders the development of large-scale quantum computers. 

In this thesis, we devise a protocol to estimate the average fidelity, a global property of the strength of the noise associated with a quantum channel. We will see that it is sufficient to use so-called mutually-unbiased bases (MUBs), as they will lead to the same average as the uniform measure over all quantum states. We re-discovered the previously known result that the states in a complete set of MUBs are a $2$-design for quantum states. The contribution in this area is an explicit construction of circuits that generate the MUB states.

From a different point of view, this thesis is concerned with the generation of quantum states and unitary operations that are ``random'' in certain respects. A \emph{truly} random quantum state on $n$ qubits can be regarded as a uniformly distributed $2^n$-dimensional unit complex vector  (where two vectors are regarded as equivalent if one is a multiple of the other). This uniform distribution is the Fubini-Study measure (Definition \ref{def:fubini-study}), and is defined by the property that it is invariant under unitary transformations. Since the space of possible states has $2^{n+1} - 2$ real degrees of freedom, it is infeasible to generate a distribution of states that is statistically close to a good approximation of this distribution with a polynomial number of operations (Section \ref{sec:fid est random states}). On the other hand, there are very efficient methods for simulating random states that are equivalent to this in certain restricted contexts.

The following sections present an introduction to the fundamental concepts of quantum mechanics and quantum computing. As we will make use of concepts from Linear Algebra, the Dirac Notation, Group Theory, Functional Analysis, Topology, Harmonic Analysis, Finite Fields, and Finite Rings, we present some background on those areas in Appendix \ref{ch:background} and refer to the appropriate literature in the respective area.

Chapter \ref{ch:noise} introduces measures to characterize noise and the average fidelity in particular. It also shows current approaches to estimate the average fidelity. The Decomposition Lemma \ref{lem:repn-of-superop} will be used in the subsequent chapters. After that, the concept of mutually-unbiased bases is introduced formally and the known constructions are presented in chapter \ref{ch:mubs}. The chapter concludes with interesting open problems in that area. The main contributions of this thesis are presented in Chapters \ref{ch:fidelity estimation} and \ref{ch:unitary 2-designs}, where we present an alternate proof for $2$-designs for quantum states and quantum operations, respectively. The contribution in both chapters are the different proof techniques and circuit constructions for the $2$-designs. Chapter \ref{ch:fidelity estimation} gives the  explicit construction for MUB states which are already known to be a $2$-design. Chapter \ref{ch:unitary 2-designs}, in contrast, shows that a subset of the Clifford group is already a unitary $2$-design. An efficient construction of the elements of that subset is presented. Finally, we summarize our conclusions and outline interesting future areas of research in Chapter \ref{ch:conclusion}.

\section{Quantum Mechanics Framework}

At any point in time, the state of a classical system is well-defined, say the position of a car on a street at a given time $t$. In quantum mechanics, however, a system is not just in a single state, but in a superposition of potentially more than one states. Formally, the state of a quantum mechanical system is given by
\be \label{eqn:general state} \ket{\psi} = \alpha_1 \ket{\psi_1} + \alpha_2 \ket{\psi_2} + \dots + \alpha_n\ket{\psi_n},\ee where $\alpha_j$ denotes the ``amplitude'' with which the system is in its basis state $\ket{\psi_j}$. For example, these could be the position or polarization angle of a photon, the energy of an electron, or the spin of an atom. Although the system is in this superposition of several states, an observation will force the system into a single state $\ket{\psi_j}$ with probability $|\alpha_j|^2$. The amplitudes must satisfy \[\sum_{j=1}^n |\alpha_j|^2 = 1\] to give a probability distribution over the states $\ket{\psi_j}$ upon an observation of the system. More formally, the state space of a quantum mechanical system is a complex inner product space $\Hilbert$, commonly referred to as a ``Hilbert space''. A valid state of a quantum mechanical system is described by a unit vector in $\Hilbert$.

The elements of the state space $\Hilbert$ are so-called ``ket'' vectors $\ket{\psi}$. Denote by $\bra{\psi}$ the dual of $\ket{\psi}$, which is a linear functional on $\Hilbert$ such that \[\bra{\psi}(\ket{\phi}) = (\ket{\psi}, \ket{\phi})\] is the inner product between $\ket{\psi}$ and $\ket{\phi}$. This notation is further shortened by letting \[\bra{\psi}(\ket{\phi}) = \bracket{\psi}{\phi}.\] Now, we can refine \Ref{eqn:general state} by specifying that the $\ket{\psi_i}$ have to be pairwise orthogonal, i.e.\@ $\bracket{\psi_i}{\psi_j} = 0$ whenever $i \neq j$, such that $\{\ket{\psi_i}\,|\,i = 1, 2, \dots, n\}$ forms a basis for $\Hilbert$. For the purpose of this work, we will not worry about infinite dimensional state spaces and will assume any state space $\Hilbert$ of finite dimension.

The time evolution of a quantum mechanical system is either unitary if the system is isolated, or a measurement if the system is observed.

\paragraph{Unitary Evolution}

The state of an isolated quantum systems evolves according to a linear function $U$ that acts on the state as a vector. Therefore, we can think of $U$ as a matrix that maps 
\[\bv\alpha_1 \\ \alpha_2 \\ \vdots \\ \alpha_n \ev \mapsto 
\bv\beta_1 \\ \beta_2 \\ \vdots \\ \beta_n \ev.\] The resulting state $\ket{\psi'} = \beta_1 \ket{\psi_1} + \dots + \beta_n\ket{\psi_n}$ must satisfy the normalization constraint as well, which requires $U$ to be unitary and leads to the name of this evolution. It also implies that the evolution is reversible with $U^{-1} = U^{\dag}$ (where $U\dag$ denotes the complex conjugate transpose of $U$).

\paragraph{Measurement}

While an isolated quantum system evolves unitarily, different laws hold when the system is inspected by an observer. This process is called a measurement of the quantum system. In general, a measurement is given by a set of measurement operators $\{M_m\}$ on the state space of the system. The probability that the measurement of a state $\ket{\psi}$ yields outcome $m$ is given by \[p(m) = \bra{\psi}M_m\dag M_m \ket{\psi}.\] After the measurement, the state ``collapses'' to \[\frac{M_m\ket{\psi}}{p(m)}.\] The measurement operators satisfy the completeness relation \[\sum_m M_m\dag M_m = \Id\] so that the outcome probabilities form a probability distribution \[\sum_m p(m) = \sum_m \bra{\psi}M_m\dag M_m \ket{\psi} = \bra{\psi} \left( \sum_m M_m\dag M_m \right) \ket{\psi} = \bracket{\psi}{\psi} = 1.\]

There is a different view on the measurement process called ``Positive Operator-Valued Measure'', abbreviated as POVM, that is most often employed by physicists. It reduces the general measurement operators $M_m$ to a set of positive operators \[E_m = M_m\dag M_m\] such that the probability of observing outcome $m$ is \[p(m) = \bra{\psi} E_m \ket{\psi}.\] The completeness relation now reads \[\sum_m E_m = \Id.\] The complete set of operators $\{E_m\}$ is usually referred to as a POVM with POVM elements $E_m$. A POVM is especially well suited for the analysis of the measurement statistics when the post-measurement state is of no interest. It is simpler than the general measurement description yet still powerful enough to describe the complete statistics of any quantum measurement.

A special kind of measurement is a \emph{projective measurement} or \emph{von Neumann measurement}. A projective measurement is given by an observable $M$, which is required to be a Hermitian operator so that its spectral decomposition \[M = \sum_m m P_m\] exists, where $P_m$ is a projector onto the eigenspace of $M$ with eigenvalue $m$. The eigenvalues $m$ of $M$ represent the possible outcomes of the experiment. The probability of measuring $m$ is given by \[p(m) = \bra{\psi} P_m \ket{\psi}\] and the state after a measurement with outcome $m$ is \[\frac{P_m \ket{\psi}}{\sqrt{p(m)}}.\] A projective measurement can be described as a general measurement with measurement operators $M_m = P_m$, which leads to simplified calculations as $P_m\dag = P_m$, $P_m^2 = P_m$ and thus $P_m\dag P_m = P_m$.

The easiest example of a projective or von Neumann measurement is a measurement in the standard basis. Say we are given the state \Ref{eqn:general state} and measure with respect to the basis $\{\ket{\psi_1}, \dots, \ket{\psi_n}\}$. The measurement is given by the projectors $P_m = \ket{\psi_m}\bra{\psi_m}$ that project onto the subspace spanned by $\ket{\psi_m}$. As the $\ket{\psi_m}$ are pairwise orthogonal, it follows that the projectors $P_m$ are pairwise orthogonal, too. Hence $\sum_m m \ket{\psi_m}\bra{\psi_m}$ is a quantum measurement that yields $m$ with probability $|\bracket{\psi_m}{\psi}|^2 = |\alpha_m|^2$, leaving the system in the post-measurement state $\f{\alpha_m}{|\alpha_m|^2} \ket{\psi_m}$ which is equivalent to $\ket{\psi_m}$.

\section{Quantum Computing}

The basic unit of information in classical computation is the bit, which can either take the value $0$ or $1$. In quantum computation, the basic unit is a \emph{qubit}, which can be in a superposition of $0$ and $1$. We usually identify the basis states of a qubit with $\ket{0}$ and $\ket{1}$. Using the notation of quantum mechanics, we can say more precisely that the state of a single qubit is the superposition \[\alpha_0 \ket{0} + \alpha_1 \ket{1}\] for $\alpha_0, \alpha_1 \in \C$ such that $|\alpha_0|^2 + |\alpha_1|^2 = 1$. Hence the state space of a single qubit is a two-dimensional Hilbert space $\Hilbert_2$.

The state space of a system with $n$ qubits is described by the $n$-fold tensor product of a single-qubit system \[\Hilbert_{2^n} = \underbrace{H_2 \otimes \dots \otimes H_2}_{n \text{ times}}.\] For example, a system consisting of two qubits has the four basis states $\ket{0}\ket{0}$, $\ket{0}\ket{1}$, $\ket{1}\ket{0}$, and $\ket{1}\ket{1}$, where the state $\ket{0}\ket{1}$ means that the first qubit is in state $\ket{0}$ and the second qubit is in state $\ket{1}$. In general, an $n$-qubit system has basis states which correspond to all binary strings of length $n$. Instead of writing \[\ket{b_0} \ket{b_1} \dots \ket{b_{n-1}}\] we will write $\ket{b_0 b_1 \dots b_{n-1}}$ or sometimes even shorter using a base-10 representation of the binary number $\left(b_0 b_1 \dots b_{n-1}\right)_2 =  \sum_{i=0}^{n-1} 2^i b_i$. Therefore, we can write the basis states of a register of $n$ qubits as $\ket{0}, \ket{1}, \dots, \ket{2^n - 1}$. The general state of such a register is given by \[\ket{\psi} = \alpha_0 \ket{0} + \alpha_1 \ket{1} + \dots + \alpha_{2^n-1} \ket{2^n - 1}\] where \[\sum_{i=0}^{2^n-1} |\alpha_i|^2 = 1.\] Hence the state is described by $2^n$ complex amplitudes. Taking into account the normalization condition, we still seem to have $2^n - 1$ complex degrees of freedom. Therefore it seems that a system of $n$ qubits contains a huge amount of information that is encoded in its complex amplitudes, as opposed to $n$ bits of information in a classical $n$-bit system. However, this is only true in a restricted sense: for the generally accepted definition of Holevo information \cite{NielsenChuang2000}, it is known that a qubit contains not more than one classical bit of information. For a deeper elaboration of quantum information theory, we refer the reader to \cite[Ch.~12]{NielsenChuang2000}. Despite those negative results, quantum computation and quantum information does offer provable advances over classical computation and information. Using de Wolf's words, it is ``the art of quantum computing to use this information for interesting computational purposes'' \cite{deWolf1999}.

\subsection{Turing Machine Model}

Classical computation can be described using a variety of different models. Two very prominent ones are the Turing machine and the circuit model. We will briefly address the quantum version of the Turing machine and go into the circuit model of quantum computation in more detail.

In analogy to classical probabilistic Turing machines, quantum Turing machines (QTM) were defined by Benioff \cite{Benioff1982} and Deutsch \cite{Deutsch1985}. We adopt the notation set by Benioff \cite{Benioff1998} in a fairly recent survey. See \cite{Meglicki2005} for an account on the history of the QTM. 

A QTM consists of a one-dimensional infinite tape with cells labelled by the integers $\Z$, a head, and a unitary step operator. Associated to each cell is a finite state space which we will usually define to be two-dimensional and therefore each cell will be one qubit. The head can be in a superposition of a finite number of orthogonal internal states $\ket{l}$, $l \in \{1, 2, \dots, L\}$, and a position $j$ on the tape. In analogy to the classical Turing machine, we define the elementary actions of a QTM as moving of the head one step to the left or one step to the right, changing the state of the qubit at the position of the head, and changing the internal part $\ket{l}$ of the head state. 

Let $\Hilbert$ be the state space of the QTM and write the QTM's computational basis as $\ket{l,j,\underline{s}}$, where $\ket{l,j}$ denotes the position of the head and the head's internal state. $\ket{\underline{s}} = \otimes_{m=-\infty}^{\infty} \ket{\underline{s}_m}$ is a basis state of the tape, where $\underline{s}_m$ is a computational basis state of a single qubit. In order to avoid technical complications, we will assume $\Hilbert$ to have a countable basis. Therefore, a common requirement is that $\underline{s_m} \neq 0$ for at most a finite number of $m$.

The computation of the QTM is given by an initial state of the head and tape and the action of the QTM on each basis state $\ket{l,j,\underline{s}_m}$. The action is specified by a step operator $T$ that obeys certain locality constraints: The head must not move by more than one position at a time, and the operation on the basis state $\ket{l,j,\underline{s}_m}$ may only depend on and change the state of the $j$-th qubit and the head's internal state. 

Although these definitions let a QTM seem analogous to a classical probabilistic Turing machine, it has not been found useful in the development of algorithms and in quantum complexity theory. Therefore, we will stick to the circuit model as our primary model to describe quantum algorithms.

\subsection{Circuit Model}

\begin{figure}[htbp]
	\centering
		\psfrag{a}[t]{$a$}
		\psfrag{b}[t]{$b$}
		\psfrag{axorb}{$a \oplus b$}
		\includegraphics[width=0.50\textwidth]{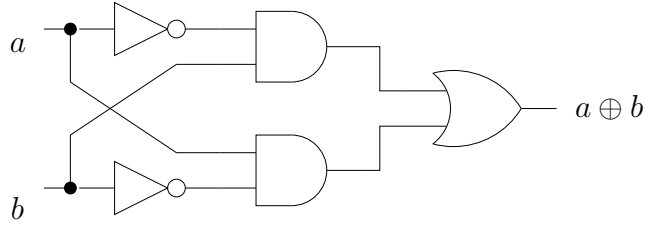}
	\caption{Classical circuit computing $a \oplus b$}
	\label{fig:classicalcircuit}
\end{figure}

A classical Boolean circuit is a directed acyclic graph with input nodes, internal nodes, and output nodes. The circuit has $n$ input nodes, $n \geq 0$. The internal nodes are the gates AND, OR, and NOT, but generally any universal set of gates will work equally well. There are $m$ designated output nodes, $m \geq 1$. The input bits $x$ are fed into the input nodes, and after all gates have been applied, the output nodes assume a value $y$. The circuit computes a Boolean function $f: \{0, 1\}^n \mapsto \{0,1\}^m$, if the output nodes assume the value $f(x)$ for all inputs $x \in \{0,1\}^n$. Figure \ref{fig:classicalcircuit} shows a simple classical circuit that computes $f(a,b) = a \oplus b$.

The circuit model is linked to the Turing machine model by the idea of circuit families. A circuit family is a set $\mathcal{C} = \{C_n\}$ of circuits, one circuit for each input length $n$. A circuit family decides a language $L \subset \{0,1\}^*$ if for any $n$ and any input $x \in \{0,1\}^n$, the circuit $C_n$ outputs $1$ if $x \in L$ and $1$ if $x \notin L$. A circuit family $\mathcal{C}$ is uniform if $C_n$ can be computed by a Turing machine given input $n$. A uniformly polynomial circuit family is a uniform circuit family that can be computed by a Turing machine using space logarithmic in $n$, which implies a run-time polynomial in $n$. It also implies that the number of gates in $C_n$ is at most polynomial in $n$ as well. The link between Turing machines and circuit families is given by the following theorem.

\begin{theorem}\cite{Papadimitriou1994} A language $L \subset \{0,1\}^*$ can be computed by a uniformly polynomial circuit familiy iff $L \in \mathbf{P}$.
\end{theorem}

\begin{table}[htbp]
	\centering
		\begin{tabular}{ccc}
			Hadamard & \begin{minipage}[c]{2em} \Qcircuit{ & \gate{H} & \qw} \end{minipage} & $ \OneOverSqrtTwo \bv 1 & 1 \\ 1 & -1 \ev$ \\[1.5em]
			Pauli-$X$ & \begin{minipage}[c]{2em} \Qcircuit{ & \gate{X} & \qw} \end{minipage} & $ \bv 0 & 1 \\ 1 & 0 \ev$ \\[1.5em]
			Pauli-$Y$ & \begin{minipage}[c]{2em} \Qcircuit{ & \gate{Y} & \qw} \end{minipage} & $ \bv 0 & -i \\ i & 0 \ev$ \\[1.5em]
			Pauli-$Z$ & \begin{minipage}[c]{2em} \Qcircuit{ & \gate{Z} & \qw} \end{minipage} & $ \bv 1 & 0 \\ 0 & -1 \ev$ \\[1.5em]
			Phase & \begin{minipage}[c]{2em} \Qcircuit{ & \gate{S} & \qw} \end{minipage} & $ \bv 1 & 0 \\ 0 & i \ev$ \\[1.5em]
			$\pi/8$ & \begin{minipage}[c]{2em} \Qcircuit{ & \gate{T} & \qw} \end{minipage} & $ \bv 1 & 0 \\ 0 & e^{i\pi/4} \ev$ \\[1.5em]
			C-NOT & \begin{minipage}[c]{2em} \Qcircuit{ & \ctrl{1} & \qw \\ & \targ & \qw} \end{minipage} & $ \bv 1 & 0  & 0 & 0\\ 0 & 1 & 0 & 0 \\ 0 & 0 & 0 & 1 \\ 0 & 0 & 1 & 0 \ev$ 
		\end{tabular}
	\caption{Elementary Quantum Gates from \cite[p. 177]{NielsenChuang2000}}
	\label{tab:ElementaryQuantumGates}
\end{table}

A quantum circuit is a directed acyclic graph with input nodes, internal nodes, and output nodes. The inputs are qubits that are prepared in the state $\ket{0}$ or $\ket{1}$. The internal nodes are quantum gates, which are unitary transformations that act on a finite number of qubits. Restricting these gates to finitely many inputs allows for a comparison of the complexities of classical and quantum circuits. Usually, we allow for one and two-qubit gates. Table \ref{tab:ElementaryQuantumGates} shows some elementary quantum gates together with their circuit symbol and the corresponding unitary matrix. The transformation described by such a quantum circuit can be computed by taking tensor products of gates applied in parallel on disjoint sets of qubits and ordinary product of gates applied in series. A quantum circuit can then be viewed as a single unitary transformation on its $n$ input qubits. 

Usually, some auxilliary qubits are needed during the computation, which are taken as needed and assumed to be initialized to $\ket{0}$. We will call them ``ancillas'' and require them to be in state $\ket{0}$ at the end of the computation. Otherwise, the result of the quantum algorithm might be corrupted by applying local unitary transformations on the ancillas. We will call this process ``uncomputation'', as it is effectively achieved by reversing that part of the computation that made use of the ancilla. 

\begin{figure}[htbp]
	\centerline{
	\Qcircuit @C=.5em @R=.7em
	{
		\lstick{\ket{a}} & \ctrl{1} \qw & \rstick{\ket{a}} \qw \\
		\lstick{\ket{b}} & \targ \qw & \rstick{\ket{a \oplus b}} \qw \\
	} }
	\caption{A quantum circuit that computes $a \oplus b$}
	\label{fig:example quantum circuit}
\end{figure}
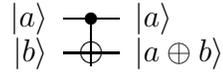

After all gates have been applied to the input qubits and the ancillas, the output nodes will assume a state $\ket{\phi}$. It is measured in the computational basis to produce a classical output string. Note that without loss of generality, all measurements can be placed at the end of the quantum circuit \cite{NielsenChuang2000}, where classically controlled gates are replaced by quantumly controlled gates. Figure \ref{fig:example quantum circuit} shows a simple quantum circuit that computes the exclusive-OR of its input states, which is the controlled-NOT quantum operation.

A distinct feature of a quantum mechanical unitary evolution is its reversability. A unitary operator $U$ is invertible with inverse $U^{-1} = U\dag$. This implies that ancillas are needed in order to compute functions that are not bijections. For example, the transformation $\ket{x} \mapsto \ket{PARITY(x)}$ is not unitary as its inverse does not exist. In order to provide quantum circuits with the ability to calculate those functions as well, we have to introduce additional qubits. It is known that for any classical Boolean function $f$ from $n$ to $m$ bits, there is a reversible function $U_f$ on $n+m$ input bits that computes \[U_f: (x, y) \mapsto (x, y \oplus f(x)),\] where $x \in \{0,1\}^n$ and $y \in \{0,1\}^m$. Furthermore, if a circuit for $f$ uses $T$ gates, there is a circuit for $U_f$ that uses $O(T)$ gates and $O(T)$ ancillas. We will understand that $U_f$ computes $f$ reversibly and will use $U_f$ instead of $f$ when we want to implement $f$ with a quantum circuit.

A universal set $\mathcal{U}$ of quantum gates is a set of single and two-qubit gates such that any quantum circuit can be built using gates from $\mathcal{U}$. It is known that the controlled-NOT gate and all single qubit gates form such a universal gate set, albeit one that is continuously parametrized. To end up with a finite set of single and two-qubit gates, we will loosen the requirements a bit and allow for approximations of unitary operations and call a finite set $\mathcal{U}$ universal if we can approximate every gate using only elements from $\mathcal{U}$. Denote the error if we try to approximate $U$ by $V$ by 
\[E(U,V) = \max_{\ket{\psi}} \|(U-V)\ket{\psi}\|,\] where $\|\cdot\|$ denotes the norm in the state Hilbert space, which is the Euclidean norm if the state space is finite-dimensional. It is known that the probability distributions obtained by a POVM on $U\ket{0}$ and $V\ket{0}$ are close in the following sense: Let $M$ be a POVM and let $p_U$ and $p_V$ be the probabilities of observing $m$ upon measurement of $U\ket{0}$ and $V\ket{0}$, respectively. Then 
\[|p_U(m) - p_V(m)| < 2E(U,V).\] 

It turns out that the discrete set \[\mathcal{U} = \{CNOT, H, P, \pi/8\}\] generates all quantum gates with an error $\epsilon > 0$ as small as desired. This is known as the Solovay-Kitaev Theorem  \cite{NielsenChuang2000}, which states that we can approximate any quantum circuit consisting of $m$ CNOT and single-qubit gates within an error of $\epsilon > 0$ using \[O\left(m \log^c \left(\frac{m}{\epsilon}\right) \right)\] gates from $\mathcal{U}$ with $c \approx 2$.

To conclude the circuit model, we want to state an obvious extension of this standard model. So far we assumed that the elementary system of our quantum computer are two-level systems which we called qubits. It is possible to use $d$-level systems as the elementary building blocks of a quantum computer for some finite $d$. We will refer to them as ``qudits'', and we can define the circuit model for them in an analogous fashion. The only difference is the size of the matrix representation of individual gates. A single qudit gate $U$ corresponds to a complex $d \times d$ matrix, and two-qubit gates have a matrix of size $d^2 \times d^2$.

\subsection{Quantum Algorithms}

It seems that the quantum mechanical principle of superpositions could be used to speed up information processing a lot. Imagine a boolean function \[f: \{0,1\}^n \mapsto \{0, 1\}\] on an $n$-bit string. With a quantum computer, we could compute $f(x)$ for all $x \in \{0,1\}^n$ in parallel, thus leading to an amazing speed-up over any classical computer. More formally, given a quantum algorithm $U_f$ that computes $f$ reversibly and the input state 
\[\ket{\psi} = \sum_{x=0}^{2^n-1} \ket{x}\ket{0} = H^{\otimes n}\ket{0^{\otimes n}} \otimes \ket{0},\] we can compute
\[U_f \ket{\psi} = \sum_{x=0}^{2^n-1} \ket{x}\ket{f(x)}.\] However, when measuring the final state, we will end up with a single answer $\ket{x}\ket{f(x)}$ chosen uniformly at random with probability $2^{-n}$. This demonstrates that the speed-up achieved by quantum computers does not naively stem from the superposition principle. It rather stems from using interference effects between different states in a superposition to obtain a global property of a function. This abstract idea is what lies behind the quantum algorithm that factor large integers or solve the discrete logarithm problem efficiently.

There are two basic ingredients that give rise to the speed-up of quantum algorithms: the quantum Fourier transform (QFT) and amplitude amplification \cite{NielsenChuang2000}. The QFT is heavily used in Peter Shor's celebrated algorithm that factors $N = pq$ in time polynomial in $\log N$, as well as the discrete logarithm problem in any abelian group. We will only present amplitude amplification as this is the building block needed for the algorithms presented in this thesis.

\subsection{Amplitude Amplification}
\label{sec:amplitude amplification}


Suppose we are given an algorithm $A$ that acts on a Hilbert space $\Hilbert$ of $N$ qubits, including all work qubits. Let \[A\ket{0} = \ket{\psi} = \sum_{x \in \{0,1\}^N} \alpha_x \ket{x}.\] We can think of $A$ as an algorithm that tries to guess the correct output and succeeds with a reasonably high probability. Let $X_{\text{good}}$ denote the set of desired outputs $x$ and let $X_{\text{bad}}$ be its complement, the set of undesired outputs. Thus we can rewrite \[A\ket{0} = \sum_{x \in X_{\text{good}}} \alpha_x \ket{x} + \sum_{x \in X_{\text{bad}}} \alpha_x  \ket{x}.\] The success probability of $A$ is given by \[p_{\text{good}} = \sum_{x \in X_{\text{good}}} |\alpha_x|^2,\] which is the probability of measuring a state from the set of good states $X_{\text{good}}$. Conversely, let \[p_{\text{bad}} = \sum_{x \in X_{\text{bad}}}|\alpha_x|^2 = 1 -  p_{\text{good}}\] denote the probability of measuring a bad state. 

If $p_{\text{good}} = 0$, amplification is useless. If $p_{\text{good}} = 1$, the algorithm is already exact and amplification is not necessary. Therefore, amplitude amplification may be used if $0 < p_{\text{good}} < 1$. In that case, the good and bad components can be renormalized to
\bess
\ket{\psi_{\text{good}}} & = & \frac{1}{\sqrt{p_{\text{good}}}} \sum_{x \in X_{\text{good}}}|\alpha_x|^2 \text{ and } \\
\ket{\psi_{\text{bad}}} & = & \frac{1}{\sqrt{p_{\text{bad}}}} \sum_{x \in X_{\text{bad}}}|\alpha_x|^2,
\eess
such that
\[ A\ket{0} = \ket{\psi} = \sin \theta \ket{\psi_{\text{good}}} + \cos \theta \ket{\psi_{\text{bad}}},\] where $\theta = \arcsin \frac{1}{\sqrt{p_{\text{good}}}}$. Note that the imaginary parts are contained in $\ket{\psi_{\text{good}}}$ and $\ket{\psi_{\text{bad}}}$, therefore we can indeed use real coefficients $\sin \theta$ and $\cos \theta$, respectively.

Inspired by Grover's search algorithm \cite{NielsenChuang2000}, amplitude amplification uses two reflections in the two-dimensional plane spanned by $\ket{\psi_{\text{good}}}$ and $\ket{\psi_{\text{bad}}}$ to amplify the amplitude of the good state (\cite{Mosca1999} and references therein). First of all, given any quantum state $\ket{\phi} \in \Hilbert$, let $U_{\ket{\phi}}^{\perp}$ be a unitary such that $U_{\ket{\phi}}^{\perp} \ket{\phi} = \ket{\phi}$ and $U_{\ket{\phi}}^{\perp} \ket{\phi^{\perp}} = - \ket{\phi^{\perp}}$ for any state $\ket{\phi^{\perp}}$ that is orthogonal to $\ket{\phi}$. 

\begin{figure}[htbp]
	\centering
	 	\psfrag{theta}{$\theta$}
	 	\psfrag{psigood}{$\ket{\psi_{\text{good}}}$}
	 	\psfrag{psibad}{$\ket{\psi_{\text{bad}}}$}
	 	\psfrag{psi}{$\ket{\psi}$}
	 	\psfrag{psibar}{$\ket{\overline{\psi}}$}
		\includegraphics[width=0.45\textwidth]{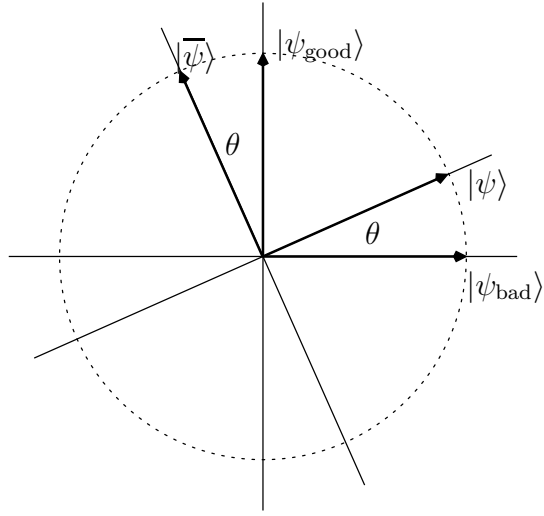}
	\caption{The Subspace of Good and Bad States in Amplitude Amplification.}
	\label{fig:ampl amplif subspace}
\end{figure}

Now define \[\ket{\overline{\psi}} = \cos \theta \ket{\psi_{\text{good}}} - \sin \theta \ket{\psi_{\text{bad}}}\] and observe that both $\{\ket{\psi}, \ket{\overline{\psi}}\}$ and $\{\ket{\psi_{\text{good}}}, \ket{\psi_{\text{bad}}}\}$ form an orthonormal basis for a two-dimensional subspace $\Hilbert_2 \subset \Hilbert$ shown in Figure \ref{fig:ampl amplif subspace}.

\begin{df} Define the amplitude amplification operator $Q = A U_{\ket{0}}^{\perp} A\dag U_{\text{bad}}^{\perp}$.
\end{df}

\begin{lemma} \label{lem:ampl aplif} The amplificaton operator rotates the input state $\ket{\psi}$ by an angle of $2 \theta$ in the two-dimensional subspace, i.e. \[Q \ket{\psi} = \sin (3\theta) \ket{\psi_{\text{good}}} + \cos (3\theta) \ket{\psi_{\text{bad}}}.\]
\end{lemma}
\begin{proof}
To show this, we first note that
\be \label{eqn:ampl amplif u_bad} U_{\text{bad}}^{\perp}\ket{\psi} = - \sin \theta \ket{\psi_{\text{good}}} + \cos \theta \ket{\psi_{\text{bad}}}.\ee Then we claim that $A U_{\ket{0}}^{\perp} A\dag = U_{A\ket{0}}^{\perp} = U_{\ket{\psi}}^{\perp}$. To see this, consider an arbitrary quantum state $\ket{\varphi} = \alpha \ket{\psi} + \beta \ket{\psi^{\perp}}$ for some state $\ket{\psi^{\perp}}$ orthogonal to $\ket{\psi}$ and complex numbers $\alpha, \beta$. Now
\bess
A U_{\ket{0}}^{\perp} A\dag \ket{\varphi} & = & A U_{\ket{0}}^{\perp} A\dag \left(\alpha \ket{\psi} + \beta \ket{\psi^{\perp}} \right) \\
& = & A U_{\ket{0}}^{\perp} \left( \alpha \ket{0} + A\dag \beta \ket{\psi^{\perp}} \right) \\
& = & A \left(\alpha \ket{0} - A\dag \beta \ket{\psi^{\perp}} \right) \\
& = & \alpha \ket{\psi} - \beta \ket{\psi^{\perp}} \\
& = & U_{A\ket{0}}^{\perp} \left( \alpha \ket{\psi} + \beta \ket{\psi^{\perp}} \right).
\eess

\begin{figure}[htbp]
	\centering
	 	\psfrag{theta}{$\theta$}
	 	\psfrag{2theta}{$2\theta$}
	 	\psfrag{psigood}{$\ket{\psi_{\text{good}}}$}
	 	\psfrag{psibad}{$\ket{\psi_{\text{bad}}}$}
	 	\psfrag{psi}{$U_{\text{bad}}^{\perp} \ket{\psi}$}
	 	\psfrag{Ufpsi}{$\ket{\psi}$}
	 	\psfrag{Qpsi}{$U_{\ket{\psi}}^{\perp}U_{\text{bad}}^{\perp}\ket{\psi}$}
		\includegraphics[width=0.45\textwidth]{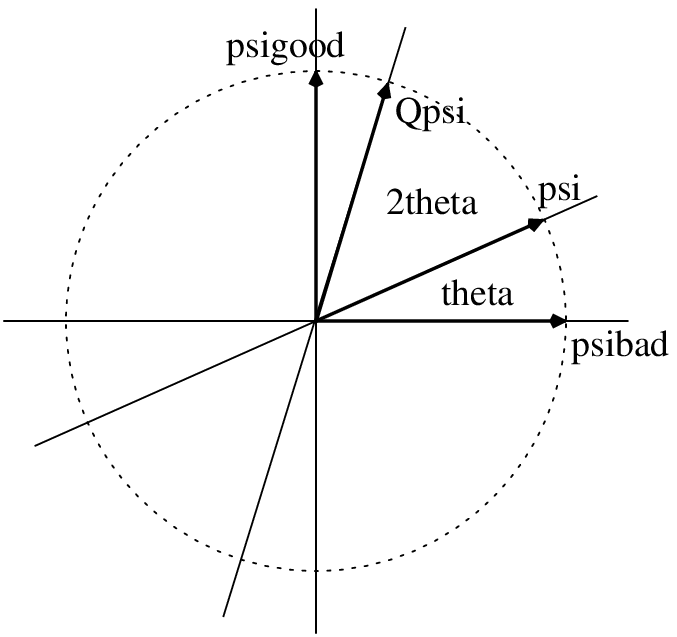}
		\hspace{1em}
		\includegraphics[width=0.45\textwidth]{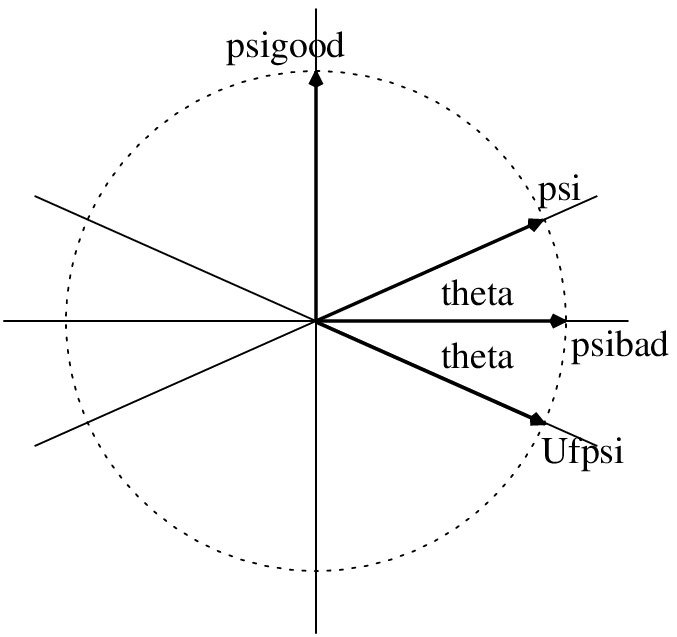}
	\caption{Amplitude Amplification as Two Reflections.}
	\label{fig:ampl amplif lemma}
\end{figure}

To conclude the proof, we rewrite the right-hand side of \Ref{eqn:ampl amplif u_bad} in the $\{\ket{\psi}, \ket{\overline{\psi}}\}$ basis:
\bess
U_{\text{bad}}^{\perp}\ket{\psi} & = & - \sin \theta \ket{\psi_{\text{good}}} + \cos \theta \ket{\psi_{\text{bad}}} \\
& = & \cos(2\theta) \ket{\psi} - \sin(2\theta) \ket{\overline{\psi}}.
\eess
Therefore
\bess
Q \ket{\psi} & = & U_{A\ket{0}}^{\perp} \left( \cos(2\theta) \ket{\psi} - \sin(2\theta) \ket{\overline{\psi}} \right) \\
& = & \cos(2\theta) \ket{\psi} + \sin(2\theta) \ket{\overline{\psi}} \\
& = & \cos (3\theta) \ket{\psi_{\text{good}}} + \cos (3\theta) \ket{\psi_{\text{bad}}}.
\eess
Figure \ref{fig:ampl amplif lemma} shows the geometrical interpretation of the action of $Q$.
\end{proof}

More generally, we can show that $Q$ rotates any input state in the subspace $\Hilbert_2$ by an angle of $2\theta$. We see that 
\[ U_{\text{bad}}^{\perp} \left(\sin \phi \ket{\psi_{\text{good}}} + \cos \phi \ket{\psi_{\text{bad}}} \right) = \left(-\sin \phi \ket{\psi_{\text{good}}} + \cos \phi \ket{\psi_{\text{bad}}} \right),\] hence $U_{\text{bad}}^{\perp}$ is a reflection in $\Hilbert_2$ about the axis defined by $\ket{\psi_{\text{bad}}}$. Analogously 
\[ U_{A\ket{0}}^{\perp} \left(\sin \phi \ket{\psi} + \cos \phi \ket{\overline{\psi}} \right) = \sin \phi \ket{\psi} - \cos \phi \ket{\overline{\psi}}\] is a reflection about the axis defined by $\ket{\psi}$. The main result of amplitude amplification now follows.

\begin{theorem}
The application of $k$ amplitude amplification rounds yields the final state \[Q^k\ket{\psi} = \sin ((2k+1)\theta) \ket{\psi_{\text{good}}} + \cos ((2k+1)\theta) \ket{\psi_{\text{bad}}}.\]
\end{theorem}

\clearpage
\thispagestyle{empty}
\cleardoublepage

\chapter{Noise in Quantum Computation}
\label{ch:noise}
\markright{Noise in Quantum Computation}

This chapter will introduce the concept of noise in quantum computing and the quantum operations formalism. The notion of the average fidelity is established and current procedures to measure this quantity are presented. Also, the important Decomposition Lemma \ref{lem:repn-of-superop} is proved.

\section{Noise in the Classical World}

We will start to look at noise in classical systems to establish an intuition for noise in quantum systems. Consider the simple example of a bit of information stored on the hard disk of a computer \cite{NielsenChuang2000}. Initially, this bit has the value $0$ or $1$. After a long period of time, the value of the bit can be corrupted by exposure to external magnetic fields and high temperatures. The easiest way to model this process is to assume a probability $p$ that the value of the bit is flipped.

\begin{figure}[htbp]
	\centering
		\psfrag{0}{$0$}
		\psfrag{1}{$1$}
		\psfrag{p}{$p$}
		\psfrag{1-p}{$1-p$}
		\includegraphics[width=0.50\textwidth]{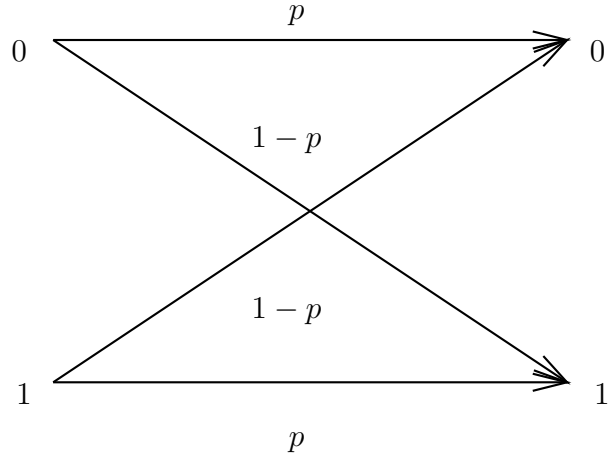}
	\caption{The Bit-Flip Error Model.}
	\label{fig:bitflipchannel}
\end{figure}

With probability $p$, the value of the bit changes from $0$ to $1$ and vice versa. With probability $1 - p$, the value of the bit remains unchanged. See Figure \ref{fig:bitflipchannel} for an illustration of that process. The value of $p$ can be estimated by sampling the external magnetic field surrounding the hard drive and the typical temperature distribution inside the computer. The value for $p$ can be derived from the sampling data by using physical models for the magnetic field and the effect of temperature on the bit.

To describe the general effect of the environment of our hard drive on the bit, we assume that we do not know its initial value exactly. We rather have knowledge about the distribution of the values of the bit. Let $p_0$ denote the probability that the initial state of the bit is $0$, and $p_1$ the corresponding probability for state $1$. The effect of the environment can now be modelled as a change of this probability distribution. Our model predicts that the probability that the bit is in final state $0$ after residing on the hard drive for a long time is $q_0 = p p_0 + (1-p) p_1$. Analogously for state $1$, we have $q_1 = (1-p) q_0 + p q_1$. If we write probability distributions as two-dimensional real column vectors, we can express the noise as a linear transformation on the probabilistic distribution of the bit's state: \be \label{eqn:noise_classical} \bv q_0 \\ q_1 \ev = \bv p & 1-p \\ 1-p & p \ev \bv p_0 \\ p_1 \ev.\ee

To see why it is useful to consider a probability distribution over the initial states of our bit, imagine a circuit that consists of two gates $A$ and $B$, both of which are noisy and either act correctly or flip the result. Although the input to the first gate is known exactly, we only know the probability distribution over the possible outcomes $0$ and $1$ after $A$ has been applied. In order to obtain information about the final state after $B$ is applied to that intermediate state, we need to consider gates as acting on probability distributions rather than definite input states.

We will make an important assumption about noise. We will assume that the noise affecting the second gate is independent from the noise affecting the first gate. This assumption turns out to be reasonable as the gates are usually physically separated in any implementation of that circuit. The assumption of the \emph{independence} of noise turns the circuit into a \emph{Markov process}. The circuit starts out with an initial bit $X$, produces an intermediate bit $Y$ and outputs a final bit $Z$. The probability distributions of the states of each two consecutive bits are linearly related by an equation similar to \ref{eqn:noise_classical}. The matrix is called the \emph{evolution matrix} and is required to be stochastic to map a probability distribution to another probability distribution. If we represent probability distributions as column vectors and the action of the evolution as left multiplication by the evolution matrix $M = (m_{i,j})$, then $M$ is stochastic if and only if $\sum_i m_{i,j} = 1$ for all $j$. In other words, the entries in each column of $M$ sum up to $1$.

\section{Noise in Quantum Computing}

The study of noise in quantum computing has identified different kinds of noise and provided a model to completely describe the effect of noise on a quantum system. We saw that classical noise is modelled using probability distributions over the classical states of the system. For quantum systems, we will use a similar approach. We will consider probability distributions over quantum states of the system, so that the concept of probability distribution is merged with the quantum mechanical principle of superpositions and complex amplitudes. It was shown that density operators can be used to completely describe probability distributions over quantum states and completely describe the statistics of any probability distribution over quantum states. We will now see how noisy quantum operations can be modelled to complete this picture. If not specified otherwise, $\Hilbert$ will denote the state space of the system in question.

\subsection{Quantum Operations Formalism}

The general evolution of the state of a quantum system can be described by a linear operator on the density operator of the system, which corresponds to the evolution matrix we have seen in the classical case. Analogous to the constraints on the evolution matrix of a classical system, we define quantum operations as the most general evolution of an open quantum system and we refer the reader to \cite[Ch.~8]{NielsenChuang2000} for an introduction to quantum operations and physical motivations.

\begin{df} \label{def:quantum operation} A \emph{quantum operation} is a linear operator \[\E: L({\Hilbert_A}) \rightarrow L({\Hilbert_B}), \rho' = \E(\rho) \] on the set of density operators on $\Hilbert$ such that
\begin{itemize}
	\item $\tr \E(\rho) = 1$.
  \item $\E$ is \emph{convex-linear}. Given a finite probability distribution $\{p_1, p_2, \dots, p_n\}$ over states $\rho_1$, \dots, $\rho_n$, \[\E \left( \sum_{i=1}^n p_i \rho_i \right) = \sum_{i=1}^n p_i \E(\rho_i).\]
  \item $\E$ is \emph{completely positive}. $\E(\sigma^A)$ must be positive for any positive operator $\sigma^A \in L({\Hilbert_A})$. Also, for any additional system $R$, $(\Id \otimes \E) (\sigma^{AR})$ must be positive for any positive operator $\sigma^{AR}$ on the joint system $AR$.
\end{itemize}
\end{df}

\begin{figure}[htbp]
	\centerline{
	\Qcircuit  @C=.7em @R=.7em
	{
  	 & \qw & \multigate{5}{\hspace{1em}U\hspace{1em}} & \qw & \qw \\
  	 \lstick{\rho} & \vdots & & \vdots & \rstick{\tr_{\text{env}} \left( U (\rho \otimes \rho_{\text{env}}) U\dag \right) = \E(\rho)}\\
  	 & \qw & \ghost{\hspace{1em}U\hspace{1em}} & \qw & \qw \\
  	 & \qw & \ghost{\hspace{1em}U\hspace{1em}} & \qw & \multimeasureD{2}{\tr_{\text{env}}}\\
  	 \lstick{\rho_{\text{env}}} & \vdots &  & \vdots & \\
  	 & \qw & \ghost{\hspace{1em}U\hspace{1em}} & \qw & \ghost{\tr_{\text{env}}}
	} }
	\caption{A Quantum Operation As Unitary Evolution in a Larger System}
	\label{fig:quantum op as unitary in larger system}
\end{figure}
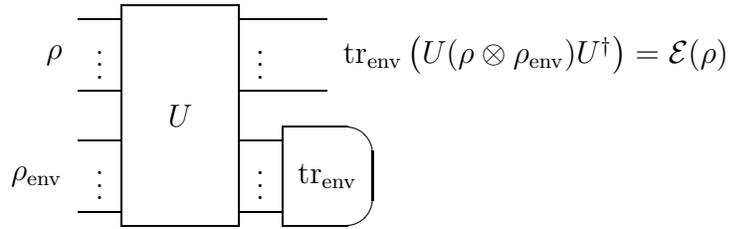

We recall that the dynamics of a closed quantum system are described by a unitary $U$, which has the corresponding quantum operation $\E(\rho) = U \rho U\dag$. However, we will have to deal with \emph{open} systems in general. In that setting the system is denoted the \emph{principal system} and is surrounded by an \emph{environment}. The environment includes everything that will interact with our principal system. Without loss of generality we may assume that the system and environment start out in a product state $\rho \otimes \rho_{\text{env}}$. As illustrated in Figure \ref{fig:quantum op as unitary in larger system}, the evolution of the joint system of principal system and the environment is unitary with some operator $U$. As we only regard the principal system, we have to trace out the environment after the interaction to get the final state of the principal system alone.

\begin{fact} The operation \[\E(\rho) = \tr_{\text{env}} \left( U (\rho \otimes \rho_{\text{env}}) U\dag \right)\] is a quantum operation. If the Hilbert space of the principal system had dimension $d$, it is sufficient to consider an environment of dimension $d^2$.
\end{fact}

\begin{fact} \label{fact:kraus decomp} Every quantum operation can be written in an \emph{operator-sum} or \emph{Kraus operator} representation \[\E(\rho) = \sum_{k=1}^{\leq d^2} A_k \rho A_k\dag\] where the $A_k$ are the \emph{operation elements} or \emph{Kraus operators} and are operators on the Hilbert space of the principal system. They satisfy the completeness condition \be \label{eqn:cond kraus operators} \sum_k  A_k\dag A_k = \Id. \ee The converse also holds: any such operator sum gives rise to a quantum operation.
\end{fact}

The Kraus operators reveal information about the structure of the noise as we will say in the following section. For that reason, determining the Kraus operators is an important goal for experimentalists \cite{WeinsteinEtAl2004}.


Sometimes non-trace-preserving quantum operations are considered. Then Definition \ref{def:quantum operation} is changed such that $0 \leq \tr \E(\rho) \leq 1$ and $\tr \E(\rho)$ is understood as the probability that the process represented by $\E$ occurs. The condition \Ref{eqn:cond kraus operators} on the Kraus operators changes to \[0 \leq \sum_k  A_k\dag A_k \leq \Id.\] Non-trace-preserving quantum operations occur when one distinguishes between measurement outcomes that occur in the middle of a process. In our model, the system-environment interaction could be described by a unitary evolution followed by a measurement $\{M_m\}$ and the quantum operations could be separated according to the outcome $m$ of the measurement. Then the operation $\E_m$ corresponding to outcome $m$ would not be trace-preserving. However, we typically do not distinguish between the outcomes of a possible measurement on the joint system-environment state and thus we only need to consider trace-preserving quantum operations.

Although not physically motivated, it will turn out to be of mathematical interest to consider general linear operators on $L(\Hilbert)$, which we will call superoperators later on. It seems to be easier to obtain certain results for this general setting and deduce them for quantum operations later on. We are interested in superoperators that can be described by up to $d^2$ Kraus operators $A_k$ which do not need to satisfy any constraints. These are called completely-positive superoperators.

\bfact Any set of up to $d^2$ operators $A_k \in L(\Hilbert)$ define a completely-positive superoperator
\[\E(\rho) = \sum_{k=1}^{\leq d^2} A_k \rho A_k\dag.\] The reverse also holds. Any completely-positive superoperator has a Kraus decomposition.
\efact

\subsection{Single Qubit Noise}
\label{sec:single qubit noise}

We will illustrate how the quantum operations formalism is useful in characterizing noise by showing how the different kinds of errors on a single-qubit system translate into the quantum operations formalism and how the Kraus operators reveal the structure of the noise. 

The \emph{bit-flip} channel is the quantum analog of the classical bit flip error. It has operation elements \[E_0 = \sqrt{p} \Id, E_1 = \sqrt{1-p} X,\] from which we see that the channel either acts as the identity with probability $p$, or as a NOT gate with probability $1-p$.

The \emph{phase-flip} channel randomly applies a certain phase with a fixed probability $1-p$. The operation elements are \[E_0 = \sqrt{p} \Id, E_1 = \sqrt{1-p} Z.\]

We can also model a combined phase and bit-flip channel, which gives a \emph{bit-phase flip} channel. It is characterized by its operation elements \[E_0 = \sqrt{p} \Id, E_1 = \sqrt{1-p} Y.\]

These examples show how the error model corresponds to the Kraus operators of a actual quantum operation implemented by a quantum computer. An even more interesting error model is the depolarizing channel, of which we will make heavy use later on. Although this is an error model that does not seem to reveal much information about the error at all, it will prove very useful. The \emph{depolarizing channel} is a channel that either sends the input state to the completely mixed state $\f{\Id}{2}$ with probability $p$, or leaves it untouched with probablity $1-p$. This channel is most naturally described as a quantum operation \[\E(\rho) = p \f{\Id}{2} + (1-p) \rho.\] To find its Kraus operator decomposition, we observe that \[\f{\Id}{2} = \f{\rho + X \rho X + Y \rho Y + Z \rho Z}{4}\] and thus \[\E(\rho) = \left(1 - \f{3p}{4}\right) \rho + \f{p}{4}(X \rho X + Y \rho Y + Z \rho Z).\] Therefore the operation elements are \[E_0 = \sqrt{1 - \f{3p}{4}} \Id, E_1 = \f{\sqrt{p}}{2} X, E_2 = \f{\sqrt{p}}{2} Y, \text{ and } E_3 = \f{\sqrt{p}}{2} Z.\] Parameterizing the channel in a slightly different way, we end up with \[\E(\rho) = (1-p)\rho + \f{p}{3}(X\rho X + Y \rho Y + Z \rho Z)\] and we can think of the channel as if it acts as the identity with probability $1-p$ and as a random Pauli gate with probability $p$.

Note that the depolarizing channel can be generalized to a $d$-dimensional system as well and reads \[\E(\rho) = p \f{\Id}{d} + (1-p) \rho.\] Note that we will consider a slightly more general notion later on.

\section{Measuring the Impact of the Noise}

Determining the structure of noise is necessary to design efficient error-correction schemes. We cannot go into the details of fault-tolerant quantum computing here, but refer to \cite{NielsenChuang2000} for an introduction to quantum error correction and fault-tolerant quantum computing. In this section, we will show how information about the structure of the noise can be revealed using current techniques. However, only one of them seems to be efficient as the number of required experiments for all other methods scales polynomial in the dimension $d = 2^N$ of the system Hilbert space $\Hilbert$, which is exponential in the number of qubits $N$. 

We will first describe how noise is assessed in general and proceed to methods that specifically determine a certain property of the noise.

\subsection{Quantum Process Tomography}
\label{sec:QPT}

Quantum process tomography is a combination of experimental and mathematical techniques to determine the elements of a matrix representation of a quantum operation $\E$ and/or the corresponding Kraus operators $A_k$. We will first introduce quantum state tomography, a prerequisite necessary to perform process tomography. See \cite{NielsenChuang2000} for a general description of quantum process tomography. \cite{Havel2003} provides the tools for converting between different representations of the quantum operation. For a description of an actual experimental determination of $\E$ of an implementation of the Quantum Fourier Transform, see \cite{WeinsteinEtAl2004}.


\subsubsection{State Tomography}
\label{sec:QST}

Quantum state tomography is a procedure to experimentally determine an unknown quantum state. Suppose we are given many copies of an unknown state $\rho$ and our task is to determine the matrix entries of $\rho$ in the computational basis. Note that it is essential to have many copies of $\rho$, as it is not possible to determine $\rho$ given just a single copy for it is not possible to distinguish non-orthogonal states.

We will first look at the case of a single qubit as it provides good insight into the general procedure. Pick an orthonormal basis for $L({\Hilbert})$, say $\{\OneOverSqrtTwo \Id, \OneOverSqrtTwo X, \OneOverSqrtTwo Y, \OneOverSqrtTwo Z\}$. Any density operator $\rho$ can be written as 
\bess \rho & = & (\OneOverSqrtTwo \Id, \rho) \OneOverSqrtTwo \Id + (\OneOverSqrtTwo X, \rho) \OneOverSqrtTwo X + (\OneOverSqrtTwo Y, \rho) \OneOverSqrtTwo Y + (\OneOverSqrtTwo Z, \rho) \OneOverSqrtTwo Z \\
& = & \frac{(\Id, \rho) \Id + (X, \rho) X + (Y, \rho) Y + (Z, \rho) Z}{2} \\
& = & \frac{ \Id  + \tr(X \rho) X + \tr (Y \rho) Y + \tr (Z \rho) Z}{2},
\eess
where we used the fact that the inner product on $L({\Hilbert})$ is the Hilbert-Schmidt or trace inner product, that the Pauli operators are self-adjoint, and that density operators have trace $1$. Quantum state tomography works because $\tr (A \rho)$ can be determined experimentally using a projective measurement of the observable $A$, which can be any Hermitian operator. It turns out that the Pauli operators are observables that are easy to measure for physical systems of interest. In general, any basis for $L({\Hilbert})$ comprised of easily measurable observables is sufficient.

Let $M$ be an observable of a von Neumann measurement with spectral decomposition $M = \sum_m m P_m$ with orthogonal projectors $P_m$. The expected value of a measurement of this observable on a state $\rho$ is given by 
\bess E(M) & = & \sum_m m p(m) = \sum_m m \tr (P_m\dag P_m \rho) = \sum_m \tr (m P_m \rho) = \tr M \rho \eess using that $P_m = P_m\dag$ and $P_m^2 = P_m$. The coefficients in the representation of $\rho$ in the Pauli basis for $L({\Hilbert})$ can be interpreted as expected values of the Pauli observables.

It is easy to estimate $\tr (X \rho)$, for example. Suppose we are given $k$ copies of $\rho$ and we measure the observable $X$ for each $\rho_i$. Given the spectral decomposition \[X = \ket{+}\bra{+} - \ket{-}\bra{-} \] we see that the outcomes of the experiments $x_i$ are $+1$ or $-1$. The average value of these $k$ experimental outcomes \[\bar{x} = \frac{1}{k} \sum_{i=1}^k x_i\] is a reasonable estimate for $\tr (X\rho)$. By the central limit theorem, we have that the random variable $\bar{x}$ is almost Gaussian distributed with expected value $\tr (X \rho)$ and standard deviation at most $\frac{1}{\sqrt{k}}$. Analogously, $\tr (Y \rho)$ and $\tr (Z \rho)$ can be determined within a desired confidence. One might of course use other standard statistical techniques to estimate the expected value of a random variable.

In the case of $N$ qubits one makes use of the following fact.
\bfact \label{fact:generalized Paulis}
The set \[\{ \frac{1}{\sqrt{2^N}} \sigma_{s_1} \otimes \sigma_{s_2} \otimes \dots \otimes \sigma_{s_N} \,|\, s_i \in \{I, X, Y, Z\}, 1 \leq i \leq N \} \] is an orthonormal basis for the space of linear operators on the $2^N$ dimensional Hilbert space $\Hilbert_{2^N}$ of $n$ qubits. We will call this basis the \emph{product basis} for the linear operators on $\Hilbert_{2^N}$. We will call the operators \emph{tensor product of Pauli operators}.
\efact
By measuring all $4^N$ observables according to the procedure layed out above, we can get complete knowledge of the state $\rho$ of an $N$ qubit system. 

\subsubsection{Process Tomography}

To determine the matrix elements of the quantum operation $\E$, we choose a basis for the space of linear operators on the state space of the system $\Hilbert$. Let $\Hilbert$ be a $d$-dimensional Hilbert space. As $\E$ is a linear operator on density operators on $\Hilbert$, it is completely characterized by its action on a basis of density operators. One possible basis is the product basis. However, in many experimental settings it seems to be more natural to use a different basis. Pick an orthonormal basis $\{\ket{\psi_1}, \dots, \ket{\psi_d}\}$ for $\Hilbert$, say the computational basis $\{\ket{0}, \ket{1}, \dots, \ket{d}\}$. Then the set of density operators $B_{\sigma} = \{ \sigma^{(i,j)} = \ket{i}\bra{j}: 1 \leq i, j \leq d\}$ forms an orthonormal basis for $L({\Hilbert})$. Prepare the input states $\sigma^{(i,j)}$ and determine the resulting state $\E\left(\sigma^{(i,j)}\right)$ using quantum state tomography. 

This gives us the matrix elements of $\E$ as a supermatrix. Using the orthonormal basis for linear operators on $\Hilbert$ introduced above, we can represent a density operator $\rho$ as a $d$-by-$d$ complex-valued matrix with matrix entries $\rho_{i,j}$ such that \[\rho = \sum_{i,j} \rho_{i,j} \sigma^{(i,j)}.\] However, we can also represent $\rho$ as a column vector \[\rho = \bv \rho_{1,1} \\ \rho_{1,2} \\ \vdots \\ \rho_{1,d} \\ \rho_{2,1} \\ \vdots \\ \rho_{d,d} \ev.\] Then $\E$ has a so-called supermatrix representation as a linear operator on a $d^2$-dimensional vector space of column vector representations of $\rho$. We will use the notation $\hat{\E}$ when we refer to this representation of the operator $\E$. Hence $\E$ can be represented as a $d^2$-by-$d^2$ \emph{supermatrix} $\hat{\E}$. The experimental setup outlined above will give the matrix elements of $\E$ in the supermatrix basis. Each of the input states $\sigma^{(i,j)}$ will yield a column $\hat{\E}_k$ of the supermatrix $\hat{\E} = (\hat{\E}_k)_{k=1}^{d^2}$ where we used a total order on the two-index structure $(i,j)$ to map it onto the single index $k$, say $(1,1), (1,2), \dots, (1,d), (2,1), \dots, (d,1), \dots, (d,d)$. It can be seen that this is exactly what gives the vector representation of $\rho$.

The supermatrix representation $\hat{\E}$ is not very convenient in the study of noise. For most applications, the Kraus operator representation is more useful as it reveals the structure of the noise \cite{NielsenChuang2000}, which we have seen in the examples for single-qubit noise in Section \ref{sec:single qubit noise}. Let $E^{(i,j)}$ be the matrix representation of $\sigma^{(i,j)}$ in the $B_{\sigma}$ basis, hence it is the matrix with a $1$ in the $i$-th row and $j$-th column and zeros everywhere else. 
\begin{df} The \emph{Choi} matrix associated to a supermatrix $\hat{\E}$ is the matrix \[\Xi = \sum_{i,j = 1}^d (E^{(i,j)} \otimes \Id_d) \hat{\E} (\Id_d \otimes E^{(i,j)}).\]
\end{df}
\begin{fact}\cite{Havel2003,WeinsteinEtAl2004} The Choi matrix $\Xi$ of a supermatrix $\hat{\E}$ is Hermitian with spectral decomposition \[\Xi = \sum_{k=1}^{d^2} \lambda_k \sigma_k \sigma_k\dag \] with all eigenvalues $\lambda_k > 0$ as $\E$ is completely-positive. Then the Kraus operator-sum representation of $\E$ is given by the Kraus operators \[A_k = \sqrt{\lambda_k} \sigma_k\] for $1 \leq k \leq d^2$.
\end{fact}

It is also possible to convert a Kraus operator-sum representation of a quantum operation $\E$ to the superoperator representation $\hat{\E}$.
\begin{fact}\cite{EmersonEtAl2005_Noise} The supermatrix representation of $\E$ with Kraus operators $\{A_k\}$ is given by \be \label{eqn:Kraus-to-Supermatrix} \hat{E} = \sum_k A_k^* \otimes A_k. \ee
\end{fact}

We will use $\E$ to denote the superoperator and $\hat{\E}$ to denote its representation as a supermatrix using some fixed basis for $L(\Hilbert)$ that will be clear from the context. Note that a superoperator is a linear operator on $L(\Hilbert)$ and hence is more general than a quantum operation, which is a special case of superoperators. In the following section, we will make use of superoperators to obtain more general results that will prove crucial later on.

\subsection{Noise Estimation Scenario}

There are two very common scenarios where noise estimation in quantum computing is important: quantum algorithms and quantum channels. We will introduce both settings and explain what noise estimation means in both contexts.

\subsubsection{Quantum Channel}

\begin{figure}[htbp]
	\centering
		\psfrag{E}{$\E$}
		\psfrag{rho}{$\rho$}
		\psfrag{Erho}{$\E(\rho)$}
		\includegraphics[width=0.50\textwidth]{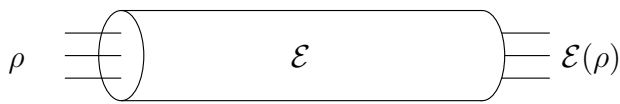}
	\caption{A Quantum Channel.}
	\label{fig:quantumchannel}
\end{figure}

From a theoretical point of view a quantum channel is a quantum gate that implements the identity transformation. However, a physical realisation of a quantum channel will usually be noisy and will implement a quantum operation that is not exactly the identity operation. We are interested in the ``distance'' between the identity and the operation the channel actually implements.

\subsubsection{Quantum Circuit}

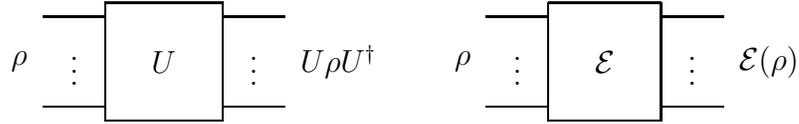
\begin{figure}[htbp]
	\centerline{
	\Qcircuit  @C=1em @R=1em
	{
 	 	& \qw & \multigate{2}{\hspace{1em}U\hspace{1em}} & \qw & \qw \\
  	\lstick{\rho} & \vdots &  & \vdots & \rstick{U \rho U\dag}  \\
	 	& \qw & \ghost{\hspace{1em}U\hspace{1em}} & \qw & \qw \\
	}
	\hspace{1in}
	\Qcircuit  @C=1em @R=1em
	{
 	 	& \qw & \multigate{2}{\hspace{1em}\E\hspace{1em}} & \qw & \qw \\
  	\lstick{\rho} & \vdots &  & \vdots & \rstick{\E(\rho)}  \\
	 	& \qw & \ghost{\hspace{1em}\E\hspace{1em}} & \qw & \qw \\
	} }
	\caption{An Ideal versus an Actual Implementation of a Quantum Algorithm $U$.}
	\label{fig:quantum circuit ideal vs actual}
\end{figure}

A quantum circuit on $N$ qubits is a unitary transformation $U$ in the $2^N$-dimensional Hilbert space $\Hilbert$. We are interested in how close a physical realization of a quantum computer implements $U$. We will call the physical implementation $\E$, where $\E$ is the actual quantum operation our quantum computer carries out when we try to implement $U$, as shown in Figure \ref{fig:quantum circuit ideal vs actual}. We will later see that we can always think of the implementation $\E$ as a perfect implementation of $U$ followed by a noisy quantum channel $\tilde{\E}$. 

\subsection{Distance Measures}

We are interested in the distance between a desired transformation $U$ and the actual operation $\E$. We have already seen that $U$ can be expressed as a quantum operation with exactly one Kraus operator, $U$ itself. We could employ the transformation \Ref{eqn:Kraus-to-Supermatrix} and try to find a distance measure between the supermatrix representations $U^* \otimes U$ and $\E$. However, it has been proven more useful to define a distance measure on density operators and characterize how noise affects a single output state. 

We will start with a metric that is derived from the Hilbert-Schmidt inner product on the space of linear operators on $\Hilbert$.

\begin{df} The \emph{trace distance} between quantum operators $\rho$ and $\sigma$ is defined as \[D(\rho, \sigma) = \ahalf \tr | \rho - \sigma |\] where $|A| = \sqrt{A\dag A}$ is the positive square root of $A\dag A$.
\end{df}

\begin{fact} The trace distance is a metric on $L({\Hilbert})$.
\end{fact}

Although this measure gives rise to a metric on the space of density operators, it is not typically used in the context of noise estimation. It is more common to use a measure called ``fidelity'' that also characterizes how similar two states are,   hence it will give rise to a real number between $0$ and $1$. It seems that the fidelity is more suitable for analysis and is thus preferred over the trace distance.

\begin{df} The \emph{Uhlmann fidelity} between two states $\rho$ and $\sigma$ is defined as \[F(\rho, \sigma) = \left( \tr \sqrt{\sqrt{\rho} \sigma \sqrt{\rho}} \right)^2 .\]
\end{df}

This measure does not give rise to a metric, but it is symmetric and turns into a simple expression when one of the states is pure. Let $\rho = \ket{\psi}\bra{\psi}$ and observe
\bess
F(\ket{\psi}\bra{\psi}, \sigma) & = & \left( \tr \sqrt{\sqrt{\ket{\psi}\bra{\psi}} \sigma \sqrt{\ket{\psi}\bra{\psi}}} \right)^2 \\ 
& = & \left( \tr \sqrt{\ket{\psi}\bra{\psi} \sigma} \right)^2 \\
& = & \left( \tr \sqrt{\bra{\psi} \sigma \ket{\psi}} \right)^2 \\
& = & \bra{\psi} \sigma \ket{\psi}
\eess
using the cyclic property of the trace. As unitary transformations $U$ map pure states to pure states, we have that \[F(U \ket{\psi}\bra{\psi} U\dag, \sigma) = \bra{\psi} U\dag \sigma U \ket{\psi}.\]

We can now define the fidelity of a quantum channel and the gate fidelity as the Uhlmann fidelity between the desired and the actual outcome of a channel or a gate, respectively. 
\begin{df} Let $\E$ denote the actual quantum operation representing a quantum channel. The \emph{channel fidelity} for an input state $\ket{\psi}$ is \[F\left(\ket{\psi}\bra{\psi}, \E(\ket{\psi}\bra{\psi}) \right) = \bra{\psi} \E(\ket{\psi}\bra{\psi}) \ket{\psi}.\]
\end{df}

\begin{df} Let $U$ be the unitary operator corresponding to a quantum gate. Let $\E$ denote the quantum operation of the actual implementation of $U$. The \emph{gate fidelity} for an input state $\ket{\psi}$ is \[F_{\ket{\psi}} (U, \E) = F\left(U \ket{\psi}\bra{\psi} U\dag, \E(\ket{\psi}\bra{\psi})\right) = \bra{\psi} U\dag \E(\ket{\psi}\bra{\psi}) U \ket{\psi}.\] We will denote the gate fidelity by $F_U(\ket{\psi})$ if $\E$ is clear from the context.
\end{df}

In order to simplify our discussion we will treat a quantum channel as a quantum algorithm that implements the identity transformation. From the following definitions and results for quantum algorithms one obtains the corresponding definitions and results for quantum channels by replacing the operation $U$ by the identity operation $\Id$. 

It is not very convenient to have the fidelity of a quantum channel or a quantum algorithm defined for a single state. There are two ways \cite{NielsenChuang2000} to proceed towards a fidelity measure that is independent of the input state. Analogous to the study of the worst-case and average case behaviour of algorithms in theoretical computer science, we will look at the minimum and average gate fidelities. The minimum gate fidelity corresponds to the worst-case behaviour of our implementation of a unitary $U$, whereas the average gate fidelity is a number associated to the average behaviour of our implementation. 

\begin{df} The \emph{minimum gate fidelity} is the minimum of the gate fidelity taken over all input states $\ket{\psi}$. Hence \[F_{\min}(U, \E) = \min_{\ket{\psi}} F_U (\ket{\psi}) = \min_{\ket{\psi}} \bra{\psi} U\dag \E(\ket{\psi}\bra{\psi}) U \ket{\psi}.\]
\end{df}

\begin{df}\label{def:avg fidelity}  The \emph{average gate fidelity} of a quantum algorithm $U$ with implementation $\E$ is defined as \[F_{\text{avg}}(U, \E) = \int_{\text{F-S}} F_{\ket{\psi}} (U, \E) d \ket{\psi} = \int_{\text{F-S}} \bra{\psi} U\dag \E(\ket{\psi}\bra{\psi}) U \ket{\psi} d\ket{\psi}\] where the integration is with respect to the Fubini-Study measure (Definition \ref{def:fubini-study}).
\end{df}

\subsection{Introduction to Fidelity Estimation}

The following sections and the main result in this thesis will be devoted to estimating the average gate fidelity.  Let $\E$ denote the quantum operation that represents the actual implementation of a quantum algorithm $U$. Let \[\E(\rho) = \sum_{k=1}^{d^2} A_k \rho A_k\dag.\] We can factor out $U$ from the Kraus operators to get a quantum operation $\tilde{\E}$ that does not depend on $U$, i.e.\@ an operation such that $\tilde{\E}(U\rho U\dag) = \E(\rho)$. We can think of $\tilde{\E}$ as the quantum operation that just characterizes the cumulative noise introduced by the implementation of $U$ and the experimental control. Let $E_k = A_k U\dag$ be the Kraus operators of $\tilde{\E}$ and observe
\bess 
\tilde{\E}(U \rho U \dag) & = & \sum_{k=1}^{d^2} E_k U \rho U\dag E_k\dag \\
&  = & \sum_{k=1}^{d^2} A_k U\dag U \rho U\dag U A_k\dag \\
& = & \sum_{k=1}^{d^2} A_k \rho A_k\dag = \E(\rho).
\eess
We will see later that the average fidelity will not depend on $U$ but only on the cumulative noise described by $\tilde{\E}$. 

\subsection{Fidelity Estimation using Quantum Process Tomography}
\label{sec:fid est random states}

For the easiest case, assume we already have all the matrix elements of $\tilde{\E}$. Now the average fidelity can be computed from that using a direct calculation \cite{EmersonEtAl2005_Noise}. We will show a general formula for averages over the Fubini-Study measure and derive an explicit formula for the average gate fidelity as a corollary. 

\begin{df} \label{def:unitarily inv} Define the representation $\hat{U}$ of $U \in U(d)$ on $L(\Hilbert)$ as $\hat{U} \rho = U \rho U \dag$ for all $\rho \in L(\Hilbert)$. Note that this is the usual action of $U(d)$ on density operators which we extend to all linear operators. Furthermore, we will call a superoperator $\Lambda$ unitarily invariant if $\hat{U} \hat{\Lambda} \hat{U}\dag = \hat{\Lambda}$ for all $U \in U(d)$. 
\end{df}

\begin{lemma} \label{lem:repn-of-superop} Let $\Lambda$ be a unitarily invariant superoperator and $X \in L(\Hilbert)$. Then \[\Lambda(X) = \frac{\tr \hat{\Lambda} - \frac{\tr \Lambda(\Id)}{d}}{d^2 - 1} \left(X - \tr X \frac{\Id}{d} \right) + \frac{\tr \Lambda(\Id)}{d} \tr X \frac{\Id}{d}.\] 
\end{lemma}
\begin{proof}
The representation $\hat{U}$ is reducible. Denote $M_d^0 \subset L(\Hilbert)$ the space of traceless linear operators, and let $M_d^1 = \{c \Id_d\,|\, c \in \C\} \subset L(\Hilbert)$ be the subspace of multiplies of the identity operator. There is no non-trivial subspace of $M_d^0$ that is invariant under $U(d)$ \cite{Boerner1967, Boerner1970} and $M_d^1$ is the smallest subspace that contains the identity. Hence both subspaces are irreducible. Observe that every linear operator $X \in L(\Hilbert)$ has a decomposition into a traceless part and a multiple of the identity: $X = (X - \frac{\tr X}{d} \Id) + \frac{\tr X}{d} \Id$, hence $L(\Hilbert) = M_d^0 \oplus M_d^1$. Hence the sets \[\mathcal{U}_0 = \{\hat{U}|_{M_d^0} \,|\, U \in U(d)\}\] and \[\mathcal{U}_1 = \{\hat{U}|_{M_d^1} \,|\, U \in U(d)\},\] with $X|_S$ meaning the restriction of $X$ onto the subspace $S$, are irreducible with respect to $L(\Hilbert)$.

$\Lambda$ is unitarily invariant and it follows that $\hat{\Lambda} \hat{U} = \hat{U} \hat{\Lambda}$ for all $U \in U(d)$. This commutation relation is also true for the operators restricted to the subspaces $M_d^0$ and $M_d^1$ of $L(\Hilbert)$. Schur's Lemma (Fact \ref{fact:schurs lemma}) implies that the restriction of $\Lambda$ onto each subspace is a multiple of the identity operator. Hence for $X \in L(\Hilbert)$, \[\Lambda(X) = p \left(X - \tr X \frac{\Id}{d} \right) + q \tr X \frac{\Id}{d} \] for complex numbers $p$ and $q$. These can be determined by evaluating the superoperator $\Lambda$ for certain operators. 
\bess
\Lambda(\Id) & = & p \left(\Id - \tr \Id \frac{\Id}{d} \right) + q \tr \Id \frac{\Id}{d} \\
& = & q \Id
\eess
This gives $q = \frac{\tr \Lambda(\Id)}{d}$. Now evaluate $\bra{i} \Lambda(\sigma^{(i,j)}) \ket{j}$ for the elements of the orthonormal basis $\sigma^{(i,j)}$ of $L(\Hilbert)$. 
\bess
\bra{i} \Lambda(\sigma^{(i,j)}) \ket{j} & = & p \bra{i} \left(\sigma^{(i,j)} - \delta_{i,j} \frac{\Id}{d} \right) \ket{j} + q \delta_{i,j} \frac{\Id}{d} \bracket{i}{j} \\
& = & p - \frac{\delta_{i,j}}{d} (p - q)
\eess
With the inner product $(X, Y) = \tr (X\dag Y)$ on $L(\Hilbert)$ and the cyclic property of the trace, we compute the value for $p$:
\bess 
\tr \hat{\Lambda} & = & \sum_{i,j = 1}^d \left(\sigma^{(i,j)}, \Lambda(\sigma^{(i,j)}) \right) \\
& = & \sum_{i,j = 1}^d \tr \left((\sigma^{(i,j)})\dag \Lambda(\sigma^{(i,j)}) \right)\\
& = & \sum_{i,j = 1}^d \tr \left(\ket{j}\bra{i} \Lambda(\sigma^{(i,j)}) \right) \\
& = & \sum_{i,j = 1}^d \bra{i} \Lambda(\sigma^{(i,j)}) \ket{j} \\
& = & \sum_{i,j = 1}^d p - \frac{\delta_{i,j}}{d} (p - q) \\
& = & d^2 p - p + q = (d^2 - 1)p + q
\eess 
\end{proof}

We will represent the previous lemma in a slightly re-arranged form to ease further calculations.

\begin{cor} \label{cor:repn-of-superop nicer} Let $\Lambda$ be a unitarily invariant superoperator and $X \in L(\Hilbert)$. Then \be \label{eqn:repn-of-superop nicer} \Lambda(X) = p X + q \tr X \frac{\Id}{d},\ee  where \[p = \frac{\tr \hat{\Lambda} - \frac{\tr \Lambda(\Id)}{d}}{d^2 - 1}\] and \[q = \frac{\tr \Lambda(\Id)}{d} - p.\]
\end{cor}

We can simplify this expression if we assume more structure on the superoperator.

\begin{cor} \label{cor:depol channel} A trace-preserving unitarily invariant quantum operation $\Lambda$ is a depolarizing channel \[\Lambda(\rho) = p \rho + (1 - p) \frac{\Id}{d}\] where \[p = \frac{\tr \hat{\Lambda} - 1}{d^2 - 1}.\]
\end{cor}
\begin{proof}
By rearranging terms in Lemma \Ref{lem:repn-of-superop}, we can see that $\Lambda$ is a depolarizing channel if $\Lambda$ is trace-preserving and restricted to density operators $\rho$. Using $\tr(\Lambda(\Id)) = \tr \Id = d$ and $\tr \rho = 1$,  \[\Lambda(\rho) = p \rho + (1 - p) \frac{\Id}{d}.\]
\end{proof}

We can now show that the average of a certain quartic function over the Fubini-Study measure can be explicitly calculated. We will need another lemma first to connect general superoperators to unitarily-invariant superoperators.

\begin{df} \label{def:haar twirling} Let $\Lambda$ be a superoperator on $L(\Hilbert)$. Define the \emph{twirled} superoperator \[\Lambda_T = \int_{U(d)} \hat{V} \hat{\Lambda} \hat{V}\dag d V\] where $\Lambda_T(X) = \int_{U(d)} V \Lambda( V\dag X V) V\dag dV$.
\end{df}

Now we can show that twirling leads to a unitarily invariant superoperator.

\begin{lemma} \label{lem:unitarily inv superop} Let $\Lambda$ be a superoperator on $L(\Hilbert)$. Then the twirled superoperator $\Lambda_T$ is unitarily invariant.
\end{lemma}
\begin{proof} Pick $U \in U(d)$. With the change of variables $V = U\dag V' $ and the invariance of the Haar measure on $U(d)$, we derive
\bess
(\hat{U} \Lambda_T \hat{U}\dag) (X) & = & \left(\hat{U} \int_{U(d)} \hat{V} \hat{\Lambda} \hat{V}\dag d V \hat{U}\dag \right) (X) \\
& = & \int_{U(d)} U V \Lambda(V\dag U\dag X U V) V\dag U\dag d V \\
& = & \int_{U(d)} V' \Lambda((V')\dag X V') (V')\dag dV \\
& = & \Lambda_T(X).
\eess
\end{proof}

\begin{theorem} \label{thm:Haar avg explicit} Let $M, N \in L(\Hilbert)$. Then 
\be \int_{\text{F-S}} \bra{\psi} M \ket{\psi}\bra{\psi} N \ket{\psi} d \ket{\psi} = \frac{1}{d(d+1)} (\tr MN + \tr M \tr N).
\ee
\end{theorem}
\begin{proof} Define the superoperator $\Lambda(X) = M X N$. We start with the unitary invariance of the Fubini-Study measure. It follows that we can replace integration over the set of all pure states by integration over the set of all unitary operators in $U(d)$. By the invariance of the Haar measure over $U(d)$, we can use any fixed pure state $\ket{\psi_0}$. 
\bess
\int_{\text{F-S}} \bra{\psi} \Lambda(\ket{\psi}\bra{\psi}) \ket{\psi} d \ket{\psi} & = & \int_{U(d)} \bra{\psi_0} V\dag \Lambda( V \ket{\psi_0}\bra{\psi_0} V\dag) V \ket{\psi_0} dV \\
& = & \int_{U(d)} \bra{\psi_0} V \Lambda( V\dag \ket{\psi_0}\bra{\psi_0} V) V\dag \ket{\psi_0} dV \\
& = & \bra{\psi_0} \int_{U(d)} V \Lambda( V\dag \ket{\psi_0}\bra{\psi_0} V) V\dag dV \ket{\psi_0} \\
& = & \bra{\psi_0} \Lambda_T (\ket{\psi_0}\bra{\psi_0}) \ket{\psi_0} 
\eess
where the second equation follows from the fact that the map $\dag: U(d) \rightarrow U(d)$ is an homeomorphism of $U(d)$ onto itself as $U(d)$ is a topological group.
Now we use the representation Lemma \ref{lem:repn-of-superop} and the unitary invariance of $\Lambda_T$ from Lemma \ref{lem:unitarily inv superop}. To directly apply Lemma \ref{lem:repn-of-superop}, we need to show that $\tr \hat{\Lambda}$ and $\tr \Lambda(\Id)$ are $U(d)$-invariant, i.e.\@ they are not changed by twirling. Observe that $\hat{U} = U \otimes U\dag$ is a unitary operator on $\Hilbert \otimes \Hilbert$. With the linearity of the integral and the linearity and the cyclic property of the trace we thus have that 
\bess
\tr \hat{\Lambda}_T & = & \tr \int_{U(d)} \hat{U} \hat{\Lambda} \hat{U}\dag dU 
 =  \int_{U(d)} \tr \hat{U} \hat{\Lambda} \hat{U}\dag dU \\
& = & \int_{U(d)} \tr \hat{\Lambda} \hat{U}\dag \hat{U} dU 
 = \int_{U(d)} \tr \hat{\Lambda} dU 
 = \tr \hat{\Lambda} \\
& = & \sum_{i,j} \tr M \ket{i}\bra{j} N \ket{j}\bra{i} 
 = \sum_{i,j} \tr \bra{i} M \ket{i}\bra{j} N \ket{j} 
 = \sum_i \bra{i} M \ket{i} \sum_j \bra{j} N \ket{j} \\
& = & \tr M \tr N.
\eess
Furthermore
\bess
\tr {\Lambda_T}(\Id) & = & \tr \int_{U(d)} U \Lambda(U\dag \Id U) U\dag dU 
 = \int_{U(d)} \tr U \Lambda(U\dag \Id U) U\dag dU \\
& = & \int_{U(d)} \tr \Lambda(\Id) U\dag U dU 
 = \int_{U(d)} \tr \Lambda(\Id) dU 
 = \tr \Lambda(\Id) \\
& = & \tr MN.
\eess
Therefore 
\bess
\int_{\text{F-S}} \bra{\psi} M \ket{\psi}\bra{\psi} N \ket{\psi} d \ket{\psi} & = & \bra{\psi_0} \Lambda_T(\ket{\psi_0}\bra{\psi_0}) \ket{\psi_0} \\
& = & \frac{\tr M \tr N + \frac{\tr MN}{d}}{d^2-1}\left(1-\f{1}{d}\right) + \frac{\tr MN}{d} \f{1}{d} \\
& = & \frac{1}{d(d+1)} (\tr MN + \tr M \tr N).
\eess
This finishes the proof.
\end{proof}

Using Theorem \ref{thm:Haar avg explicit}, the average gate fidelity can be calculated given the superoperator or Kraus operator representation of the actual implementation $\E$.

\begin{corollary}[\cite{EmersonEtAl2005_Noise}] \label{cor:avg-fidelity-explicit} The average fidelity of a trace-preserving quantum operation is 
\[F_{\text{avg}}(U, \E) = \frac{\sum_{k=1}^{d^2} |\tr (E_k)|^2 + d}{d^2 + d}\] where $E_k$ are the Kraus operators of $\tilde{\E}$ where $U$ was factored out.
\end{corollary}
\begin{proof}
By the unitary invariance of the Fubini-Study measure we can introduce the change of variables $\ket{\psi} = U \ket{\psi'}$.
\bess
F_{\text{avg}}(U, \E) & = & \int_{\text{F-S}} \bra{\psi} U\dag \E(\ket{\psi}\bra{\psi}) U \ket{\psi} d\ket{\psi} \\
& = & \int_{\text{F-S}} \bra{\psi'} \E(U \ket{\psi'}\bra{\psi'} U\dag) \ket{\psi'} d \ket{\psi'}\\
& = & \int_{\text{F-S}} \bra{\psi'} \tilde{\E}(\ket{\psi'}\bra{\psi'}) \ket{\psi'} d \ket{\psi'}
\eess
This shows that the average gate fidelity does not depend on $U$ but rather on the cumulative noise $\tilde{\E}$ introduced by the implementation of $U$ and the overall experimental control. By Theorem \ref{thm:Haar avg explicit} and the linearity of the integral, we can rewrite the average gate fidelity for the trace-preserving noise operation $\tilde{\E}$ using its Kraus operator decomposition as follows.
\bess
F_{\text{avg}}(\tilde{\E}) & = & \int_{\text{F-S}} \bra{\psi'} \tilde{\E}(\ket{\psi'}\bra{\psi'}) \ket{\psi'} d \ket{\psi'} \\
& = & \sum_{k=1}^{d^2} \frac{1}{d(d+1)}(\tr E_k E_k\dag + \tr E_k \tr E_k\dag) \\
& = & \frac{1}{d(d+1)} \left( \tr \left(\sum_{k=1}^{d^2} E_k E_k\dag\right) + \sum_{k=1}^{d^2} \tr E_k \overline{\tr E_k} \right) \\
& = & \frac{1}{d(d+1)} \left( \tr \Id +  \sum_{k=1}^{d^2} \left|\tr E_k \right|^2 \right) \\
& = & \frac{\sum_{k=1}^{d^2} \left|\tr E_k \right|^2 + d}{d(d+1)}
\eess
\end{proof}

Although Corollary \ref{cor:avg-fidelity-explicit} provides an easy way to compute the average gate fidelity, the Kraus operator or superoperator representation of the cumulative noise operation $\tilde{\E}$ needs to be known. This will in general require quantum process tomography to be conducted. We will provide two alternate approaches to estimating the average fidelity that do not require explicit knowledge of $\tilde{\E}$. 

\subsection{Fidelity Estimation using Quantum State Tomography}
\label{sec:avg fidelity using entanglement fidelity}

The straightforward method to get complete knowledge of a quantum operation $\E$ is to perform quantum process tomography. However, this requires quantum state tomography on $d^2$ states, which requires of order $d^4$ experiments. Extending an earlier observation for a single qubit by Bowdrey et al.\@ \cite{BowdreyEtAl2002}, Nielsen \cite{Nielsen2002} showed how to estimate the average gate fidelity using quantum state tomography on fewer states. We will assume that we have a trace-preserving quantum operation $\E$.

\subsubsection{Single Qubit Case}

\cite{BowdreyEtAl2002} analytically evaluated the average gate fidelity for a single qubit. They described the average fidelity as a sum over four mixed states which seem to arise naturally in certain experimental setups:
\be
F_{\text{avg}}(U, \E) = \ahalf + \frac{1}{3} \sum_{j \in \{x, y, z\}} \tr \left( U \frac{\sigma_j}{2} U\dag \E\left( \frac{\sigma_j}{2} \right) \right)
\ee
They also gave a characterization using the six pure states corresponding to the axes of the Bloch sphere (see \Ref{sec:bloch sphere}), which we will denote by $\rho_{\pm x}, \rho_{\pm y}, \rho_{\pm z}$:
\be
F_{\text{avg}}(U, \E) = \frac{1}{6} \sum_{j \in \{\pm x, \pm y, \pm z\}} \tr \left( U \rho_j U\dag \E(\rho_j) \right)
\ee

In both cases the states $\E\left(\frac{\sigma_j}{2}\right)$ and $\E\left(\rho_j\right)$ need to be determined experimentally using quantum state tomography.

\subsubsection{General Case}

\cite{Horodecki1999} showed a relationship between the so-called entanglement fidelity and the average gate fidelity. 

\begin{df} Let $\E$ be a quantum operation on a Hilbert space $\Hilbert$ of a system $R$. Now consider a second system $Q$ with the same state space. Let $\rho$ be a maximally entangled state on $RQ$. The \emph{entanglement fidelity} of $\E$ is defined as \[F_e(\E) = \left(\rho, (\Id \otimes \E)(\rho) \right) = \tr \left( \rho\dag (\Id \otimes \E)(\rho) \right).\]
\end{df}

The entanglement fidelity is a measure of how well entanglement is preserved by the operation $\E$. The definition is sound as all maximally entangled states are related by a unitary transformation on $R$ which does not change the value of $F_e(\E)$. 

\begin{theorem} \label{thm:avg-fidelity and ent fidelity} Let $\E$ be a trace-preserving quantum operation and let $U$ be a unitary operator. The following relationship holds between  the average gate fidelity and the entanglement fidelity:
\be \label{eqn:avg-fidelity and ent fidelity} F_{\text{avg}}(\E, U) = \frac{d F_e(\E \hat{U}\dag) + 1}{d + 1}\ee
\end{theorem}
\begin{proof}
We will consider the case of a quantum channel first, i.e.\@ $U = \Id$. Furthermore, we consider a depolarizing channel $\E_D$ with channel parameter $p$. We can show \Ref{eqn:avg-fidelity and ent fidelity} directly.
\bes
\nn F_{\text{avg}}(\E_D, \Id) & = & \int_{F-S} \bra{\psi} \E_D(\ket{\psi}\bra{\psi}) \ket{\psi} d \ket{\psi} \\
\nn & = & \int_{F-S} \bra{\psi} \left( p \ket{\psi}\bra{\psi} + (1-p) \frac{\Id}{d} \right) \ket{\psi} d \ket{\psi}  \\
\nn & = & \int_{F-S} p + (1-p) \frac{1}{d} d \ket{\psi} \\
\label{eqn:fidelity depol channel} & = & p + (1-p) \frac{1}{d}.
\ees
Using the maximally entangled state $\ket{\phi} = \sum_{x=1}^d \ket{x}\ket{x}$, we can compute the entanglement fidelity of the depolarizing channel as follows.
\bes
\nn F_{\text{ent}}(\E_D) & = & \frac{1}{\sqrt{d}} \sum_{w=1}^d \bra{w} \otimes \bra{w} \left( (\Id \otimes \E_D)\left( \frac{1}{d} \sum_{x,y =1}^d (\ket{x} \otimes \ket{x}) (\bra{y} \otimes \bra{y}) \right) \right) \frac{1}{\sqrt{d}} \sum_{z=1}^d \ket{z} \otimes \ket{z} \\
\nn & = & \frac{1}{d^2} \sum_{w,x,y,z = 1}^d \bra{w} \otimes \bra{w} \left( (\ket{x} \bra{y}) \otimes (p \ket{x}\bra{y} + (1-p) \frac{\Id}{d} )  \right) \\
\nn & = & \frac{1}{d^2} \sum_{w,x,y,z = 1}^d \delta_{w,x} \delta{y,z} \left( p \delta_{w,x} \delta{y,z} + (1-p) \frac{1}{d} \delta_{w,z} \right) \\
\nn & = & \frac{1}{d^2} \left( d^2 p + (1 - p) \right) \\
\label{eqn:ent fidelity depol channel}& = & p + (1-p) \frac{1}{d^2}
\ees
From the explicit formulas \Ref{eqn:fidelity depol channel} and \Ref{eqn:ent fidelity depol channel}, \[F_{\text{avg}}(\E_D, \Id) = \frac{d F_e(\E_D) + 1}{d + 1}\] follows.

Using the technique of twirling from Definition \ref{def:haar twirling}, we can extend this proof from depolarizing channels to general channels. From Lemma \ref{lem:unitarily inv superop}, it follows that $\E_T$ is unitarily invariant and Corollary \Ref{cor:depol channel} shows that $\E_T$ is a depolarizing channel. It remains to show that the average channel fidelity is invariant under twirling:
\bess
F_{\text{avg}}(\E_T, \Id) & = & \int_{F-S} \bra{\psi} \dag \left( \int_{U(d)} U\dag \E(U \ket{\psi} \bra{\psi} U\dag) U dU \right) \ket{\psi} d \ket{\psi} \\
& = & \int_{U(d)} \int_{F-S} \bra{\psi} U\dag \E(U \ket{\psi'} \bra{\psi'} U U\dag) U \ket{\psi} dU d \ket{\psi} \\
& = & \int_{U(d)} \int_{F-S} \bra{\psi'} \E(\dag \ket{\psi'} \bra{\psi'}) \ket{\psi'} d\ket{\psi'} dU \\
& = & \int_{U(d)} F_{\text{avg}}(\Id, \E) dU \\
& = & F_{\text{avg}}(\Id, \E).
\eess
We have substituted $\ket{\psi'} = U \ket{\psi}$ and used the unitary invariance of the Fubini-Study measure. For the entanglement fidelity, we use the fact that $\ket{\phi'} = U \ket{\phi} $ is also a maximally entangled state for any unitary transformation $U$. Therefore
\bess
F_e(\E_T) & = & \bra{\phi} \int_{U(d)} U\dag \E(U \ket{\psi} \bra{\psi} U\dag) U dU \ket{\phi} \\
& = & \int_{U(d)} \bra{\phi'} \E(\ket{\psi'} \bra{\psi'}) \ket{\phi'} dU \\
& = & \E(\ket{\psi'} \bra{\psi'}) \ket{\phi'} = F_e(\E).
\eess

Thus \Ref{eqn:avg-fidelity and ent fidelity} holds for general channels as well. It is extended to the gate fidelity using 
\bess 
F_{\text{avg}}(U, \E) & = & \int_{F-S} \bra{\psi} U\dag \E(\ket{\psi}\bra{\psi}) U \ket{\psi} d {\psi} \\
& = & \int_{F-S} \bra{\psi'} U U\dag \E(U\dag \ket{\psi} \bra{\psi} U) \ket{\psi} d \ket{\psi} \\
& = & F_{\text{avg}}(\Id, \E \hat{U}\dag)
\eess
where we have substituted $\ket{\psi} = U\dag \ket{\psi'}$.
\end{proof}

\cite{Nielsen2002} utilized this connection and showed how to calculate the entanglement fidelity experimentally using state tomography on $\E(\sigma^{(i,j)})$ for a set of $d^2$ states $\sigma^{(i,j)}$ that form an operator basis. This yields the entanglement fidelity using $O(d^4)$ experiments and classical post-processing of $d^2 \times d^2$ complex matrices. Using \Ref{eqn:avg-fidelity and ent fidelity}, we can subsequently compute the average fidelity.

\subsection{Fidelity Estimation using Random States}
\label{sec:random circuits}


\subsubsection{Introduction}

\begin{figure}[htbp]
	\centerline{
	\Qcircuit @C=1em @R=1em
	{
  	\lstick{\ket{0}} & \gate{V} & \gate{U} & \gate{U\dag} & \gate{V\dag} & \meter & \cw
	} }
	\caption[Circuit to Estimate the Average Fidelity using Random Unitaries]{Circuit to Estimate the Average Fidelity using Random Unitaries $V \in_R U(d)$}
	\label{fig:haar random estimation circuit}
\end{figure}
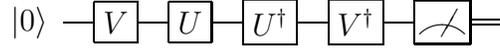

Another approach to calculating the average fidelity is to sample over uniformly distributed random quantum states. Given the definiton of the average fidelity in Definition \ref{def:avg fidelity}, Emerson et al.\@ \cite{EmersonEtAl2005_Noise} suggested a ``motion-reversal'' experiment shown in Figure \ref{fig:haar random estimation circuit}. As shown in Corollary \ref{cor:avg-fidelity-explicit}, the average fidelity does not depend on the actual algorithm $U$ in question. It only depends on the cumulative noise $\tilde{\E}$. Assuming that noise introduced by reversing $U$ will not increase the fidelity as the motion reversal does not cancel out noise, we will get a lower bound for the fidelity by implementing $U U\dag = \Id$.

\begin{theorem} Let $p$ denote the probability to measure zero at the end of the estimation circuit. Then \[p = F_{\text{avg}}(\tilde{\E}).\]
\end{theorem}
\begin{proof}
The theorem follows directly from the unitary invariance of the Fubini-Study measure. We see that
\bess
p & = & \int_{U(d)} \bra{0} V U U\dag \tilde{\E}(U\dag U V \ket{0} \bra{0} V\dag U\dag U) V \ket{\psi} d V \\
& = & \int_{\text{F-S}} \bra{\psi} \tilde{\E}(\ket{\psi}\bra{\psi}) \ket{\psi} d \ket{\psi} \\
& = & F_{\text{avg}}(\tilde{\E}).
\eess
\end{proof}

$p$ can be estimated up to an arbitrary precision using standard tools from statistics as seen in Section \ref{sec:QST}. Although this approach seems promising and would drastically reduce the amount of classical postprocessing that is needed for the other approaches, it requires the implementation of random circuits. It is known \cite{NielsenChuang2000} for the case of $d = 2^N$ that the decomposition of most unitary operators in $V \in U(d)$ requires \be \label{eqn:arbitrary state hard to approx} \Omega\left( \f{d \log (1/\epsilon)}{\log\log d} \right)\ee one and two-qubit gates to approximate to within $\epsilon$ in the $2$-norm for linear operators. Thus most unitary operators cannot be efficiently realized on a quantum computer. Therefore, uniformly distributed random unitaries are generally not feasible practically.

\cite{EmersonEtAl2005_RandomUnitaries} presents an approach to efficiently approximate Haar-distributed unitaries. The idea is to start with a probability distribution $f$ over a subset of $S \subset U(d)$ that either generates the full set $U(d)$ or a dense subset. In the first case, $S$ will be continuously parametrized, whereas a discrete gate set $S$ will be sufficient in the second case. The key idea is to choose a gate $V_i \in S$ according to the distribution $f$ for each step $i = 1, 2, \dots, m$. Then the resulting probability distribution over $V = \prod_{i=1}^m V_i$ will converge to the Haar measure either uniformly or in the weak topology according to a given test function. See sections \ref{sec:rep theory} and \ref{sec:fourier analysis} for an introduction to Fourier analysis on compact groups. For the remainder of this section, let $G$ denote the compact topological group $U(d)$ with elements $g \in G$. 

\subsubsection{Exponential Uniform Convergence to the Haar Measure}

Let $\mu_f \in M(G)$ be an absolutely continuous probability measure on $G$ over a subset $S \subset G$ that generates $G$. This enables us to consider $\mu_f$ both as a measure and as a function $f \in L^1(G)$. We will further restrict ourselves to nice positive-definite $f$ (see Definition \ref{def:nice pd}), such that we do not need to worry about the pointwise convergence of its Fourier series. For convenience, we will consider the function $f$ for the remainder of this section, where $f$ is the probability distribution of a single element $g \in G$. If we pick two elements $g_1, g_2 \in G$ independently according to $f$, the distribution of $g = g_1 g_2$ is given by the convolution of $f$ with itself. Thus, the distribution over random circuits that consist of two gates which were chosen indepently according to $f$ is given by $f \star f$. We will repeat this process $m$ times and have that the group elements $g = g_1 g_2 \dots g_m \in G$ are distributed according to \[f^{\star m} = \underbrace{f \star f \star \dots \star f}_{m}.\]

In order to show uniform convergence of $f^{\star m}$ to the Haar measure on $G$, we need two technical lemmas.

\begin{lemma} \label{lem:fourier coeff haar} For the Haar measure, we have the Fourier coefficients \[\hat{m}(D^s) = \begin{cases}1 & s = 0 \\ 0 & s \geq 1 \end{cases}.\]
\end{lemma}
\begin{proof}
Using the unitary invariance of the Haar measure, we observe that
\bess
\hat{m}(D^s) & = & \int_G D^s(g) dg = \int_G D^s(hg) dg = \int_G D^s(h)D^s(g) dg \\
& = & D^s(h) \int_G D^s(g) dg
\eess
for arbitrary $h \in G$. It follows that $D^s(h) \hat{m}(D^s) = \hat{m}(D^s)$ for all $h \in G$, which implies $D^s(h) = \Id$ for all $h \in G$ or $\hat{m}(D^s) = 0$. This implies $\hat{m}(D^0) = 1$ and $\hat{m}(D^s) = 0$ for $s \geq 1$ as $D^s(h) \neq \Id$ provided $h \neq e \in G$.
\end{proof}

\begin{lemma} \label{lem:rand circ uniform conv}  $\left\|\hat{f}(D^s)\right\|_2 < 1$ for $s \geq 1$. $\hat{f}(D^0) = \left\|\hat{f}(D^0)\right\|_2 = 1$.
\end{lemma}
\begin{proof}
The case $s = 0$ follows immediately from \[\left\| \hat{f}(D^0) \right\|_2 = \left\| \int_G f(g) dg \right\|_2 = 1\] as $f$ is a probability measure. For the case $s \geq 1$, let $\v{x} \in \Hilbert_{d_s}$ be a vector in the representation Hilbert space for the $s$ representation. Now
\bess
\left\| \hat{f}(D^s) \v{x} \right\|_2 & = & \left\| \int_G f(g) D^s(g) dg \v{x} \right\|_2 \\
& \leq & \int_G  \left\| f(g) D^s(g) \v{x} \right\|_2 dg \\
& = & \int_G  f(g) \left\|D^s(g) \v{x} \right\|_2 dg \\
& = & \int_G  f(g) \left\|\v{x}\right\|_2 dg \\
& = & \int_G  f(g) dg \left\|\v{x}\right\|_2 \\
& = & \left\|\v{x}\right\|_2
\eess
where we used that $f$ is a probability measure and that $D^s$ is a unitary representation of $G$. To show that $\left\| \hat{f}(D^s) \v{x} \right\|_2 < 1$, we assume \[\left\| \int_G f(g) D^s(g) dg \v{x} \right\|_2 = \int_G  \left\| f(g) D^s(g) \v{x} \right\|_2 dg.\] It follows from the triangle inequality of the norm and the linearity of the integral that there is a vector $\v{y} \in \Hilbert_{d_s}$ such that for all $g \in G: D^s(g) \v{x} = \xi(g) \v{y}$ for $\xi \in C(G)$ a continuous bounded complex-valued function on $G$. This implies that $\xi$ is a one-dimensional representation embedded in the irreducible representation $D^s$ of dimension $d_s > 1$. Contradiction and the lemma follows.
\end{proof}

First we note that $f^{\star m}$ converges uniformly to the Haar measure over $G$ if $f^{\star m}$ converges uniformly to the constant function $1 \in L^1(G)$. However, we do not just consider convergence with respect to the $L^1$-norm, but rather pointwise uniform convergence. To analyse the convergence, we will consider the Fourier transform of $f^{\star m}$ as the convolution of two functions $\phi, \psi \in L^1(G)$ turns into a simple product in its Fourier representations \[\what{\phi \star \psi}(D^s) = \what{\phi}(D^s) \what{\psi}(D^s).\] We have \[\what{f^{\star m}}(D^s) = \left(\hat{f}(D^s)\right)^m,\] which is an $m$-fold product of $d_s \times d_s$ complex matrices. Lemmas \ref{lem:rand circ uniform conv} and \ref{lem:fourier coeff haar} already show that the Fourier coefficients of $f^{\star m}$ converge to the Fourier coefficients of the Haar measure as $m$ approaches infinity. However, this does not prove \emph{uniform} convergence yet. We need to show that the Fourier approach is valid and that we have uniform convergence indeed.

\begin{theorem} The probability measure $f^{\star m}$ converges uniformly to the Haar measure over $G$.
\end{theorem}
\begin{proof}
We note that $f^{\star m}$ is nice positive-definite if $f$ is. Therefore for any $m$, the Fourier series of $f^{\star m}$ converges pointwise \cite{Edwards1972}, where the limit is taken over finite subsets $P \subset \hat{G}$ of irreducible representations of $G$. 


Uniform convergence is understood in the $L^{\infty}$ norm meaning that for almost all $g \in G$,
\[ \lim_{m \rightarrow \infty} f^{\star m}(g) = 1\] where the limit is uniform, i.e.\@ we want that for any $\epsilon > 0$ there is a maximum number of convolutions $M$ such that for almost all $g \in G$ and all $m \geq M$:
\[ | f^{\star m}(g) - 1| < \epsilon.\] In other words, we want that 
\[ \lim_{m \rightarrow \infty} \| f^{\star m} - 1\|_{\infty} = 0.\]

In order to get this uniform convergence, we will consider the convergence of the Fourier coefficients of $f^{\star m}$ and show that we can restrict ourselves to a finite number of Fourier coefficients $\what{f^{\star m}}(D^s)$. The Peter-Weyl approximation from Fact \ref{fact:peter weyl approx} guarantees that for any real $\epsilon > 0$ there is an $N_{\epsilon}$ such that for almost all $g \in G$,
\[ \left| f(g) - \sum_{s=0}^{N_{\epsilon}} d_s \tr \hat{f}(D^s) D^s(g)\dag \right| < \epsilon.\] This ``representation cut-off'' \cite{EmersonEtAl2005_RandomUnitaries} enables us to consider the truncated function \[ f_{N_{\epsilon}}(g) = \sum_{s \leq N_{\epsilon}} d_s \tr \hat{f}(D^s) D^s(g)\dag\] for further analysis. For almost all $g \in G$ and $m \geq 2$, we have from the triangle inequality and the Peter-Weyl approximation that
\bes
\nn |f^{\star m}(g) - 1|  & \leq & |f^{\star m}(g) - f^{\star m}_{N_{\epsilon}}(g)| + |f^{\star m}_{N_{\epsilon}}(g) - 1| \\
\nn & \leq & \left|f^{\star m}(g) - f^{\star m}_{N_{\epsilon}}(g)\right| + \left| \left( \sum_{s=0}^{N_{\epsilon}} d_s \tr \what{f^{\star m}}(D^s) D^s(g)\dag\right) - 1\right| \\
\label{eqn:rand circ norm to bound} & = & \left|f^{\star m}(g) - f^{\star m}_{N_{\epsilon}}(g)\right| + \left| \sum_{s=1}^{N_{\epsilon}} d_s \tr \what{f^{\star m}}(D^s) D^s(g)\dag \right|
\ees
where the last line follows from the case $s = 0$ in Lemma \ref{lem:rand circ uniform conv}.

To bound the first term we claim that $f_{N_{\epsilon}} \star f_{N_{\epsilon}} = f \star f_{N_{\epsilon}}$ almost everywhere. Using the Uniqueness Theorem (see Fact \ref{fact:uniqueness thm}), it suffices to show that their Fourier coefficients are equal. Let $D_s \in \hat{G}$ and observe
\[ \what{f_{N_{\epsilon}} \star f_{N_{\epsilon}}}(D^s) = \hat{f}_{N_{\epsilon}}(D^s)^2 = 
\begin{cases} \hat{f}(D^s)^2 & s \leq N_{\epsilon} \\ 0 & s > N_{\epsilon} \end{cases}\] and
\[ \what{f \star f_{N_{\epsilon}}}(D^s) = \hat{f}\hat{f}_{N_{\epsilon}}(D^s) = 
\begin{cases} \hat{f}(D^s)^2 & s \leq N_{\epsilon} \\ 0 & s > N_{\epsilon} \end{cases}\] and the claim follows. 

Therefore for almost all $g \in G$,
\bess
\left|f^{\star m}(g) - f^{\star m}_{N_{\epsilon}}(g)\right| & = & \left|f(g)\left(f^{\star (m-1)}(g) - f^{\star (m-1)}_{N_{\epsilon}}\right)(g) \right| \\
& \leq & \left| \int_G f(h) \left(f^{\star (m-1)}(g) - f^{\star (m-1)}_{N_{\epsilon}}\right)(g^{-1}h) dh \right| \\
& \leq & \int_G \left| f(h) \left(f^{\star (m-1)}(g) - f^{\star (m-1)}_{N_{\epsilon}}\right)(g^{-1}h) \right| dh \\
& \leq & \int_G f(h) \left\| f^{\star (m-1)} - f^{\star (m-1)}_{N_{\epsilon}} \right\|_{\infty} dh \\
& = & \left\| f^{\star (m-1)} - f^{\star (m-1)}_{N_{\epsilon}} \right\|_{\infty}
\eess
where we used that $f$ is a positive function with integral $1$ and that the convolution is associative. It follows by induction that
\[ \left\| f^{\star m} - f^{\star m}_{N_{\epsilon}} \right\|_{\infty} \leq  \left\| f - f_{N_{\epsilon}} \right\|_{\infty} \leq \epsilon.\]

The second term in \Ref{eqn:rand circ norm to bound} can be bound using the notation $D^s(g)_j\dag$ to denote the single-column matrix that consists of the complex-conjugates and transposed $j$-th row of $D^s(g)$. We have
\bes
\nn \left| \tr \what{f^{\star m}}(D^s) D^s(g)\dag \right| & = & \sum_{j=1}^{d_s} \left(\hat{f}(D^s)^m D^s(g)_j\dag\right)_j \\
\nn & \leq & \sum_{j=1}^{d_s} \left\|\hat{f}(D^s)^m\right\| \\
\label{eqn:fourier coefficient norm bound} & = & d_s \left\|\hat{f}(D^s)^m\right\|
\ees
where we used the definition of the matrix norm, the unitarity of $D^s(g)$, and that $|x_i| \leq \|\v{x}\|_2$ for any column vector $\v{x} = (x_i)_{i=1}^d$ and any $i = 1, 2, \dots, d$. It follows with the triangle inequality and the definition of 
\[ \alpha_{\epsilon} = \max_{1 \leq s \leq N_{\epsilon}} \| \hat{f}(D^s) \| \] that
\bess
\left| \sum_{s=1}^{N_{\epsilon}} d_s \tr \what{f^{\star m}}(D^s) D^s(g)\dag \right| 
& \leq & \sum_{s=1}^{N_{\epsilon}} d_s \left| \tr \what{f^{\star m}}(D^s) D^s(g)\dag \right| \\
& \leq & \sum_{s=1}^{N_{\epsilon}} d_s^2 \left\|\hat{f}(D^s)^m\right\| \\
& \leq & \alpha_{\epsilon}^m \sum_{s=1}^{N_{\epsilon}} d_s^2.
\eess
Now we can choose a constant $M_{\epsilon}$ that will only depend on $\epsilon$ such that for all $m \geq M_{\epsilon}$, \[ \alpha_{\epsilon}^m \sum_{s=1}^{N_{\epsilon}} d_s^2 \leq \epsilon. \]

Putting the bounds for both terms in \Ref{eqn:rand circ norm to bound} together, we arrive at
\[\|f^{\star m} - 1\|_{\infty} \leq 2 \epsilon,\] thus proving uniform convergence.
\end{proof}

We can improve the previous theorem to give the explicit convergence rate for nice positive-definite probability measures $f \in L^2(G)$.

\begin{theorem} For a fixed dimension $d$ and additionally $f \in L^2(G)$, $f$ nice positive-definite, $f^{\star m}$ converges exponentially to the constant function $1$.
\end{theorem}
\begin{proof}
From the Parseval formula (see Fact \ref{fact:parseval}) and the fact that the induced matrix norm from $\Hilbert_{d_s}$ is the Frobenius norm $\|A\| = \sqrt{\tr A A\dag}$, we conclude
\bess
\| \hat{f}(D^s) \| & \leq & \sqrt{\tr \hat{f}(D^s) \hat{f}(D^s)\dag} \\
& \leq & \f{1}{\sqrt{d_s}} \sqrt{d_s \tr \hat{f}(D^s) \hat{f}(D^s)\dag} \\
& \leq & \f{1}{\sqrt{d_s}} \sum_{s \geq 0} \sqrt{d_s \tr \hat{f}(D^s) \hat{f}(D^s)\dag} \\
& = & \f{1}{\sqrt{d_s}} \| f \|_2 = \f{1}{\sqrt{d_s}}
\eess 
as $f$ is a probability measure with integral $1$.

Hence we have that for any $\epsilon > 0$ there is a ``representation cut-off'' $S_{\epsilon}$ such that for all $s > S_{\epsilon}$ \[\|\hat{f}(D^s)\| \leq \f{1}{\sqrt{d_s}} \epsilon.\] As $f^{\star m}$ is a nice positive-definite function for $f$ is nice positive-definite, we can use the uniform convergence of its Fourier series. Define \be \label{eqn:max rep norm} \alpha_{\epsilon} = \max_{1 \leq s \leq S_{\epsilon}} \| \hat{f}(D^s) \| \ee and use \Ref{eqn:fourier coefficient norm bound} to bound
\bess
|f^{\star m}(g) - 1| & = & \sum_{s \geq 1} d_s \tr \hat{f}(D^s)^m D^s(g)\dag \\
& \leq & \sum_{s \geq 1} d_s^2 \left\|\hat{f}(D^s)^m\right\| \\
& \leq & \sum_{s=1}^{S_{\epsilon}} d_s^2 \left\|\hat{f}(D^s)^m\right\| + \sum_{s > S_{\epsilon}} d_s^{2 - m/2} \epsilon^m\\
& \leq & \alpha_{\epsilon}^m \sum_{s=1}^{S_{\epsilon}} d_s^2 + \epsilon^m \sum_{s > S_{\epsilon}} d_s^{-(m/2 - 2)}
\eess
for all $g \in G$. Using known formulas for the dimensions $d_s$ of the irreducible representations of $U(d)$ from \cite{VilenkinKlimyk1, VilenkinKlimyk2} (see Fact \ref{fact:dim irreps u(d)}), Emerson et al.\@ \cite{EmersonEtAl2005_RandomUnitaries} showed that \[\sum_{s > S_{\epsilon}} d_s^{-(m/2 - 2)}\] converges as long as $m > 6$. Hence exponential convergence of $\|f^{\star m} - 1\|_{\infty}$ follows. 
\end{proof}

This proves that by choosing an arbitrary single qubit or two-qubit gate according to the initial distribution $f$ in each step, the circuit comprised of the composition of these gates will converge to a random circuit with a rate exponential in the number of steps. However, it is not clear how the convergence rate \Ref{eqn:max rep norm} depends on the dimension $d = 2^N$ of the $N$ qubit system Hilbert space. In order for the random circuit construction to be efficient, \[\alpha_{\epsilon} = 1 - O\left(\f{1}{poly(N, \f{1}{\epsilon})}\right).\] It is not clear whether this can be achieved, as no reasonable bounds on the norm $\| \hat{f}(D^s) \|$ could be established so far.

Provided an efficient pseudorandom distribution of circuits $V$ with $\|f^{\star m} - 1\|_{\infty} < \epsilon$ exists, the average fidelity 
\[ \int_{U(d)} \bra{\psi_0} V \tilde{\E}(V \ket{\psi_0}\bra{\psi_0} V\dag) V\dag \ket{\psi_0} f^{\star m}(V) dV \] 
could be estimated using the circuit shown in Figure \ref{fig:haar random estimation circuit} within a precision of $\epsilon$.

\subsubsection{Weak Convergence to the Haar Measure}

For many practical applications, pseudorandom unitaries need not be drawn from a measure that converges uniformly to the Haar measure. Also, if $\mu_f$ is not an absolutely continuous probability measure that gives rise to a nice positive-definite function $f \in L^1(G)$, uniform convergence could not be shown so far. This is the case if the initial probability measure does not have a support over a continuously parametrized gate set $S \subset U(d)$, but rather a discrete set. Then $\mu_f$ will be a weighted sum of $\delta$-functions over the elements $g_i \in S$.

In that case, the random unitary approach can still give convergence, but in a weaker sense. Specifically, convergence to the Haar measure can be guaranteed with respect to certain test functions $\phi(g)$ in the weak topology: \[ \lim_{m \rightarrow \infty} \int_G \phi(g) d \mu_f^{\star m} = \int_G \phi(g)dg.\] The most accessible test functions are trigonometric polynomials which are functions $\phi$ such that $S_{\phi} = \{s \in \hat{G}\,|\, \hat{f}(D^s) \neq 0\}$ is finite. 

Using the orthogonality relations (see Fact \ref{fact:ortho rel}), it follows that we need only consider those irreducible representations $D^s$ for which $\hat{\phi}(D^s) \neq 0$. However, it remains an open problem to actually calculate the convergence rate and to pick a suitable initial distribution $f$ in this setting.

\subsection{Alternative Approach using Many Additional Qubits}

If additional qubits can be added to the system, there is an easy way to determine the entanglement fidelity of a quantum operation $\E$ \cite{PoulinEtAl2004}. Using the approach described in Section \ref{sec:avg fidelity using entanglement fidelity}, a motion-reversal experiment can be used to determine the average fidelity of $\tilde{\E}$ by implementing $U\dag U = \Id$ and assuming that the noise will not cancel out for it will be non-unitary. However, it is conceivable that this process might reduce the average fidelity as two operations have to be implemented. Nonetheless, this will provide a lower bound, at least.

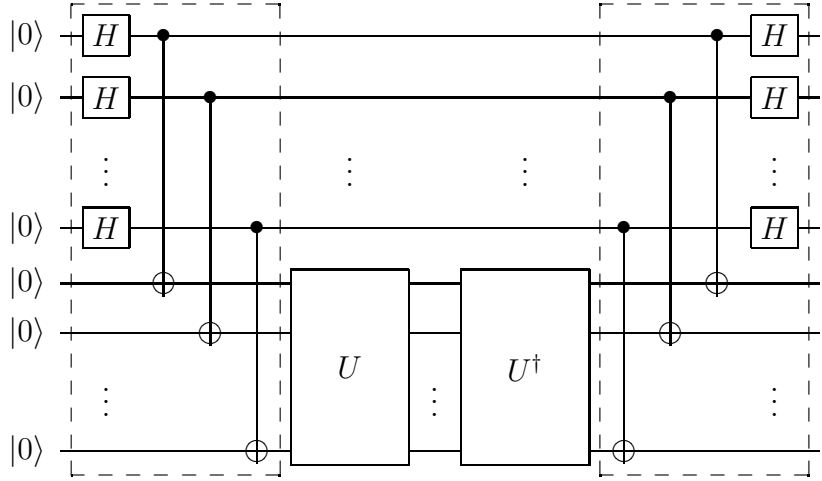
\begin{figure}[htbp]
	\centerline{
	\Qcircuit  @C=.7em @R=.7em
	{
		\lstick{\ket{0}} & \gate{H} \qw & \ctrl{4} \qw & \qw & \qw & \qw & \qw & \qw & \qw & \qw & \ctrl{4} \qw & \gate{H} \qw & \qw \\
		\lstick{\ket{0}} & \gate{H} \qw & \qw & \ctrl{4} \qw & \qw & \qw & \qw & \qw & \qw & \ctrl{4} \qw & \qw & \gate{H} \qw & \qw \\
		& \qvdots & & & & \qvdots & & \qvdots & & & & \qvdots & \\
		\lstick{\ket{0}} & \gate{H} \qw & \qw & \qw & \ctrl{4} \qw & \qw & \qw & \qw & \ctrl{4} \qw & \qw & \qw & \gate{H} \qw & \qw \\
		\lstick{\ket{0}} & \qw & \targ \qw & \qw & \qw & \multigate{3}{\hspace{1em} U \hspace{1em}} \qw & \qw & \multigate{3}{\hspace{1em} U\dag \hspace{1em}} \qw & \qw & \qw & \targ \qw & \qw & \qw \\
		\lstick{\ket{0}} & \qw & \qw & \targ \qw & \qw & \ghost{\hspace{1em} U \hspace{1em}} \qw & \qw & \ghost{\hspace{1em} U\dag \hspace{1em}} \qw & \qw & \targ \qw & \qw & \qw & \qw \\
		& \qvdots & & & & & \qvdots & & & & & \qvdots & \\
		\lstick{\ket{0}} & \qw & \qw & \qw & \targ \qw & \ghost{\hspace{1em} U \hspace{1em}} \qw & \qw & \ghost{\hspace{1em} U\dag \hspace{1em}} \qw & \targ \qw & \qw & \qw & \qw & \qw \gategroup{1}{2}{8}{5}{.7em}{--} \gategroup{1}{9}{8}{12}{.7em}{--}
	} }
	\caption{Circuit to Estimate the Average Fidelity using $N$ ancillas.}
	\label{fig:estimation circuit using n addtl qubits}
\end{figure}

Consider the circuit in Figure \ref{fig:estimation circuit using n addtl qubits}. The first part creates the maximally entangled state \[\ket{\phi} = \sum_{x=0}^{2^N - 1} \ket{x}\ket{x},\] and the third part of the circuit is the inverse of that computation. Thus measuring the final state in the computational basis enables us to measure the entanglement fidelity. To see this, denote $p$ the probability to measure $\ket{0^{\otimes 2N}}$ at the end of the computation and see that \[p = \bra{\phi} (\Id \otimes \tilde{\E}) \ket{\phi} = F_e(\tilde{\E}).\] $p$ is the probability of a binary random variable and can thus be estimated to an arbitrary precision using the techniques outlined in Section \ref{sec:QST}. Using \Ref{eqn:avg-fidelity and ent fidelity} and an estimate for $p$, we can calculate the average fidelity $F_{\text{avg}}(\tilde{\E}, \Id) = F_{\text{avg}}(\E, U)$.

\subsection{Discussion}

Both process tomography and Nielsen's approach using the entanglement fidelity work without any additional qubits, but require $O(d^4)$ experiments. This is exponential in the number of qubits $N = \log d$, hence these methods are deemed inefficient. Furthermore, it requires classical processing of either $d^4 \times d^4$ or $d^2 \times d^2$ complex matrices, which is also inefficient.

The random circuit approach seems promising as it does not require additional qubits and only relies on standard statistical techniques to estimate the probability $p$ of observing $0$ at the end of the experiment. However, the convergence rate of the pseudorandom circuit construction as a function of the Hilbert space dimension $d$ is not clear yet. It is a promising technique and further work should investigate the convergence condition for a test function like the average gate fidelity. 

The last approach estimates the average fidelity using $N$ additional qubits and similar classical postprocessing as in the random circuit case is required. Provided additional qubits do not introduce too much additional noise and are experimentally feasible, this is the preferred construction. However, in many practical settings, the number of qubits is still strictly limited and each additional qubit is quite expensive \cite{NielsenChuang2000}. It seems to be necessary to actually gain information about the structure of the noise before additional qubits can be realised. Also, both the random circuit and the last approach assume that the fidelity of implementing the motion-reversal experiment $U\dag U$ does not differ significantly from the average fidelity of an implementation of $U$.

\clearpage
\thispagestyle{empty}
\cleardoublepage

\chapter{Mutually-Unbiased Bases}
\label{ch:mubs}
\markright{Mutually-Unbiased Bases}

In this chapter, we will formally introduce the concept of mutually-unbiased bases and present the easiest constructions of these bases known so far. We will show that there are interesting open problems and nice applications beyond the context of this thesis.

\section{Introduction}


Two orthonormal bases $B_1$ and $B_2$ of a Hilbert space $\Hilbert$ of dimension $d$ are called mutually unbiased if \be \label{eqn:def mub}| \bracket{\psi_1}{\psi_2} | = \frac{1}{\sqrt{d}}.\ee In a 1960 paper, Schwinger \cite{Schwinger1960} realized that if a state $\ket{\psi}$ is prepared as a basis state of $B_1$ and measured with respect to the basis $B_2$, it is just an equally weighted superposition over all basis states of $B_2$ and vice versa. Hence no information can be gained about a state $\ket{\psi}$ that is created as a basis state of either $B_1$ or $B_2$ with the choice of basis unknown. This idea also underlies the famous BB84 quantum key distribution protocol \cite{BB84}.

\begin{figure}[htbp]
	\centering
	 	\psfrag{x}{$x$}
	 	\psfrag{y}{$y$}
	 	\psfrag{z}{$z$}
	 	\psfrag{ket0}{$\ket{0}$}
	 	\psfrag{ket1}{$\ket{1}$}
	 	\psfrag{ket+}{$\ket{+}$}
	 	\psfrag{ket-}{$\ket{-}$}
	 	\psfrag{ket+i}{$\ket{+i}$}
	 	\psfrag{ket-i}{$\ket{-i}$}
		\includegraphics[width=0.50\textwidth]{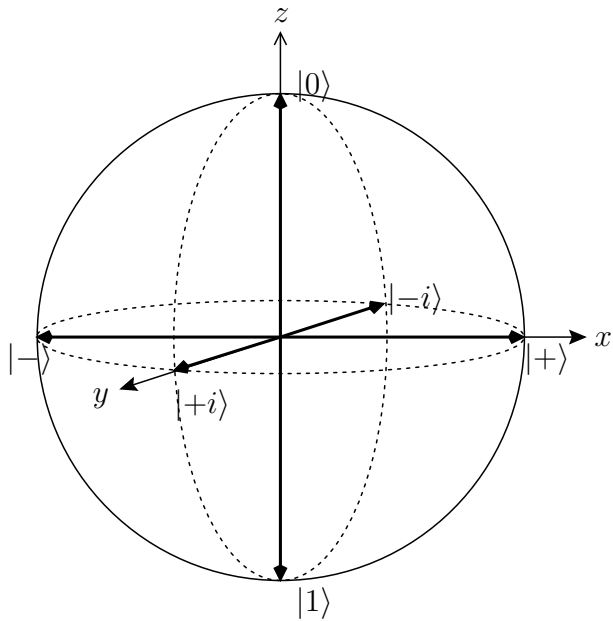}
	\caption{Mutually-Unbiased Bases on the Bloch Sphere}
	\label{fig:MUBs on the Bloch Sphere}
\end{figure}

For a single qubit system, the three bases
\bes 
\nn B_1 & = & \{\ket{0}, \ket{1}\},\\
\label{eqn:mubs single qubit} B_2 & = & \{\ket{+} = \frac{\ket{0} + \ket{1}}{\sqrt{2}}, \ket{-} = \frac{\ket{0} - \ket{1}}{\sqrt{2}} \text{ and }\\
\nn B_2 & = & \{\ket{+i} = \frac{\ket{0} + i \ket{1}}{\sqrt{2}}, \ket{-i} = \frac{\ket{0} - i \ket{1}}{\sqrt{2}}
\ees
form a set of pairwise mutually-unbiased bases or just ``mutually-unbiased bases'' for short. Sometimes, this is abbreviated by ``MUB''. The absolute value of the inner product between two vectors from different bases is $\frac{1}{\sqrt{2}}$ which corresponds to an angle of $\f{\pi}{4}$. On the Bloch sphere (see Section \ref{sec:bloch sphere}) the angles double and hence the vectors are orthogonal in the geometry of the three-dimensional Euclidean space. Figure \Ref{fig:MUBs on the Bloch Sphere} shows the layout of $B_1$, $B_2$, and $B_3$ on the Bloch sphere.

From the Bloch sphere it is apparent that we cannot find a fourth basis that is mutually unbiased to $B_1$, $B_2$, and $B_3$. It was already suggested by \cite{Ivanovic1981} that there can be at most $d+1$ mutually-unbiased bases in a Hilbert space of dimension $d$.

\section{History and Applications}

The concept of mutually-unbiased bases seems to have emerged in 1960 in a work by Schwinger \cite{Schwinger1960,KlappeneckerRoetteler2004,KlappeneckerRoetteler2005,Ivanovic1981,WoottersFields1989}. Schwinger considered the problem of determining an unknown, possibly mixed state $\rho$ provided sufficiently many copies of $\rho$ are given. He introduced the term ``complementarity'' between two measurement operators. Given a system prepared in a basis state of a basis $B_1$, a measurement with respect to a mutually-unbiased basis $B_2$ gives no information about the state but just an equal distribution over all states in $B_2$. Although this fact has been known long before Schwinger \cite{Schwinger1960}, he showed that the measurement operators corresponding to measurements in $d+1$ mutually-unbiased bases form an operator basis and he called such measurement operators ``maximally non-commutative''. It was not until 20 years later that Ivanovi\'{c} \cite{Ivanovic1981} explicitly showed how these measurements can be used to completely determine the unknown state $\rho$, thereby introducing the term ``mutually 'orthogonal' '' operators. Wootters and Fields \cite{WoottersFields1989} seemed to have coined the term ``mutually-unbiased bases''. They also showed that there are at most $d+1$ mutually-unbiased bases in a Hilbert space of dimension $d$ and gave the first explicit construction for such a complete set in case of prime power dimensions $d = p^k$ for $p > 2$.

The applications of mutually-unbiased bases are diverse. Firstly, the obvious application was quantum state determination \cite{Ivanovic1981}, where measurements with respect to mutually-unbiased bases are sufficient for quantum state tomography (see Section \ref{sec:QST}). Then, they have an application in quantum key distribution because of their nice information-theoretical property that a closely localized state in a basis $B_1$ looks like an equal superposition in a basis $B_2$ that is unbiased with respect to $B_1$. The BB84 protocol \cite{BB84} made use of the fact that 
\bess
\ket{+} & = & \frac{\ket{0} + \ket{1}}{\sqrt{2}}, \\
\ket{-} & = & \frac{\ket{0} - \ket{1}}{\sqrt{2}}, \\
\ket{0} & = & \frac{\ket{+} + \ket{-}}{\sqrt{2}}, \text{ and }\\
\ket{1} & = & \frac{\ket{+} - \ket{-}}{\sqrt{2}} \\
\eess
and hence an eavesdropper cannot obtain any information about a state prepared as a basis state of either $B_1$ or $B_2$ if the choice of basis is unknown to them. This was also generalized to $d$-dimensional systems \cite{CerfEtAl2002}. Buhrman et al.\@ \cite{BuhrmanEtAl2005} have recently shown how mutually-unbiased bases can be used to implement a quantum string commitment protocol. 

There is also an interesting application to the so-called Mean King's Problem, which amounts to determining the outcome of a measurement chosen randomly from a set of complementary observables. See \cite{KlappeneckerRoetteler2005_MK} for an overview of the state-of-the-art of the problem and how mutually-unbiased bases play a role. Besides showing another application of  mutually-unbiased bases, this article is fun to read.

This thesis will use mutually-unbiased bases to estimate averages over the uniform measure of all pure states of a quantum system. We will present a result similar to \cite{KlappeneckerRoetteler2005}, where it was shown that mutually-unbiased bases can be used to estimate certain Fubini-Study averages. For that, we will give an alternative proof.

\section{Constructions}

The first explicit construction by Wootters and Fields \cite{WoottersFields1989} was simplified and extended by subsequent works \cite{BandyopadhyayEtAl2002, Chaturvedi2002, LawrenceEtAl2002, KlappeneckerRoetteler2004, Durt2005, PlanatRosu2005, RomeroEtAl2005}. Although significant simplifications were achieved, constructions for the cases of odd and even prime power dimensions still differ. Ref.~\cite{KlappeneckerRoetteler2005} gives a brief overview of most currently known constructions.

\subsection{Odd Prime Power Dimension}

Let $\Hilbert$ be a Hilbert space of dimension $d = p^k$, $p$ an odd prime and $k \in \N$. Denote the computational basis by $\{\ket{x}\,|\, x \in GF(p^k)\}$ assuming an arbitrary ordering of the elements of $GF(p^k)$. The following lemma will be crucial in the construction of an extremal set of mutually-unbiased bases for $\Hilbert$.

\begin{lemma} \label{lem:weil sum} Let $p > 2$ and let $\chi$ be a non-trivial additive character of $GF(p^k)$. Let \[p(X) = a_2 X^2 + a_1 X + a_0 \in GF(p^k)[X], a_2 \neq 0,\] be a polynomial of degree $2$. Then
\[\left| \sum_{x \in GF(p^k)} \chi(p(x)) \right| = \sqrt{p^k}.\]
\end{lemma}
\begin{proof}
See \cite[Ch.~5]{LidlNiederreiter1994} for a proof.
\end{proof}

Alltop \cite{Alltop1980} constructed sequences of complex numbers that exhibit very low correlations for use in spread spectrum radar and communication applications. It was not until recently that Alltop's work was rediscovered and found to give a construction for a set of $p+1$ mutually-unbiased bases in prime dimension $p$, $p \geq 5$ \cite{KlappeneckerRoetteler2004}. Ref.~\cite{KlappeneckerRoetteler2004} also gave the generalization of Alltop's construction to the case of prime power dimensions.

This construction was improved by Klappenecker and R\"{o}tteler \cite{KlappeneckerRoetteler2004} to work for any odd prime power dimension $p^k$. It is based on Ivanovi\'{c}s work for prime dimensions \cite{Ivanovic1981} that was later generalized by Wootters and Fields \cite{WoottersFields1989}. Different versions of the proof were given by Chaturvedi \cite{Chaturvedi2002} and Bandyopadhyay et al.\@ \cite{BandyopadhyayEtAl2002}. We will present the proof by Klappenecker and R\"{o}tteler as it is the shortest one known to the author.
\begin{theorem} \label{thm:mub kr}
Let $p \geq 3$. Then the sets  \[B_a = \{\ket{\psi^a_b}\,|\, b \in GF(p^k)\}, a \in GF(p^k)\] where \be \label{eqn:mub kr} \ket{\psi^a_b} = \frac{1}{\sqrt{p^k}} \sum_{x \in GF(p^k)} \omega_p^{\tr(a x^2 + bx)} \ket{x}\ee together with the computational basis are a complete set of $d + 1$ mutually-unbiased bases. 
\end{theorem}
\begin{proof}
Again, we consider the inner product
\be | \bracket{\psi^a_b}{\psi^{a'}_{b'}} |
 = \frac{1}{p^k} \left| \sum_{x \in GF(p^k)} \omega_p^{\tr( (a'-a) x^2+ (b'-b)x)} \right| \label{eqn:kr inner prod}.
\ee
In the case of vectors from the same basis, we take $a = a'$ and thus 
\[ | \bracket{\psi^a_b}{\psi^{a}_{b}} | = \frac{1}{p^k} \left| \sum_{x \in GF(p^k)} \omega_p^{\tr( (b'-b)x)} \right|
 = \begin{cases} 1 & b = b' \\ 0 & b \neq b' \end{cases}.\]
Now assume $a \neq a'$. Lemma \Ref{lem:weil sum} implies that \[| \bracket{\psi^a_b}{\psi^{a'}_{b'}} | = \frac{1}{\sqrt{d}}\] and hence $B_a$ and $B_{a'}$ are mutually unbiased. As the coefficients of the basis states of the computational basis in \Ref{eqn:mub kr} have absolute value $\frac{1}{\sqrt{p^k}}$, $B_a$ is mutually unbiased to the computational basis for any $a$.
\end{proof}

For a qutrit system with dimension $d = 3$, a complete set of mutually-unbiased bases is now given by
\bess
B_0 & = & \frac{1}{\sqrt{3}} \{(1,1,1), (1, \omega_3, \omega_3^2), (1, \omega_3^2, \omega_3)\}, \\
B_1 & = & \frac{1}{\sqrt{3}} \{(1, \omega_3, \omega_3), (1, \omega_3^2, 1), (1, 1, \omega_3^2)\}, \\
B_2 & = & \frac{1}{\sqrt{3}} \{(1, \omega_3^2, \omega_3^2), (1, \omega_3, 1), (1, 1, \omega_3)\}, \text{ and }
\eess
the computational basis, where we represented the states as column vectors with respect to the computational basis.

\subsection{Qubits}

In the case of $n$ qubits, the dimension of the state space $\Hilbert$ is $d = 2^n$, which is an even prime power. The construction in Theorem \ref{thm:mub kr} breaks down in fields of characteristic $2$, which is the case for $GF(2^n)$. Specifically, Lemma \ref{lem:weil sum} does not hold in fields of characteristic $2$.

Klappenecker and Roettler \cite{KlappeneckerRoetteler2004} came up with a solution for the qubit case by considering finite rings instead of finite fields. In particular, they employed a lemma analogous to Lemma \ref{lem:weil sum} that holds in Galois Rings (see Section \ref{sec:galois rings}). Let $GR(4^n)$ denote the Galois Ring with $4^n$ elements and let $\Tau_n$ be its Teichm\"{u}ller set (see Definition \ref{def:teichmueller set}). We assume an arbitrary ordering of the elements of $\Tau_n$ so that we can identify the elements of the computational basis with the elements of $\Tau_n$.

\begin{lemma} \label{lem:exp sum gr} The exponential sum $\Gamma: GR(4^n) \rightarrow \C,$ \[\Gamma(x) = \sum_{y \in \Tau_n} \omega_4^{\tr(xy)}\] evaluates to
\[| \Gamma(x) | = \begin{cases}
0 & \text{if } x \in 2 \Tau_n, x \neq 0 \\
2^n & \text{if } x = 0 \\
\sqrt{2^n} & \text{otherwise}
\end{cases}.
\]
\end{lemma}
\begin{proof}
See \cite[Lemma 3]{Carlet1998} for a proof.
\end{proof}

Using this lemma, the construction of a maximal set of mutually-unbiased bases in an $n$ qubit system is very simple and elegant.
\begin{theorem} \label{thm:mub qubits}
Let $\Tau_n$ be the Teichm\"{u}ller set of $GR(4^n)$. Then the sets  \[B_a = \{\ket{\psi^a_b}\,|\, b \in \Tau_n\}, a \in \Tau_n\] where \be \label{eqn:mub qubits} \ket{\psi^a_b} = \frac{1}{\sqrt{2^n}} \sum_{x \in \Tau_n} \omega_4^{\tr((a + 2b) x)} \ket{x}\ee together with the computational basis form a complete set of $2^n + 1$ mutually-unbiased bases. 
\end{theorem}
\begin{proof}
The inner product between two vectors from bases $B_a$ and $B_{a'}$ evaluates to
\be | \bracket{\psi^a_b}{\psi^{a'}_{b'}} |
 = \frac{1}{2^n} \left| \sum_{x \in \Tau_n} \omega_4^{\tr( ( (a'-a) + 2 (b'-b))x)} \right| \label{eqn:mub qubit inner prod}.
\ee
For states from the same basis $a = a'$, \Ref{eqn:mub qubit inner prod} simplifies to 
\[
| \bracket{\psi^a_b}{\psi^{a}_{b}} | 
 = \frac{1}{2^n} \left| \sum_{x \in \Tau_n} \omega_4^{\tr( 2 (b'-b)x)} \right| 
 = \frac{1}{2^n} \left| \sum_{x \in \Tau_n} (-1)^{\tr( (b'-b)x)} \right| \\
 = \begin{cases} 1 & b = b' \\ 0 & b \neq b' \end{cases},
\]
hence $B_a$ is an orthonormal basis for any $a \in \Tau_n$. For different bases $a \neq a'$, Lemma \ref{lem:exp sum gr} implies \[ | \bracket{\psi^a_b}{\psi^{a'}_{b'}} | = \frac{1}{\sqrt{2^n}}, \] thus $B_a$ and $B_{a'}$ are mutually unbiased for any $a, a' \in \Tau_n$, $a \neq a'$. Any $B_a$ is mutually unbiased to the computational basis as the absolute value of each coefficient of the basis states in $\ket{\psi^a_b}$ is $\frac{1}{\sqrt{2^n}}$.
\end{proof}

For the case of a single qubit, Theorem \ref{thm:mub qubits} recovers the well-known mutually unbiased bases \Ref{eqn:mubs single qubit}. We will now state the explicit example for two qubits.

Observe that $h(X) = X^2 + X + 1 \in \Z_4[X]$ is a primitive polynomial. Hence $GR(4^2) = \Z_4[X] / (X^2 + X + 1)$ has Teichm\"{u}ller set $\Tau_2 = \{0, 1, X, X^2 = 3X + 3\}$. The trace is given by $\tr (a+2b) = a + a^2 + 2(b + b^2)$. Therefore the mutually unbiased bases are given by
\bess
B_0 		 & = & \ahalf \{ (+1, +1, +1, +1), (+1, +1, -1, -1), (+1, -1, -1, +1), (+1, -1, +1, -1) \},\\
B_1 		 & = & \ahalf \{ (+1, -1, -i, -i), (+1, -1, +i, +i), (+1, +1, +i, -i), (+1, +1, -i, +i) \},\\
B_X 		 & = & \ahalf \{ (+1, -i, -i, -1), (+1, -i, +i, +1), (+1, +i, +i, -i), (+1, +i, -i, +1) \}, \\
B_{3X+3} & = & \ahalf \{ (+1, -i, -1, -i), (+1, -i, +1, +i), (+1, +i, +1, -i), (+1, +i, -1, +i) \},\text{ and }
\eess
the computational basis. Note that although we used the Teichm\"{u}ller elements specific to our choice of $h(X)$, Fact \ref{fact:galos rings isomorphic} guarantees that we will always get the same set of mutually-unbiased bases up to relabelling of the basis elements.

A different construction makes use of the generalized Pauli operators that were introduced in the discussion of Quantum State Tomography (Fact \ref{fact:generalized Paulis}). It was discovered, extended and simplified by several authors \cite{BandyopadhyayEtAl2002, LawrenceEtAl2002, Durt2005, RomeroEtAl2005}. The constructions are based on the following theorem.
\begin{theorem}[\cite{LawrenceEtAl2002}] The set of $4^N - 1$ non-identity generalized Pauli operators may be partitioned into $2^N+1$ sets of $2^N-1$ pairwise commuting operators. The common eigenbases are mutually-unbiased with respect to each other.
\end{theorem}

\section{Non Prime-Power Dimensions and Open Problems}

We gave several constructions for a maximal set of mutually-unbiased bases in prime power dimensions. However, the situation is quite different if the dimension is not a prime power.
\begin{df} Denote $M(d)$ the maximal number of mutually-unbiased bases in a Hilbert space of dimension $d$.
\end{df}
From \cite{WoottersFields1989} and the preceeding section, the following upper bound and lower bounds are known.
\bfact $M(d) \leq d + 1$ for all $d \in \N$. $M(p^k) = p^k + 1$ for $p$ prime and $k \in \N$.
\efact
For the case of non-prime power dimensions, only a fairly weak lower bound is known so far.
\begin{theorem}[\cite{KlappeneckerRoetteler2004}] \label{thm:mub lower bound} Let $d = p_1^{\alpha_1} p_2^{\alpha_2} \dots p_k^{\alpha_k}$ be the decomposition of $d$ into its distinct prime factors $p_i$. Then \[M(d) \geq \min_i \left(p_i^{\alpha_i} + 1 \right).\]
\end{theorem}
\begin{proof}
Let $\Hilbert_d$ be the Hilbert space of dimension $d$. Given a decomposition of $d$, we can decompose \be \label{eqn:decomp hilbert space into prime factor dim spaces} \Hilbert_d = H_{p_1^{\alpha_1}} \otimes H_{p_2^{\alpha_2}} \otimes \dots \otimes H_{p_k^{\alpha_k}}. \ee Denote $d(i) = p_i^{\alpha_i}$ the dimension of the $i$-th Hilbert space $H_{p_i^{\alpha_i}}$ in the decomposition \Ref{eqn:decomp hilbert space into prime factor dim spaces}. Let $\B^{(i)} = \{B^{(i)}_1, \dots, B^{(i)}_{d(i)}\}$ be a maximal set of $d(i) + 1 = p_i^{\alpha_i} + 1$ mutually-unbiased bases for $H_{p_i^{\alpha_i}}$. Denote the elements of the basis $B^{(i)}_j$ by \[B^{(i)}_j = \{ \ket{\psi^{(i,j)}_1}, \dots, \ket{\psi^{(i,j)}_d(i)}\} \] and define \[m = \min_i \left(p_i^{\alpha_i} + 1 \right).\] 

Now we claim that the $m$ sets \[ A_j = \left\{ \ket{\psi^{(1,j)}_{l_1}} \otimes \dots \otimes \ket{\psi^{(k,j)}_{l_k}}\,|\, l_i \in \{1, 2, \dots d(i)\} \right\}, j = 1, 2, \dots, m, \] are orthonormal bases and form a set of $m$ mutually-unbiased bases for $\Hilbert$. Remember that the inner product of a tensor product evaluates as \[\left(\ket{\phi_A} \otimes \ket{\phi_B}, \ket{\psi_A} \otimes \ket{\psi_B} \right) = \bracket{\phi_A}{\psi_A} \bracket{\phi_B}{\psi_B} \] and thus the claim follows.
\end{proof}

It is not known whether this lower bound can be improved in any way. An obvious way to extend Theorem \ref{thm:mub lower bound} is to allow for a more general construction of the sets $A_j$ in order to increase their number. This can be done by mixing states from different bases in the tensor product
\[ \ket{\psi^{(1,{j_1})}_{l_1}} \otimes \dots \otimes \ket{\psi^{(k,{j_k})}_{l_k}}\] where both the $j_i$ and $l_i$ are picked according to some combinatorial criteria. However, this will not lead to a set of mutually-unbiased bases. Suppose $d = p_1 p_2$, $p_1, p_2$ distinct primes. Any construction that yields more than $m$ bases will w.l.o.g.\@ assign states $\ket{\psi^{(1,1)}_1} \otimes \ket{\psi^{(1,1)}_1}$ and $\ket{\psi^{(1,1)}_1} \otimes \ket{\psi^{(1,2)}_1}$ to different bases. This leads to the inner product 
\bess
\left|\left( \ket{\psi^{(1,1)}_1} \otimes \ket{\psi^{(1,1)}_1}, \ket{\psi^{(1,1)}_1} \otimes \ket{\psi^{(1,2)}_1}\right) \right| & = & \left| \bracket{\psi^{(1,1)}_1}{\psi^{(1,1)}_1} \bracket{\psi^{(1,1)}_1}{\psi^{(1,2)}_1} \right| \\
& = & \frac{1}{\sqrt{p_2}} \neq \frac{1}{\sqrt{d}}.
\eess
Therefore, this naive construction cannot give us more than $m$ mutually-unbiased bases.

Furthermore, it was recently shown that the methods presented for prime power dimensions cannot be generalized to non prime-power cases \cite{Archer2005}. It is conjectured \cite{Zauner1999, KlappeneckerRoetteler2004, KlappeneckerRoetteler2005} that $M(d)$ is substantially smaller than $d + 1$ if $d$ is not a power of a prime. However, even the maximal number of mutually-unbiased bases $M(6)$ in a $6$-dimensional system is not known exactly. Theorem \ref{thm:mub lower bound} gives $M(6) \geq 3$ only, and we know that $M(6) \leq 7$. It is an interesting open problem to even determine $M(6)$.

The problem of the maximal number of mutually-unbiased bases was linked to the problem of determining the maximal number of mutually orthogonal latin squares \cite{WojcanBeth2004,KlappeneckerRoetteler2005,HayashiEtAl2005}. It seems that there are connections between both concepts that should be subject of future research. In particular, further investigation into the existence of a maximal number of mutually-unbiased bases and a maximal number of mutually orthogonal latin squares could lead to fruitful results in either area.
\clearpage
\thispagestyle{empty}
\cleardoublepage

\chapter{Scalable Efficient Noise Estimation}
\label{ch:fidelity estimation}
\markright{Scalable Efficient Noise Estimation}

This chapter contains the first main result, which shows that mutually-unbiased bases are a $2$-design for quantum states using different techniques than the proofs known so far. We will give an explicit construction of circuits that generate states from a complete set of MUBs. We will use that construction to show how the average fidelity can be estimated efficiently.

\section{Introduction}

The main result emerged in joint work with Richard Cleve, Joseph Emerson, and Etera Livine. As it turned out, a similar result was already proved using different proof techniques by \cite{KlappeneckerRoetteler2005} in general and \cite{Barnum2002} for a specific construction. Our result is purely algebraic and relies on the explicit calculation of the integral in Theorem \ref{thm:Haar avg explicit} using Schur's Lemma. Although our result follows as a corollary from \cite{KlappeneckerRoetteler2005}, the constructions of explicit circuits that generate mutually-unbiased basis states appear to be new. We will first present the main result in our language, present the earlier proof from \cite{KlappeneckerRoetteler2005}, and derive an efficient circuit that can be used to estimate the average gate fidelity.

Later on, we will generalize the notion of a design from \cite{KlappeneckerRoetteler2005, Barnum2002, Zauner1999} from states to unitary operators and present an outline for further research in that direction. We suggest that the techniques developed so far can be further generalized to derive efficient experimental protocols that reveal more information about the noise than the average fidelity can. Furthermore, this opens a new perspective on various notions of pseudo-randomness used in different quantum protocols. This unifies several applications from different areas of quantum computation in that they use the same ``amount'' of pseudo-randomness according to our classification.


\section{Calculation of Haar averages using MUB vectors}

We show that the average of some quartic function over the Haar measure over all unitary operators on a complex inner product space $\C^d$ can be calculated using only the vectors of a maximal set of mutually unbiased bases. We assume that such a maximal set of mutually unbiased bases exists. As it is only known that $d+1$ mutually unbiased bases exist for prime power dimensions, we restrict ourselves to these cases. The proof we will present below is original work and to the best of our knowledge has not been found before. We will discuss how this result can be derived as a corollary of a fairly recent result by Klappenecker and Roetteler \cite{KlappeneckerRoetteler2005} in the next section.

Let $\Hilbert = \C^d$ be a complex inner product space of dimension $d$. Denote \be B_a = \{\ket{\psi^a_b}: b = 0, \dots, d-1\}\ee  the $a$-th basis of a set of $d+1$ mutually unbiases bases for $a \in \{0, 1, \dots, d\}$. It follows that 
\[|\bracket{\psi^a_b}{\psi^{a'}_{b'}}  |= 
\begin{cases}
1 & a = a', b = b' \\
0 & a = a', b \neq b'\\
\frac{1}{\sqrt{d}} & a \neq a'
\end{cases}
\]
Recall that $L(\Hilbert)$ denotes the inner product space of all linear operators on $\Hilbert$, using the Hilbert-Schmidt inner product $(A, B) = \thr(A^{\dag} B)$. Let $W \subset L(\Hilbert)$ be the subspace of all Hermitian traceless linear operators on $\Hilbert$. Note that the inner product simplifies to the real-valued $(A, B) = \thr(AB)$. 

\begin{lemma} \label{lem:1}
Let \be W_a = \left\{\sum_{b=0}^{d-1} r_b \ket{\psi^a_b}\bra{\psi^a_b}: \sum_{b=0}^{d-1} r_b = 0, r_b \in \R \right\}.\ee Then $W = \bigoplus_{a=0}^d W_a$.
\end{lemma}
\begin{proof} Note that $W_a \perp W_{a'}$ for $a \neq a'$ as
\bess
\thr \left(\sum_{b=0}^{d-1} r_b \ket{\psi^a_b}\bra{\psi^a_b}  \sum_{b'=0}^{d-1} r_{b'} \ket{\psi^{a'}_{b'}}\bra{\psi^{a'}_{b'}} \right) & = & \sum_{b=0}^{d-1} r_b \sum_{b'=0}^{d-1} r_{b'} \thr \ket{\psi^a_b}\bracket{\psi^a_b}{\psi^{a'}_{b'}}\bra{\psi^{a'}_{b'}} \\ 
& = & \sum_{b=0}^{d-1} r_b \sum_{b'=0}^{d-1} r_{b'} \thr \bracket{\psi^{a'}_{b'}}{\psi^a_b}\bracket{\psi^a_b}{\psi^{a'}_{b'}} \\
& = & \sum_{b=0}^{d-1} r_b \sum_{b'=0}^{d-1} r_{b'} \frac{1}{\sqrt{d}} \\
& = & 0
\eess
and $\dim W_a = d-1, \dim W = d^2 - 1$ in real parameters. Thus $W$ is indeed the direct sum of its $d+1$ subspaces $W_a$ as the sum of their dimensions $d+1$ yields $d^2 - 1$.
\end{proof}

\begin{lemma} For each $a$, \be \Pi_a(V) = \sum_{b=0}^{d-1} \ket{\psi^a_b}\bra{\psi^a_b} V \ket{\psi^a_b}\bra{\psi^a_b}\ee is an orthogonal projector onto $W_a$. The operators $\{\Pi_a\,|\, a = 0, 1, \dots, d\} \subset L(W)$ form a complete set of orthogonal projectors onto $W$. 
\end{lemma}
\begin{proof} Pick an arbitrary operator $X \in W_a$ and observe that 
\bess
\Pi_a(X) & = & \sum_{b=0}^{d-1} \ket{\psi^a_b}\bra{\psi^a_b} \sum_{b'=0}^{d-1} r_{b'} \ket{\psi^a_{b'}}\bra{\psi^a_{b'}} \ket{\psi^a_b}\bra{\psi^a_b} \\
& = & \sum_{b=0}^{d-1} \sum_{b'=0}^{d-1} r_{b'} \ket{\psi^a_b} \bracket{\psi^a_b}{\psi^a_{b'}} \bracket{\psi^a_{b'}}{\psi^a_b}\bra{\psi^a_b} \\
& = & \sum_{b=0}^{d-1} r_b \ket{\psi^a_b}\bra{\psi^a_b} 
\eess and analogously for $a \neq a'$
\bess
\sum_{b=0}^{d-1} \ket{\psi^a_b}\bra{\psi^a_b} \sum_{b'=0}^{d-1} r_{b'} \ket{\psi^{a'}_{b'}}\bra{\psi^{a'}_{b'}} \ket{\psi^a_b}\bra{\psi^a_b} 
& = &
\sum_{b=0}^{d-1} \sum_{b'=0}^{d-1} r_{b'} \bracket{\psi^a_b}{\psi^{a'}_{b'}}\bracket{\psi^{a'}_{b'}}{\psi^a_b} \ket{\psi^a_b}\bra{\psi^a_b} \\
& = &
\sum_{b=0}^{d-1} \frac{r_b}{d} \ket{\psi^a_b}\bra{\psi^a_b} 
= 0
\eess
since $\sum_{b=0}^{d-1} r_b = 0$. Therefore, $\Pi_a$ is a projector onto $W_a$. Completeness and orthogonality follow from Lemma \ref{lem:1}.
\end{proof}

\begin{corollary} \label{cor:projector sum}
For $M, N \in W$, \be \sum_{a=0}^d \thr \left( \Pi_a(M) \Pi_a(N) \right) = \thr MN. \ee
\end{corollary}
\begin{proof} Observe that $\sum_{a=0}^d \thr \left( \Pi_a(M) \Pi_a(N) \right) = \sum_{a=0}^d \left( \Pi_a(M), \Pi_a(N) \right)$ and the statement follows directly from the fact that the $\Pi_a$ form a complete set of orthogonal projectors.
\end{proof}

\begin{theorem} Let $M, N \in W$. Then 
\be \label{eqn:main} \sum_{a=0}^{d} \sum_{b=0}^{d-1} \bra{\psi^a_b} M \ket{\psi^a_b}\bra{\psi^a_b} N \ket{\psi^a_b} = \thr MN. \ee
\end{theorem}
\begin{proof}
\bess 
\sum_{a=0}^{d} \sum_{b=0}^{d-1} \bra{\psi^a_b} M \ket{\psi^a_b}\bra{\psi^a_b} N \ket{\psi^a_b} & = & \sum_{a=0}^{d} \thr \left( \sum_{b=0}^{d-1} \ket{\psi^a_b}\bra{\psi^a_b} M \ket{\psi^a_b}\bra{\psi^a_b} N \ket{\psi^a_b} \bra{\psi^a_b} \right) \\
& = & \sum_{a=0}^{d} \thr \left( \left(  \sum_{b=0}^{d-1} \ket{\psi^a_b}\bra{\psi^a_b} M \ket{\psi^a_b} \bra{\psi^a_b} \right) \right. \\
& & \qquad \qquad \left. \left( \sum_{b=0}^{d-1} \ket{\psi^a_b}\bra{\psi^a_b} N \ket{\psi^a_b} \bra{\psi^a_b} \right) \right) \\
& = & \sum_{a=0}^d \thr \left( \Pi_a(M) \Pi_a(N) \right) = \thr MN
\eess
by Corollary \ref{cor:projector sum}.
\end{proof}

\begin{corollary} Let $M$, $N$ be Hermitian operators. Then 
\be \label{eqn:main-hermitian} \sum_{a=0}^{d} \sum_{b=0}^{d-1} \bra{\psi^a_b} M \ket{\psi^a_b}\bra{\psi^a_b} N \ket{\psi^a_b} = \thr MN + \thr M \thr N. \ee
\end{corollary}
\begin{proof}
Construct the traceless Hermitian operators $\widetilde{M} = M - \f{\thr M}{d}\Id, \widetilde{N} = N - \f{\thr N}{d}\Id$ and simplify the left and right hand sides of \Ref{eqn:main} using the bilinearity in the space of linear operators $L(\Hilbert)$ on $\Hilbert$ of \Ref{eqn:main-hermitian} to get the result. 
\end{proof}

\begin{corollary} \label{cor:main} Let $M$, $N$ be linear operators on $\Hilbert$. Then
\be \label{eqn:main-general}\sum_{a=0}^{d} \sum_{b=0}^{d-1} \bra{\psi^a_b} M \ket{\psi^a_b}\bra{\psi^a_b} N \ket{\psi^a_b} = \thr MN + \thr M \thr N. \ee
\end{corollary}
\begin{proof}
Construct the Hermitian operators 
\[M_1 = M + M^{\dag}, M_2 = i(M - M^{\dag}), N_1 = N + N^{\dag}, N_2 = i(N - N^{\dag}).\] Using that both sides of \Ref{eqn:main-hermitian} are bilinear forms $\< \cdot, \cdot \>$ on the space $L(\Hilbert)$ of linear operators on $\Hilbert$, observe that
\bess
\< M_1, N_1\> - i \< M_1, N_2\> - i \< M_2, N_1\> - \< M_2, N_2\> & = & 
 \< M, N\> + \< M, N^{\dag}\> + \< M^{\dag}, N\> + \< M^{\dag}, N^{\dag}\> \\
 & & \,  + \< M, N\> - \< M, N^{\dag}\> + \< M^{\dag}, N\> - \< M^{\dag}, N^{\dag}\> \\
 & & \, + \< M, N\> + \< M, N^{\dag}\> - \< M^{\dag}, N\> - \< M^{\dag}, N^{\dag}\> \\
 & & \, + \< M, N\> - \< M, N^{\dag}\> - \< M^{\dag}, N\> + \< M^{\dag}, N^{\dag}\> \\
 & = & 4 \< M, N\> 
\eess
and the statement follows.
\end{proof}

\begin{cor} \label{cor:Haar avg = MUB avg} For any linear operators $M, N \in L(\Hilbert)$, the average over the Fubini-Study measure is the same as the average over a complete set of mutually-unbiased bases:
\be \label{eqn:Haar avg = MUB avg}
\int_{F-S} \bra{\psi} M \ket{\psi}\bra{\psi} N \ket{\psi} d \ket{\psi} = \frac{1}{d(d+1)}  \sum_{a=0}^d \sum_{b=0}^{d-1} \bra{\psi^a_b} M \ket{\psi^a_b}\bra{\psi^a_b} N \ket{\psi^a_b} d \ket{\psi^a_b}
\ee
\end{cor}
\begin{proof} Combine Corollary \ref{cor:main} and Theorem \ref{thm:Haar avg explicit}.
\end{proof}

\section{Mutually-Unbiased Bases are $2$-designs}

Klappenecker and Roetteler actually showed a similar result a little earlier \cite{KlappeneckerRoetteler2005}, which we will present in this section. We will start with a little bit of notation.

The set of all quantum states forms a complex unit sphere $S^{d-1}$ in $\Hilbert_d$. As global phases have no observable effect, we define an equivalence relation on quantum states by letting $\ket{\psi} \equiv \ket{\varphi}$ if and only if there is $\theta \in [0, 2\pi)$ such that $\ket{\psi} = e^{i\theta} \ket{\varphi}$. Then $CS^{d-1} = S^{d-1} / \equiv$ can be thought of as the analog of the Bloch sphere for a $d$-dimensional quantum system.

\begin{lemma}[\cite{KlappeneckerRoetteler2005}] \label{lem:simple haar integral} For all normalized $\ket{\varphi} \in \Hilbert$ and $k \in \N$, \[\int_{F-S} \left| \bracket{\varphi}{\psi} \right|^{2k} d\ket{\psi} = \f{1}{\binom{d + k - 1}{k}}.\]
\end{lemma}

The next ingredient is the notion of homogeneous polynomials. Denote $Hom(k,l) \subseteq \C[x_1, \dots, x_d, y_1, \dots, y_2]$ the set of all polynomials of homogeneous degree $k$ in the variables $x_1, \dots, x_d$ and of homogeneous degree $l$ in the variables $y_1, \dots, y_d$. We define the restriction of $p \in Hom(k,l)$ onto the complex sphere of quantum states with different observable effects as \[p_{\circ}(\ket{\psi}) = p(\alpha_1, \dots, \alpha_d, \overline{\alpha_1}, \dots, \overline{\alpha_d})\] where $\ket{\psi} = \sum_{i=1}^d \alpha_i \ket{\psi_i}$ for an orthonormal basis $\{\ket{\psi_i}\}_{i=1}^d$ for $\Hilbert$. It follows from the equivalence relation that defined $CS^{d-1}$ that $k = l$ in order for the definition of $p_{\circ}$ to be independent of the representative $\ket{\psi} \in CS^{d-1}$. Thus we define \[Hom(k,k)_{\circ} = \{p_{\circ}\,|\, p \in Hom(k,k)\}.\] Now we can turn to the definition of complex projective designs. 

\begin{df} \label{def:2-design for states} A complex projective $t$-design is a nonempty finite subset $X \subseteq CS^{d-1}$ such that the ``cubature formula''
\be \label{eqn:cubature formula} \f{1}{|X|} \sum_{\ket{\psi} \in X} p(\ket{\psi}) = \int_{\text{F-S}} p(\ket{\psi}) d \ket{\psi} \ee
holds for any $p \in Hom_{\circ}(t,t)$, where we understand that $p(\ket{\psi})$ is a function in the coefficients of $\ket{\psi}$ in some orthonormal basis. 

We will refer to $X$ just as a $t$-design if it is clear from the context that $X \subseteq CS^{d-1}$.
\end{df}

\begin{theorem} \label{thm:t-design stuff} Let $X$ be a finite subset of $CS^{d-1}$. The following statements are equivalent:
\begin{enumerate}
	\item $X$ is a $t$-design.
	\item For all $\ket{\psi} \in \Hilbert$ and all $0 \leq k \leq t$, 
	\be \label{eqn:kr t-design 2} \f{\bracket{\psi}{\psi}^k}{\binom{d + k - 1}{k}} = \f{1}{|X|} \sum_{\ket{\varphi} \in X} \left| \bracket{\psi}{\varphi}\right|^{2k}.\ee
	\item For $0 \leq k \leq t$, 
	\be \label{eqn:kr t-design 3} \f{1}{|X|^2} \sum_{\ket{\psi}, \ket{\varphi} \in X} \left| \bracket{\psi}{\varphi}\right|^{2k} = \f{1}{\binom{d + k - 1}{k}}. \ee
\end{enumerate}
\end{theorem}
\begin{proof}
We will show that (1) implies (2). Let $\ket{\psi} \in \Hilbert$ and observe that $p(\ket{\varphi}) = \left| \bracket{\psi}{\varphi}\right|^{2k} = \bracket{\psi}{\varphi}^k\bracket{\varphi}{\psi}^k$ is a homogeneous polynomial in $Hom(k,k)_{\circ}$. $X$ is a $t$-design, therefore \[\f{1}{|X|} \sum_{\ket{\varphi} \in X} \left|\bracket{\psi}{\varphi}\right|^{2k} = \int_{\text{F-S}} \left| \bracket{\psi}{\varphi}\right|^{2k} d \ket{\varphi} \] holds for all $0 \leq k \leq t$. Dividing by  $\left|\bracket{\psi}{\psi}\right|^k$ enables us to use Lemma \ref{lem:simple haar integral} and thus the right-hand side evaluates to \[\f{\left|\bracket{\psi}{\psi}\right|^k}{\binom{d + k - 1}{k}}\] and \Ref{eqn:kr t-design 2} follows.

Next we will show that (2) implies (3). Summing \Ref{eqn:kr t-design 2} over all $\ket{\psi} \in X$ and using that $X$ consists of normalized unit vectors, we have 
\bess \sum_{\ket{\psi} \in X} \f{\bracket{\psi}{\psi}^k}{\binom{d + k - 1}{k}} & = & \sum_{\ket{\psi} \in X} \f{1}{|X|} \sum_{\ket{\varphi} \in X} \left| \bracket{\psi}{\varphi}\right|^{2k} \\
\f{|X|}{\binom{d + k - 1}{k}} & = & \f{1}{|X|} \sum_{\ket{\psi},\ket{\varphi} \in X} \left| \bracket{\psi}{\varphi}\right|^{2k}
\eess
for all $0 \leq k \leq t$ and \Ref{eqn:kr t-design 3} follows.

Now we will show how (1) follows from (3). We will use that for the $k$-fold tensor product, the inner product evaluates to  $\bracket{\psi^{\otimes k}}{\varphi^{\otimes k}} = \bracket{\psi}{\varphi}^k$. Define the vector 
\[\ket{\upsilon} = \f{1}{|X|} \sum_{\ket{\psi} \in X} \ket{\psi}^{\otimes k} \otimes \overline{\ket{\psi}}^{\otimes k} - \int_{\text{F-S}} \ket{\psi}^{\otimes k} \otimes \overline{\ket{\psi}}^{\otimes k} d\ket{\psi}\] 
where integration is understood with respect to the coordinate functions of $\ket{\psi}^{\otimes k} \otimes \overline{\ket{\psi}}^{\otimes k}$. The inner product evaluates to 
\[ \bracket{\upsilon}{\upsilon} = \f{1}{|X|^2} \sum_{\ket{\psi}, \ket{\varphi} \in X} \left| \bracket{\psi}{\varphi}\right|^{2k} - \int_{\text{F-S}} \int_{\text{F-S}} \left| \bracket{\psi}{\varphi}\right|^{2k} d\ket{\varphi} d\ket{\psi}.\]
From \Ref{eqn:kr t-design 3}, Lemma \ref{lem:simple haar integral} and from the normalization of the Fubini-Study measure follows that
\bess
\f{1}{|X|^2} \sum_{\ket{\psi}, \ket{\varphi} \in X} \left| \bracket{\psi}{\varphi}\right|^{2k} - \int_{\text{F-S}} \int_{\text{F-S}} \left| \bracket{\psi}{\varphi}\right|^{2k} d\ket{\varphi} d\ket{\psi} & = &
\f{1}{\binom{d + k - 1}{k}} - \int_{\text{F-S}} \f{1}{\binom{d + k - 1}{k}} d\ket{\psi} \\
& = & \f{1}{\binom{d + k - 1}{k}} - \f{1}{\binom{d + k - 1}{k}} = 0
\eess
for all $0 \leq k \leq t$. 

From $\bracket{\upsilon}{\upsilon} = 0$ it follows that $\ket{\upsilon} = \v{o}$, thus \Ref{eqn:cubature formula} holds for every monomial in $Hom(k,k)_{\circ}$ as $\ket{\psi}^{\otimes k} \otimes \overline{\ket{\psi}}^{\otimes k}$ gives all monomials in $Hom(k,k)_{\circ}$ with coefficient $1$ in its coordinate functions. By linearity, we conclude that the cubature formula holds for all polynomials in $Hom(k,k)_{\circ}$ and thus $X$ is a $t$-design.
\end{proof}

Some more notation is needed. The \emph{``angle'' set} $A$ of a subset $X \subseteq CS^{d-1}$ is defined as \[A = \{\left|\bracket{\psi}{\varphi}\right|^2\,|\, \ket{\psi}, \ket{\varphi} \in X, \ket{\psi} \neq \ket{\varphi}\}.\] For $\ket{\psi} \in X$ and ``angle'' $\alpha \in A$, the subdegree of $\ket{\psi}$ with respect to $\alpha$ is \[d_{\alpha}(\ket{\psi}) = \left| \{\ket{\varphi} \in X\,|\,\left|\bracket{\psi}{\varphi}\right|^2 = \alpha\} \right|.\] If for all $\alpha \in A$, $d_{\alpha}(\ket{\psi})$ is the same for all $\ket{\psi} \in X$, $X$ is called a \emph{regular scheme}. The states of a set of  mutually-unbiased bases form a regular scheme.

\begin{theorem} \label{thm:mubs are 2-design} The states of a complete set of mutually-unbiased based form a $2$-design $X$ in $CS^{d-1}$ with ``angle'' set $\{0, \f{1}{d}\}$ and $d(d+1)$ elements.
\end{theorem}
\begin{proof}
The number of elements and the ``angle'' set follow from the definiton of mutually-unbiased bases \Ref{eqn:def mub}. We use statement (3) in Theorem \ref{thm:t-design stuff} and show that \Ref{eqn:kr t-design 3} holds for $k = 0, 1, 2$. The $k = 0$ case is immediate.

For $k = 1$, we see
\bess
\f{1}{d^2(d+1)^2} \sum_{\ket{\psi}, \ket{\varphi} \in X} \left| \bracket{\psi}{\varphi}\right|^{2} & = & 
\f{1}{d^2(d+1)^2} \left( (d+1)d d^2 \f{1}{d} + d(d+1) \right) \\
& = & \f{1}{d(d+1)} (d + 1) = \f{1}{d} = \f{1}{\binom{d + 1 - 1}{1}}.
\eess

For $k = 2$, we have
\bess
\f{1}{d^2(d+1)^2} \sum_{\ket{\psi}, \ket{\varphi} \in X} \left| \bracket{\psi}{\varphi}\right|^{4} & = & 
\f{1}{d^2(d+1)^2} \left( (d+1)d d^2 \f{1}{d^2} + d(d+1) \right) \\
& = & \f{1}{d(d+1)} (1 + 1) = \f{2}{d(d+1)} = \f{1}{\binom{d + 2 - 1}{2}}.
\eess
\end{proof}

\cite{KlappeneckerRoetteler2005} also showed the converse, which we state without a proof.
\begin{theorem} A $2$-design in $CS^{d-1}$ with ``angle'' set $\{0, \f{1}{d}\}$ is a union of $d+1$ mutually-unbiased bases.
\end{theorem}

\subsection{Equivalence to Our Approach}

The main result in Corollary \ref{cor:Haar avg = MUB avg} from the previous section now follows directly from Theorem \ref{thm:mubs are 2-design} as \[p(\ket{\psi}) = \bra{\psi} M \ket{\psi}\bra{\psi} N \ket{\psi} \] is a homogeneous polynomial in $Hom(2,2)_{\circ}$. The other direction follows as well. To see this, pick a monomial $m(\ket{\psi}) = x_a x_b \overline{x_c} \overline{x_d} \in Hom(2,2)_{\circ}$, where $x_i$ denotes the component of the $i$-th computational basis state in some state $\ket{\psi} = \sum_{i=1}^d x_i \ket{i}$. Let $M = \ket{c}\bra{a}$, $N = \ket{b}\bra{d}$ and observe that 
\bess
\bra{\psi} M \ket{\psi}\bra{\psi} N \ket{\psi} & = & \bra{\psi} \ket{c}\bra{a} \ket{\psi}\bra{\psi} \ket{b}\bra{d} \ket{\psi}  \\
& = & \overline{x_c} x_a \overline{x_d} x_b = m(\ket{\psi}).
\eess
This extends to all homogeneous polynomials by linearity of \Ref{eqn:Haar avg = MUB avg}, thus it also shows that a complete set of mutually-unbiased bases is a $2$-design.

\section{Efficient Fidelity Estimation}

\subsection{Introduction}

\begin{figure}[htbp]
	\centerline{
	\Qcircuit  @C=.5em @R=.7em
	{
  	\lstick{\ket{0}} & \gate{V^a_b} & \gate{U} & \gate{U\dag} & \gate{(V^a_b)\dag} & \meter & \cw
	} }
	\caption{Circuit to Estimate the Average Fidelity}
	\label{fig:estimation circuit}
\end{figure}
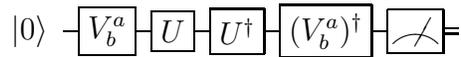

We are now ready to show that the average gate fidelity (see Corollary \ref{cor:avg-fidelity-explicit})
\[ F_{\text{avg}}(U, \E) = \f{\sum_k | \thr E_k |^2 + d}{d(d+1)}\] can be estimated using a simple experimental setup. The $E_k$ denote the Kraus operators of $\tilde{\E}$ (see Corollary \ref{cor:avg-fidelity-explicit}). Figure \ref{fig:estimation circuit} shows the circuit that can be used to estimate the fidelity of an implementation of $U$. 

However, we will need to make certain assumptions to end up with a circuit as simple as that. First of all, we need to assume that the cumulative noise characterized by $\tilde{\E}$ is independent of the actual quantum algorithm $U$ that is implemented in the quantum computer in question. Although $\tilde{\E}$ can be thought of as to cover the noise induced by our experimental control, it is clear that the cumulative noise will usually depend on the gate being implemented. It seems natural that an implementation of the Quantum Fourier Transform will introduce more noise than the implementation of the identity operation $\Id$. Furthermore, we need to assume that the additional pieces of the circuit used to measure the average fidelity introduce no additional noise. For our purposes, it would already be helpful if we could get a lower bound on the average fidelity. This is what our procedure will lead to, as the fidelity cannot increase when we implement $U$ and the additional operation $U\dag$.

\subsection{Using Mutually-Unbiased Bases}

We start with the basis state $\ket{0}$ and map it to a random vector $\ket{\psi^a_b}$ in one of the mutually-unbiased bases $B_a$ chosen at random. The parameters $a$ and $b$ are chosen classically at random. Then we apply the ``motion-reversal procedure'' $U U\dag$ \cite{EmersonEtAl2005_Noise} and measure the result in the $B_a$ basis. To implement this, we will show how to construct a unitary $V^a_b$: $\ket{0} \mapsto \ket{\psi^a_b}$, apply it to $\ket{0}$ in the beginning and apply $\left(V^a_b\right)^{\dag}$ at the end and measure with respect to $\ket{0}$ and $\ket{0^{\perp}}$. Let $p$ be the probability that the outcome is $\ket{0}$.  According to our assumptions, the cumulative noise is characterized by $\tilde{\E}$ and the quantum operation of the implementation is given as \[\E(\rho) = \sum_k E_k U\dag U \rho U\dag U E_k^{\dag} = \sum_k E_k \rho E_k\dag. \] 

The construction covers quantum channels as well. Set $U = \Id$ and we obtain the corresponding circuits that estimate the fidelity of a quantum channel.

\begin{theorem} \label{thm:probablity to measure 0 explicit} The probability to measure $\ket{0}$ is \be p = \f{\sum_k | \thr E_k |^2 + d}{d(d+1)} \ee
\end{theorem}
\begin{proof}
Using the main result in Corollary \ref{cor:main}, 
\bess
p & = & \f{1}{d(d+1)} \sum_{a=0}^{d} \sum_{b=0}^{d-1} \bra{\psi^a_b} \E( \ket{\psi^a_b}\bra{\psi^a_b}) \ket{\psi^a_b} \\
& = & \f{1}{d(d+1)} \sum_{a=0}^{d} \sum_{b=0}^{d-1} \bra{\psi^a_b} \sum_k E_k \ket{\psi^a_b}\bra{\psi^a_b} E_k\dag \ket{\psi^a_b} \\
& = & \f{1}{d(d+1)} \sum_{a=0}^{d} \sum_{b=0}^{d-1} \sum_k \bra{\psi^a_b} E_k \ket{\psi^a_b}\bra{\psi^a_b} E_k\dag \ket{\psi^a_b} \\
& \stackrel{\text{\small Cor.~\Ref{cor:main}}}{=} & \f{1}{d(d+1)} \sum_k \left( \thr E_k E_k^{\dag} + \thr E_k \thr E_k\dag \right)\\
& = & \f{1}{d(d+1)} \left(  \thr \left( \sum_k E_k E_k^{\dag} \right) + \sum_k \thr E_k \thr E_k\dag \right)\\
& = & \f{1}{d(d+1)} \left(  d + \sum_k |\thr E_k |^2 \right)
\eess
\end{proof}

\begin{cor} The probablity to measure $\ket{0}$ equals the average gate fidelity. \be p = F_{\text{avg}}(U, \E) \ee
\end{cor}
\begin{proof} Follows directly from Corollary \ref{cor:Haar avg = MUB avg}.
\end{proof}

Estimation of the average fidelity has been reduced to estimating the probability $p$. In the discussion of Quantum State Tomography in Section \ref{sec:QST}, it was shown how a probability can be estimated in $l$ trials within a standard deviation of at most $\frac{1}{\sqrt{l}}$. Hence we need a constant number of experiments to estimate the average fidelity within a fixed absolute error. 

In order to justify the assumption that the additional circuit around $U$ and $U\dag$ supporting the estimation of the average fidelity do not cause any significant additional noise, we need to find constructions using as few additional qubits and gates as possible. The idea is that additional gates and qubits generally require more experimental control which in turn introduces additional noise. In order to minimize the effect of this additional noise, we would like to keep the number of gates in the additional circuitry smaller than the number of gates used to realize $U$ and $U\dag$. We would also like to keep the number of ancillas as small as possible. 
\subsection{Prime Dimension Construction}

The construction of mutually-unbiased bases for prime power dimension was particularly intriguing and it turns into a very easy construction if the dimension $d$ is not a power of a prime but just a prime $p > 2$. Let $\Hilbert_d$ be the Hilbert space of the $N$ qubit system of dimension $d = 2^N$. Let $p$ be the smallest prime such that $p \geq 2^N$ and let $\Hilbert_p$ be a Hilbert space of dimension $p$. It is known that $p < 2^{N+1}$ \cite[Th.~5.9]{ErdosSuranyi2003}, which we will use to emulate dimension $p$ in dimension $2^{N+1}$. 

It seems tempting to just add another qubit to the circuit and embed $\Hilbert_p$ into the $2^{N+1}$-dimensional Hilbert space $\Hilbert_{2d}$. This is done by identifying $\Hilbert_p$ with the span of the first $p$ basis vectors of the computational basis of $\Hilbert_{2d}$, for example. Let $\E$ be the quantum operation on $\Hilbert_d$ in question and let \[\E(\rho) = \sum_k A_k \rho A_k\dag\] be its Kraus operator-sum decomposition from Fact \ref{fact:kraus decomp}. The map $\E'$ in the larger space $\Hilbert_{2d}$ is given by \[\E'(\rho) = (\E \otimes \Id)(\rho) = \sum_k (A_k \otimes \Id) \rho (A_k\dag \otimes \Id).\] This is not a trivial embedding of $\E$ into $\Hilbert_p$ as the tensor product structure of $\E'$ forbids the direct use of $\ket{\psi^a_b}$ on $\Hilbert_p$ as this would only make sense if $\E'$ was a direct sum $\E \oplus \Id$. However, we will show how we can still make use of the embedded prime dimension Hilbert space $\Hilbert_p$.

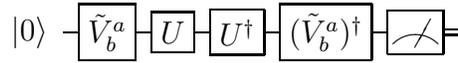
\begin{figure}[htbp]
	\centerline{
	\Qcircuit  @C=.5em @R=.7em
	{
  	\lstick{\ket{0}} & \gate{\tilde{V}^a_b} & \gate{U} & \gate{U\dag} & \gate{(\tilde{V}^a_b)\dag} & \meter & \cw
	} }
	\caption{Circuit to Estimate the Average Fidelity using MUBs.}
	\label{fig:estimation circuit modified}
\end{figure}

Suppose $P$ is a projector from $\Hilbert_{2d}$ onto $\Hilbert_d$. Clearly, it is also a projector from $\Hilbert_p$ onto $\Hilbert_d$. Let further $\tilde{p}$ denote the probability to measure zero at the end of the circuit shown in Figure \ref{fig:estimation circuit modified}, where we average over the set of states \[\wtilde{B} = \{\ket{\wtilde{\psi}^a_b} = P \ket{\psi^a_b}\,|\, a \in \{0, 1, \dots, p\}, b \in \{0, 1, \dots, p-1\}\}.\] Then $\tilde{p}$ is the average fidelity up to a constant factor. More precisely, we have the following
\begin{theorem} The probability to measure zero averaged over the the uniform distribution of $a$ and $b$ is given by \be \label{eqn: prob to measure zero with modified mubs} \tilde{p} = \f{\sum_k | \thr E_k |^2 + d}{p(p+1)}. \ee
\end{theorem}
\begin{proof} We use that $P$ also maps $U \otimes \Id$ onto $U$ by conjugation, hence \[\bra{\wtilde{\psi}^a_b} U \otimes \Id \ket{\wtilde{\psi}^a_b} = \bra{\psi^a_b} P (U \otimes \Id) P \ket{\psi^a_b} = \bra{\psi^a_b} U \ket{\psi^a_b}.\] The same argument can be used in conjunction with Corollary \ref{cor:main} and Theorem \ref{thm:probablity to measure 0 explicit} and the right-hand side of \Ref{eqn: prob to measure zero with modified mubs} follows. 
\end{proof}
\begin{cor} The average fidelity is given by $F_{\text{avg}}(\E,U) = \frac{p(p+1)}{d(d+1)} \tilde{p}$.
\end{cor}

It remains to show that we can efficiently construct the elements $\ket{\wtilde{\psi}^a_b}$ of the set $\wtilde{B}$ given the gate $V'^a_b$ that creates the state $\ket{\psi^a_b}$ from a complete set of mutually-unbiases bases in prime dimension $p$ on $N+1$ qubits. 

\begin{theorem}\label{thm:projected mubs} Let $V'^a_b$ denote the gate that maps \[V'^a_b: \ket{0} \mapsto \ket{\psi^a_b}.\] Let $C(N)$ and $D(N)$ denote its gate complexity and depth, respectively. Then $\wtilde{V}^a_b$ can be constructed using $O(N^2 + C(N))$ single and two-qubit gates in depth $O(N + D(N))$ using two ancilla qubits.
\end{theorem}
\begin{proof}

Using $V'^a_b$ on $N$ qubits and the first ancilla, we can construct $\ket{\psi^a_b} = V'^a_b\ket{0^{\otimes (N+1)}}$ on $N+1$ qubits and leave the second ancilla in the state $\ket{0}$. We can rewrite \[\ket{\psi^a_b}\ket{0} = \cos \theta \ket{\phi_0} \ket{0} + \sin \theta \ket{\phi_1} \ket{1}\] where $\ket{\phi_0}$ and $\ket{\phi_1}$ are normalized states on $N$ qubits and $\cos \theta$ and $\sin \theta$ depend on the amplitudes of the components of $\ket{\psi^a_b}$ that have a $0$ and a $1$ on the last qubit, respectively. The value $\theta$ can be determined from the construction for the mutually-unbiased bases $V'^a_b$. Observe that $\ket{\phi_0}\ket{0}$ and $\ket{\phi_1}\ket{1} $ are the renormalized projections of $\ket{\psi^a_b}$ onto the subspaces where the ancilla is in $\ket{0}$ and $\ket{1}$, respectively. 

In order to make use of just one round of amplitude amplification (Section \ref{sec:amplitude amplification}), we will rotate  the second ancilla to create a ''nice'' angle. Choose \[\alpha = \frac{\cos \f{\pi}{3}}{\cos \theta}\] and apply the rotation \[ R = \bv \alpha & -\sqrt{1-\alpha} \\ \sqrt{1-\alpha} & \alpha \ev\] to the second ancilla that is in its initial state $\ket{0}$. The state of the whole $N+2$ qubit system becomes
\[ \ket{\varphi} = \alpha \cos \theta \ket{\phi_0} \ket{00} + \sqrt{1-\alpha} \cos \theta \ket{\phi_0} \ket{01} - \sqrt{1-\alpha} \sin \theta \ket{\phi_1} \ket{10} + \alpha \sin \theta \ket{\phi_1} \ket{11}. \]
Substituting \[\ket{\phi^{\perp}} = \sqrt{1-\alpha} \cos \theta \ket{\phi_0} \ket{01} - \sqrt{1-\alpha} \sin \theta \ket{\phi_1} \ket{10} + \alpha \sin \theta \ket{\phi_1} \ket{11} \] we end up with
\[ \ket{\varphi} = \cos \f{\pi}{3} \ket{\phi_0} \ket{00} + \sin \f{\pi}{3} \ket{\psi^{\perp}}.\]

Using only one round of amplitude amplification (see Lemma \ref{lem:ampl aplif}), we can amplify the amplitude of $\ket{\phi_0} \ket{00}$ from $\cos \f{\pi}{3}$ to $\cos 3 \f{\pi}{3} = \cos \pi = 1$. As this is a product state between the $N$ qubits and the ancillas and the ancillas have been restored to their initial value, we can discard both ancillas after this step.

In order to implement the amplitude amplification step, we need to implement the reflections $U_{\text{bad}}^{\perp}$ and $U_0^{\perp}$. We will formalize these first for the computational basis on the $N+2$ qubit system. This yields \[U_{\text{bad}}^{\perp}: \ket{x_0 x_1 \dots x_{N} x_{N+1}} \mapsto (-1)^{[x < 2^N]} \ket{x_0 x_1 \dots x_{N} x_{N+1}}\] and \[U_0^{\perp}: \ket{x_0 x_1 \dots x_{N} x_{N+1}} \mapsto (-1)^{[x_1 = x_2 = \dots = x_{N+1} = 0]} \ket{x_0 x_1 \dots x_{N} x_{N+1}}.\]

$U_{\text{bad}}^{\perp}$ is just a phase-gate conditional on both ancillas being one, thus we can treat it as a two-qubit gate acting on the ancillas only. Therefore, \[U_{\text{bad}}^{\perp} = \bv -1 & 0 & 0 & 0 \\ 0 & 1 & 0 & 0 \\ 0 & 0 & 1 & 0 \\ 0 & 0 & 0 & 1 \ev.\]
$U_0^{\perp}$ is a phase gate conditional on all qubits being zero, which is equivalent to an $(N+1)$-fold controlled phase \[P_{-0} = \bv -1 & 0 \\ 0 & 1 \ev.\] It can be implemented \cite{NielsenChuang2000} using $O(N^2)$ single and two-qubit gates in depth $O(N)$. Figure \ref{fig:circuit for the mub projection} shows the complete circuit that computes  $\wtilde{V}^a_b = Q V'^a_b$, where $Q = V'^a_b U_0^{\perp} (V'^a_b)\dag U_{\text{bad}}^{\perp}$ is the amplitude amplification operator.

\begin{figure}[htbp]
	\centerline{
	\Qcircuit  @C=.7em @R=.7em
	{
  	 \push{} \gategroup{1}{1}{4}{1}{1em}{\{} & \lstick{\hspace{1em} \ket{0}} & \multigate{4}{V'^a_b} & \qw & \qw & \qw & \multigate{4}{(V'^a_b)\dag} & \qw & \gate{P_{-0}} \qw & \qw &  \multigate{4}{V'^a_b} & \qw \\
  	 & \lstick{\vspace{1em} \ket{0}} & \ghost{V'^a_b} & \qw & \qw & \qw & \ghost{(V'^a_b)\dag} & \qw & \ctrlo{-1} & \qw & \ghost{V'^a_b} & \qw \\
  	 & \vdots & & \vdots & & \vdots & & \push{\vspace{1em} \vdots \vspace{1em}} & & \vdots & &  \\  	 
   	 & \lstick{\vspace{1em} \ket{0}} & \ghost{V'^a_b} & \qw & \qw & \qw & \ghost{(V'^a_b)\dag} & \qw & \ctrlo{-2} & \qw & \ghost{V'^a_b} & \qw  \\
  	 %
   	 \push{} \gategroup{5}{1}{6}{1}{1em}{\{} & \lstick{\hspace{1em} \ket{0}} & \ghost{V'^a_b} & \qw & \multigate{1}{U_{\text{bad}}^{\perp}}  & \qw & \ghost{(V'^a_b)\dag} & \qw & \ctrlo{-1} & \qw & \ghost{V'^a_b} & \multimeasureD{1}{\text{Junk}} \\
   	 & \lstick{\vspace{1em} \ket{0}} & \gate{R} & \qw & \ghost{U_{\text{bad}}^{\perp}} & \qw & \qw & \qw & \ctrlo{-1} & \qw & \qw & \ghost{\text{Junk}} 
	} }
	\caption[Circuit that computes the projected MUBs $\wtilde{V}^a_b \ket{0^{\otimes N}}$.]{Circuit that computes the projected MUBs $\wtilde{V}^a_b \ket{0^{\otimes N}}$. The classical controls $a$ and $b$ are not shown here.}
	\label{fig:circuit for the mub projection}
\end{figure}
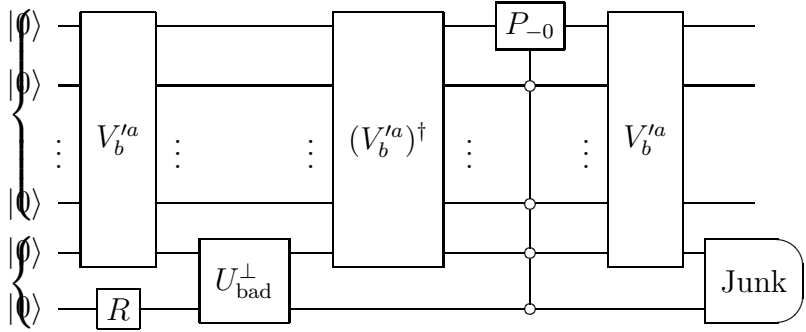
\end{proof}

Therefore adding two qubits to the system enables us to employ the prime dimension construction for mutually-unbiased bases. In the remainder of this section, we will show an explicit gate decomposition of $V'^a_b$. The arithmetic in $\F_p$ will be implemented on $N+1$ qubits, so that we will need only one additional qubit. Furthermore, this ancilla does not need to be sent through the channel or be fed to the noisy implementation $\E$ as it can be discarded after the projected mutually-unbiased state has been constructed. 

The construction from Theorem \ref{thm:mub kr} ensures that the initial angle is given by $\cos \theta = \sqrt{\f{p}{2^N}}$, therefore $\theta = \arccos \sqrt{\f{p}{2^N}}$ for all $a, b \in \F_p$. In case of the computational basis, we do not need to use any amplitude amplification. In this case the resulting state either is a computational basis state for $N$ qubits or the projection is $\ket{o}$, in which case we will define that the final measurement yields the correct value, depending on which basis state $\ket{b}$ is to be chosen.

We will now show that $V'^a_b$ has an efficient gate decomposition.

\begin{theorem}\label{thm:efficient prime power mub constrution}
$V'^a_b$ can be realized using $N+1$ qubits with $O(N^2)$ gates in depth $O(N)$.
\end{theorem}
\begin{proof}
From Theorem \ref{thm:mub kr}, the states of a maximal set of $p + 1$ mutually-unbiased bases are given by 
\be \ket{\psi^a_b} = \frac{1}{\sqrt{p}} \sum_{x \in \F_p} \omega_p^{a x^2 + bx} \ket{x}
\ee
for $a, b \in \F_p$. This can be rewritten as 
\bess \label{eqn:mub construction prime dim rewritten}
\ket{\psi^a_b} & = & \frac{1}{\sqrt{p}} \sum_{x \in \F_p} \omega_p^{a x^2 + bx} \ket{x} \\
& = &  \frac{1}{\sqrt{p}} \sum_{x \in \F_p} \left( \omega_p^a \right)^{x^2} \left( \omega_p^b \right)^x \ket{x}.
\eess
We see from \Ref{eqn:mub construction prime dim rewritten} that we need to implement the basic operations
\be \label{eqn:x to w_p^bx} \ket{x} \mapsto  \left( \omega_p^b \right)^x  \ket{x}\ee and 
\be \label{eqn:x to w_p^ax^2} \ket{x} \mapsto  \left( \omega_p^a \right)^{x^2} \ket{x}.\ee 

The implementation of \Ref{eqn:x to w_p^bx} is straightforward using a phase gate \[ P_i = \bv 1 & 0 \\ 0 & \left( \omega_p \right)^{b 2^i} \ev \] on the $i$-th qubit, where we label the $N+1$ qubits from $0$ to $N$.
\begin{lemma} For all $\ket{x}$, \[\bigotimes_{i=0}^{N} P_i \ket{x} = \left( \omega_p \right)^{bx} \ket{x}.\]
\end{lemma}
\begin{proof} We use the decomposition $x = \sum_{i=0}^N x_i 2^i$. Direct calculation gives
\bess
\bigotimes_{i=0}^{N} P_i \ket{x} & = & \bigotimes_{i=0}^{N} \left(P_i \ket{x_i} \right) \\
& = & \bigotimes_{i=0}^{N} \left( \left( \omega_p \right)^{b 2^i x_i} \ket{x_i} \right) \\
& = & \prod_{i=0}^{N}\left( \omega_p \right)^{b 2^i x_i} \ket{x_0 x_1 \dots x_N} \\
& = & \left( \omega_p \right)^{b \sum_{i=0}^{N} 2^i x_i} \ket{x} \\
& = & \left( \omega_p \right)^{bx} \ket{x}.
\eess
\end{proof}

For the implementation of \Ref{eqn:x to w_p^ax^2} we observe that for $x = \sum_{i=0}^{N} x_i 2^i$, we get
\bess x^2  & = & \left( \sum_{i=0}^{N} x_i 2^i \right)^2 = \sum_{i,j=0}^N x_i x_j 2^{i+j} \\
& = & \sum_{i<j}^N 2 x_i x_j 2^{i+j} + \sum_{i=0}^N x_i^2 2^{2i} = \sum_{i<j}^N x_i x_j 2^{i+j+1} + \sum_{i=0}^N x_i 2^{2i}
\eess
where the last equation follows from $x^2 = x$ for $x \in \{0,1\}$. Therefore 
\be \label{eqn:final form phase injection of x^2}
\left( \omega_p^a \right)^{x^2} = \prod_{i<j}^N \left( \omega_p^a \right)^{2^{i+j+1} x_i x_j} \prod_{i=0}^N \left( \omega_p^a \right)^{2^{2i} x_i}.
\ee 
The first term in \Ref{eqn:final form phase injection of x^2} is a product of \[{N+1 \choose 2} = \frac{(N+1)(N+2)}{2} = O(N^2)\] conditional phases $\left( \omega_p^a \right)^{2^{i+j+1}}$ which can be realized using a controlled-phase gate. The remaining term corresponds to a single qubit phase gate on each of the $N+1$ qubits. 

\begin{figure}[htbp]
	\centerline{
	\Qcircuit @C=.5em @R=.7em
	{
		& \ctrl{1} & \ctrl{2} & \qdots \qw & \ctrl{4} & \qw & \qdots \qw & \qw & \qdots \qw & \qw & \qdots \qw & \gate{\omega^a} & \qw \\
		& \gate{\omega^{2^2 a}} & \qw & \qdots \qw & \qw & \ctrl{1} & \qdots \qw & \ctrl{3} &  \qdots \qw & \qw & \qdots \qw & \gate{\omega^{2a}} & \qw \\
		& \qw & \gate{\omega^{2^3 a}} & \qdots \qw & \qw & \gate{\omega^{2^4 a}} & \qdots \qw & \qw & \qdots \qw & \ctrl{2} & \qdots \qw & \gate{\omega^{2^4 a}} & \qw \\
		& \vdots 	& \vdots 			&			  & 		&					&			&				 & 			 & & & \push{\vspace{1em} \vdots \vspace{1em}} & \\
		& \qw & \qw	& \qdots \qw & \gate{\omega^{2^{N+1} a}} & \qw & \qdots \qw & \gate{\omega^{2^{N+2} a}} & \qdots \qw & \gate{\omega^{2^{N+3} a}} & \qdots \qw &  \gate{\omega^{2^{2N} a}} & \qw 
	} }
	\caption{The Circuit that Maps $\ket{x} \mapsto \omega^{ax^2 + bx} \ket{x}$, where $\omega = e^{2\pi i/p}$.}
	\label{fig:circuit prime dimension construction -- phases}
\end{figure}
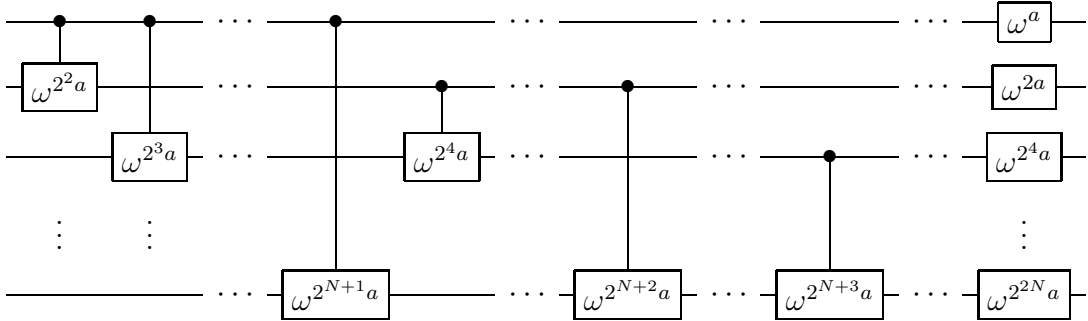

Thus we end up with the phase injection circuit in Figure \ref{fig:circuit prime dimension construction -- phases}. The phases are classically controlled by the values of $a$ and $b$. The complete circuit is shown in Figure \ref{fig:circuit prime dimension construction} and shows how the Hadamard gates that create the initial superposition are conditional on whether the basis is the computational basis or one of the $N$ other mutually-unbiased bases. We define $a = p$ to denote the computational basis and restrict $a \in \{0, 1, \dots, p\}$ and $b \in \{0, 1, \dots, p - 1\}$ and assume a binary encoding of $a$ and $b$ into $\left\lceil \log (p + 1)\right\rceil = N+1$ classical bits. Both the phase and the controlled-phase gates are conditional on the classical choices for $a$ and $b$. If $a = p$, we just select the $b$-th state of the computational basis. The last part of the complete circuit is responsible for that. The total cost is $\frac{(N+1)(N+2)}{2} + 3 (N+1) = O(N^2)$ single and two-qubit gates on $N+1$ qubits in a depth of $N + 3 = O(N)$. This circuit can easily be reversed by reversing each of the one and two-qubit gates. The Hadamard is its own inverse, whereas the inverse of a phase gate is a phase gate with the inverse phase.

\begin{figure}[htbp]
	\centerline{
	\Qcircuit @C=2em @R=1em
	{
  	\lstick{a} 			 & \control \cwx[2] \cw & \control \cwx[2]		\cw						&  \cw & \control \cwx[1] \cw & \control \cwx[1] \cw & \qdots \cw & \control \cwx[1] \cw \\
  	\lstick{b} 			 & \cw & \cw 		 									& \control \cwx[1] \cw		&  \control \cwx[1] \cw & \control \cwx[2] \cw & \qdots \cw & \control \cwx[5] \cw\\
  	\lstick{\ket{0}} & \multigate{4}{H^{\otimes(N+1)}} & \multigate{4}{\text{Phase } ax^2} & \multigate{4}{\text{Phase } bx } & \targ & \qw & \qdots \qw & \qw & \qw\\
  	\lstick{\ket{0}} & \ghost{H^{\otimes(N+1)}} & \ghost{\text{Phase } ax^2} 		& \ghost{\text{Phase } bx } 		& \qw & \targ & \qdots \qw & \qw & \qw \\
  	\lstick{\ket{0}} & \ghost{H^{\otimes(N+1)}} & \ghost{\text{Phase } ax^2} 		& \ghost{\text{Phase } bx } 		& \qw & \qw & \qdots \qw & \qw & \rstick{\ket{\psi^a_b}} \qw \\
  	 					 		\vdots	 & 	&  		&		&   & & & & \vdots \\
  	\lstick{\ket{0}} & \ghost{H^{\otimes(N+1)}} & \ghost{\text{Phase } ax^2} 		& \ghost{\text{Phase } bx } 		& \qw & \qw & \qdots \qw & \targ & \qw
	} }
	\caption{The Circuit that Creates $\ket{\psi^a_b}$ Given $a$ and $b$ Classically.}
	\label{fig:circuit prime dimension construction}
\end{figure}
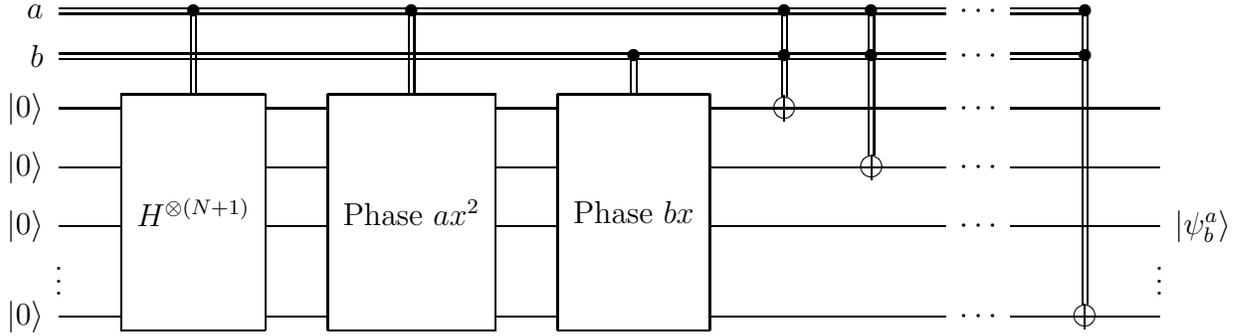

\end{proof}

\subsection{Prime Power Dimension Construction}
\label{sec:mub prime power dim constr}

The construction for a Hilbert space of prime dimension can be generalized to a Hilbert space of prime power dimension $p^N$ where $p > 2$ is a prime and $N \in \N$ is the number of qudits in the system, each a system of dimension $p$. In that case, the construction from Theorem \ref{thm:mub kr} reads
\be \label{eqn:mub kr construction repeated} \ket{\psi^a_b} = \frac{1}{\sqrt{p^N}} \sum_{x \in GF(p^N)} \omega_p^{\tr(a x^2 + bx)} \ket{x}
\ee
for $a, b \in \F_p$ plus the computational basis for $a = p$ by definition. Using the polynomial representation from Section \ref{sec:galois fields}, we may choose to represent $GF(p^N)$ in a vector space $\F_p^k$ so that each qudit encodes the coefficient of $\xi^i$, $i \in \{0, 1, \dots, N-1\}$ of $x = x_0 + x_1 \xi + \dots + x_{N-1}\xi^{N-1} \in GF(p^N)$. We use $\xi$ to denote the formal variable in the polynomial representation in order to avoid confusion with the variable $x$ that we will use to denote an element of $GF(p^N)$. We define $\v{x}$ as the column vector \[\v{x} = \bv x_0 \\ x_1 \\ \vdots \\ x_{N-1} \ev\] that represents $x$ in $\F_p^N$.

As trace is a linear functional $\tr: \F_p^N \rightarrow \F_p$, Fact \ref{fact:field trace} guarantees the existence of a vector $\v{t} \in \F_p^N$ such that $\tr x = (\v{t}, \v{x})$ where $(\cdot,\cdot)$ denotes the usual inner product on $\F_p$. Furthermore, the multiplication in $GF(p^N)$ is a linear function of $\F_p^N$ and thus for any $y \in GF(p^N)$ there is a matrix $M_y \in \F_p^{N \times N}$ such that $\v{yx} = M_y \v{x}$. Hence \be \label{eqn:trace of product as inner prod} \tr yx = (\v{t}, M_y \v{x}) = (\v{t_y}, \v{x}) \ee where $\v{t_y} = \v{t} M_y$.

This representation \Ref{eqn:trace of product as inner prod} of the trace function enables us to rewrite \Ref{eqn:mub kr construction repeated} as
\bess
\ket{\psi^a_b} & = & \frac{1}{\sqrt{p^N}} \sum_{x \in GF(p^N)} \omega_p^{(\v{t_a}, \v{x^2})} \omega_p^{(\v{t_b}, \v{x})} \ket{x}
\eess
where $\v{t_a}$ and $\v{t_b}$ are classical values that can be precomputed by the classical control. 

Beaudrap et al.\@ \cite{BeaudrapEtAl2002} showed how to implement a generalization of the Quantum Fourier Transform for qudits. 
\begin{df} The \emph{generalized Quantum Fourier Transform} on $N$ qudits of dimension $p$ relative to any nonzero linear mapping $\varphi$ on $GF(p^N)$ is defined as
\[F_{p^N, \varphi}: \ket{x} \mapsto \frac{1}{\sqrt{p^N}} \sum_{y \in GF(p^N)} \omega_p^{\varphi(xy)} \ket{y}. \]
\end{df}
\begin{theorem}[\cite{BeaudrapEtAl2002}] \label{thm:gen qft} Let $p$ be a constant, $N \in \N$, and let $\varphi: GF(p^N) \rightarrow \F_p$. Then $F_{p^N,\varphi}$ can be performed exactly by a quantum circuit of size $O(N^2)$.
\end{theorem}

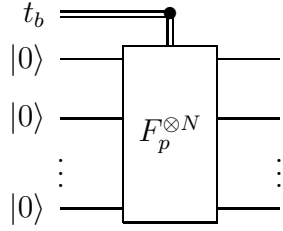
\begin{figure}[htbp]
	\centerline{
	\Qcircuit @C=2em @R=1em
	{
	 \lstick{t_b} & \control \cwx[1] \cw & \\
	 \lstick{\ket{0}} & \multigate{3}{F_p^{\otimes N}} & \qw \\
	 \lstick{\ket{0}} & \ghost{F_p^{\otimes N}} & \qw \\
		\vdots & & \vdots \\	 
	 \lstick{\ket{0}} & \ghost{F_p^{\otimes N}} & \qw
	} }
	\caption{The Circuit that Creates $\frac{1}{\sqrt{p^N}} \sum_{x \in GF(p^N)} \omega_p^{\tr bx}$.}
	\label{fig:circuit for tr bx for prime power dim}
\end{figure}

This makes use of the fact that every nonzero linear functional $\varphi$ on $GF(p^N)$ can be represented as $\varphi(x,y) = \v{x}^T M_{\varphi} \v{y}$ where $M_{\varphi} \in \F_p^{N \times N}$, which is a generalized inner product on $\F_p^N$. In the case of the trace function, we can reduce $\varphi$ to the conventional inner product $\varphi(xy) = \v{x}^T \v{y}$ and prepare the initial state as $t_b$. Figure \ref{fig:circuit for tr bx for prime power dim} shows how this can be done when $t_b$ is prepared by the classical control. Now the implementation in the proof of Theorem \ref{thm:gen qft} reduces to implementing $F_p$ on every qudit, which can be done exactly using $O(N)$ gates.

After that, the phases for the quadratic term need to be injected. Thus we need to implement the transformation
\be \label{eqn:tr ax^2} \ket{x} \mapsto \omega_p^{(\v{t_a}, \v{x^2})}. \ee Suppose the primitive polynomial that is used in the representation of $GF(p^N)$ is $h(\xi)$. First we observe that for $x \in GF^{p^N}$, $x = x_0 + x_1 \xi + \dots + x_{N-1} \xi^{N-1}$,
\bess 
x^2 & = & \sum_{i,j=0}^{N-1} x_i x_j \xi^{i+j} \bmod h(\xi) \\
\label{eqn:prime power - the square decomp} & = & \sum_{i<j}^{N-1} 2 x_i x_j (\xi^{i+j} \bmod h(\xi)) +  \sum_{i=0}^{N-1} x_i^2 (\xi^{2i} \bmod h(\xi)).
\eess
Denote $\v{\xi}^{(i,j)}$ the vector corresponding to $\xi^{i+j} \bmod h(\xi)$. Then the inner product $(\v{t_a}, \v{x^2})$ can be written as \[(\v{t_a}, \v{x}) = \sum_{i,j = 0}^{N-1} x_i x_j (\v{t_a}, \v{\xi}^{(i,j)})\] where $\v{t_a}$ only depends on the number of qudits $N$ and the classical parameter $a$, whereas $\v{\xi}^{(i,j)}$ only depends on the number of qudits $N$. Thus $t_a^{(i,j)} = (\v{t_a}, \v{\xi}^{(i,j)})$ can be computed classically for all values of $a$, $i$, and $j$. Therefore we only need to implement two-qudit gates \[ \text{Phase}_2^{(i,j)}: \ket{a} \otimes \ket{b} \mapsto \omega_p^{2 t_a^{(i,j)} a b} \ket{a} \otimes \ket{b}\]
for all cross-term qudits $(i,j)$ in \Ref{eqn:prime power - the square decomp}, and single-qudit gates 
\[ \text{Phase}_1^i: \ket{a} \mapsto \omega_p^{t_a^{(i,i)} a^2} \ket{a} \]
for the diagonal terms. Hence we need $N \choose 2$ two-qudit gates and $N$ single-qudit gates and the circuit has a depth $O(N)$, as each qudit appears $2N-1 = O(N)$ times in \Ref{eqn:prime power - the square decomp}.

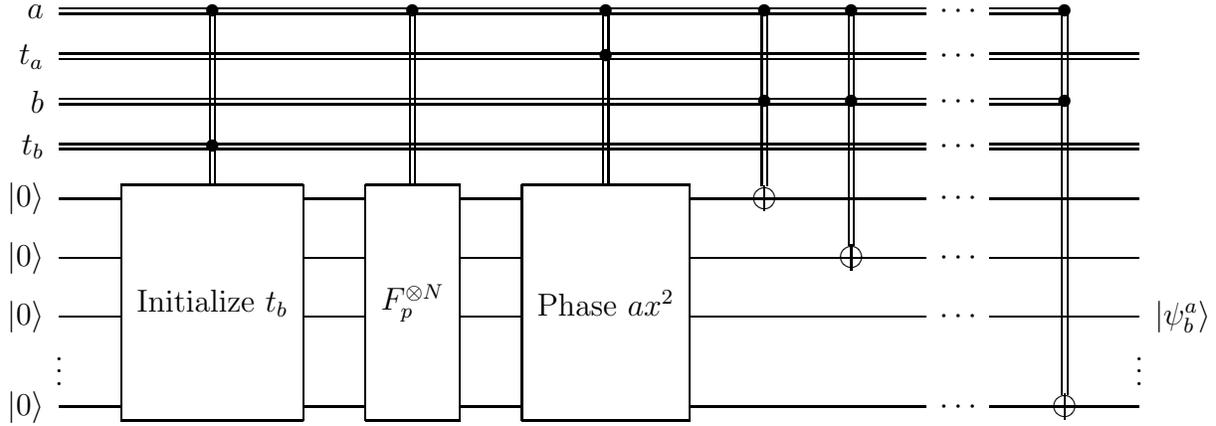
\begin{figure}[htbp]
	\centerline{
	\Qcircuit @C=2em @R=1em
	{
  	\lstick{a} & \control \cwx[4] \cw & \control \cwx[4] \cw	& \control \cwx[1] \cw & \control \cwx[2] \cw & \control \cwx[2] \cw & \qdots \cw & \control \cwx[2] \cw \\
  	\lstick{t_a} & \cw & \cw & \control \cwx[3] \cw & \cw & \cw & \qdots \cw & \cw & \cw \\
  	\lstick{b} & \cw & \cw & \cw	& \control \cwx[2] \cw & \control \cwx[3] \cw & \qdots \cw & \control \cwx[6] \cw\\
  	\lstick{t_b} & \control \cwx[1] \cw & \cw & \cw & \cw & \cw & \qdots \cw & \cw & \cw \\
  	\lstick{\ket{0}} & \multigate{4}{\text{Initialize } t_b} & \multigate{4}{F_p^{\otimes N}} & \multigate{4}{\text{Phase } ax^2}  & \targ & \qw & \qdots \qw & \qw & \qw\\
  	\lstick{\ket{0}} 	& \ghost{\text{Initialize } t_b}  & \ghost{F_p^{\otimes N}} & \ghost{\text{Phase } ax^2} 		& \qw & \targ & \qdots \qw & \qw & \qw \\
  	\lstick{\ket{0}} & \ghost{\text{Initialize } t_b} 	& \ghost{F_p^{\otimes N}} & \ghost{\text{Phase } ax^2} 	& \qw & \qw & \qdots \qw & \qw & \rstick{\ket{\psi^a_b}} \qw \\
  	 					 		\vdots	 & 	&  		&		&   & & & & \vdots \\
  	\lstick{\ket{0}} & \ghost{\text{Initialize } t_b} 	& \ghost{F_p^{\otimes N}} & \ghost{\text{Phase } ax^2} 			& \qw & \qw & \qdots \qw & \targ & \qw
	} }
	\caption{The Circuit that Creates $\ket{\psi^a_b}$ Given $a$, $b$, $t_a = \left(t_a^{(i,j)}\right)$ and $t_b$ Classically.}
	\label{fig:circuit prime power dimension construction}
\end{figure}

The computational basis can be integrated by classically controlling the phase gates on $a \neq p^N$. If $a = p^N$, we prepare the state $\ket{b}$ using classically-controlled addition gates on each qudit. This costs $N$ additional one-qudit gates. Therefore the whole circuit can be implemented using $O(N^2)$ single and two-qudit gates in depth $O(N)$. Figure \ref{fig:circuit prime power dimension construction} shows the complete circuit.

\subsection{Galois Ring Construction}

Another approach to constructing MUBs in dimension $d = 2^N$ is to use Theorem \ref{thm:mub qubits} and employ Galois ring arithmetic. Although it is known how to implement Galois ring arithmetic classically \cite{Abrahamsson2004} and thus can be realized on a quantum computer with only polynomial overhead, it is not clear how to do it with only a modest number of additional qubits. The problem is that the Galois ring has $4^N$ elements, which require $2N$ qubits for a faithful representation. 

Remember the expression for the states of the mutually-unbiased bases \Ref{eqn:mub kr} \[\ket{\psi^a_b} = \frac{1}{\sqrt{2^n}} \sum_{x \in \Tau_n)} \omega_4^{\tr((a + 2b) x)} \ket{x}.\] Although we can classically precompute the vectors such that the trace expression in \Ref{eqn:mub qubits} reduces to inner products in $\Z_4^N$ using the ideas from the previous section, the representation problem remains. Specifically, we can compute vectors $\v{t}_{X^i}$ for a basis $\{X^i\,|\,i = 0, 1, \dots, N-1\}$ for $\Z_4^N$. Thus we can reduce $\tr((a + 2b) x)$ to $\v{t}_x \v{y}$, where $\v{y}$ is the polynomial representation of $a+2b$. As $a+2b$ ranges throughout all of $GF(4^N)$ and as we do not need not partition the states \Ref{eqn:mub qubits} into bases, we can ignore the Teichm\"{u}ller decomposition. Using the precomputed values of $\v{t}_{X^i}$, all we need to compute quantumly is the polynomial representation of $x \in \Tau$. However, a naive approach will require $2N$ qubits and enough multiplication operations to realize $X \mapsto X^j \bmod h(X)$ for all $j = 1, \dots, 2^N - 2$. Even using repeated squaring, this still requires $N$ multiplications. Each multiplication seems to require at least $N$ elementary gates in the quantum setting \cite{BeauregardEtAl2003}, thus we end up with at least $O(N^2)$ gates on least $N$ ancilla qubits.

\section{Discussion}

Mutually-unbiases bases are a powerful tool in quantum information. They can be used to efficiently estimate the average fidelity in an experimental context. \cite{KlappeneckerRoetteler2005} showed that they might even be more powerful as they form a $2$-design for quantum states, for which we gave a different proof. 

We constructed efficient circuits that generate a state out of a complete set of mutually-unbiases bases. However, these circuits still need $O(N^2)$ gates, which might be reduced to only $O(N)$ gates. As these circuits should be used for noise estimation, we need to make sure that the additional noise caused be the circuits is small compared to the circuit to be measured. As the approximate Quantum Fourier Transform can be realized on $O(N \log N)$ gates, it seems that $O(N^2)$ is still too high.

\clearpage
\thispagestyle{empty}
\cleardoublepage

\chapter{Unitary $2$-Designs}
\label{ch:unitary 2-designs}
\markright{Unitary $2$-Designs}

\section{Motivation and Notation}
\label{sec:motivation 2designs}

In the previous section, we showed how to construct a set of unitary transformations $\U = \{U_k\,|\,k = 1 \dots K\}$ on a $d$-dimensional Hilbert space $\Hilbert_d$, $d$ being a prime power, such that
\be \label{eqn:2-design formula} \sum_{k=1}^{K} \bra{\psi_0} U_k\dag M U_k \ket{\psi_0} \bra{\psi_0} U_k\dag N U_k \ket{\psi_0} = \int_{\text{F-S}} \bra{\psi} M \ket{\psi} \bra{\psi} N \ket{\psi} d\ket{\psi}\ee for any linear operators $M, N \in L(\Hilbert)$. We also saw that this is equivalent to the condition that $\U$ generates a state $2$-design given a fiducial initial state ($\ket{0}$ in the previous chapter). However, \Ref{eqn:2-design formula} only holds for a very specific initial state $\ket{\psi_0}$ and the constructions in the previous sections actually required to start in the $\ket{0}$ state. Although we might choose an arbitrary initial state $\ket{\psi}$ and map it to $\ket{0}$, this might be hard to realize. From \Ref{eqn:arbitrary state hard to approx} we know that this might require a circuit of size exponential in the number of qubits $N = \left\lceil \log d\right\rceil$ needed to realize $\Hilbert_d$. Hence we are interested in more general $2$-designs which are based on a set of unitaries $\U$ that generate the states from a given initial state. The motivation is that the unitary invariance of the Fubini-Study measure enables us to turn any Fubini-Study integral into an integral over the Haar measure on $U(d)$ with an arbitrary initial state $\ket{\psi_0}$,
\[ \int_{\text{F-S}} f(\ket{\psi}) d\ket{\psi} = \int_{U(d)} f(U \ket{\psi_0}) dU.\]

Let us first define how a state $2$-design arises form a finite set of unitaries.
\begin{df} \label{def:unitary state design} A $2$-design for quantum states is generated by a set of operators $\U \subset U(d)$ if there exists $\ket{\psi_0} \in \Hilbert$ such that \Ref{eqn:2-design formula} holds for all $M, N \in L(\Hilbert)$.
\end{df}

The first extension is a set of unitary transformations that generate a $2$-design for quantum states independent of the initial state $\ket{\psi_0}$. 
\begin{df} \label{def:unitary state design with arb initial state} A set of unitary operators $\U \subset U(d)$ generates a $2$-design for states any state if \Ref{eqn:2-design formula} holds for any $M, N \in L(\Hilbert)$ and any $\ket{\psi_0} \in \Hilbert$.
\end{df}

We can further generalize this into a unitary $2$-design that gives the same average as the Haar measure on $U(d)$ for operator-valued functions on $U(d)$.
\begin{df} \label{def:unitary 2-design} A \emph{unitary $2$-design} is a set of unitary operators $\U = \{U_1, \dots, U_K\} \subset U(d)$ such that \be \label{eqn:unitary 2-design} \sum_{k=1}^K U_k\dag M U_k N U_k\dag O U_k = \int_{U(d)} U\dag M U N U\dag O U dU\ee for all linear operators $M, N, O \in L(\Hilbert)$.
\end{df}

A unitary $2$-design generates a $2$-design on quantum statesfor any initial state. We can pick an arbitrary initial state $\ket{\psi_0}$ and multiply \Ref{eqn:unitary 2-design} from both sides to get
\[ \sum_{k=1}^K \bra{\psi_0} U_k\dag M U_k N U_k\dag O U_k \ket{\psi_0} = \int_{U(d)} \bra{\psi_0} U\dag M U N U\dag O U  \ket{\psi_0} dU. \] Replace $N = \ket{\psi_0} \bra{\psi_0}$, and \Ref{eqn:2-design formula} follows.

We can also define a unitary $2$-design in analogy to the definition of $2$-design for states (Definition \ref{def:2-design for states}), where the homogeneous polynomial is of homogeneous degree $(2,2)$ in the matrix elements of $U \in U(d)$ and global phases are ignored again. From \Ref{eqn:unitary 2-design}, we can conclude that a unitary $2$-design will correctly give the integral over all monomials of degree $(2,2)$ in $U$. This is extended to all homogeneous polynomials of degree $(2,2)$ by linearity. The $2$-design condition now reads that a set $S \subseteq U(D)$ is a unitary $2$-design if for all polynomials $p(U)$ of homogeneous degree $(2,2)$, \be \label{eqn:unitary 2-design condition} \f{1}{|S|} \sum_{s \in S} p(s) = \int_{U(D)} p(U) dU. \ee For the ease of notation, we will use Definition \ref{def:unitary 2-design} but fall back to this alternate definition to support certain arguments. 

Although the term ``unitary $2$-design'' did not seem have appeared before, such an object has already been identified by \cite{BennettEtAl1996} for the case of a single qubit and by \cite{DiVincenzoLeungTerhal2001} for an arbitrary number of qubits. However, they could only give an approximate sampling algorithm. We will give a different proof and a construction that yields circuits of smaller complexity and fewer random bits, while still being exponentially close to a $2$-design in the induced superoperator norm and even in the stronger diamond norm \cite{Kitaev2002}.

We will introduce the notation of the Pauli operators and the Clifford group here. Consider a system that consists of $N$ qudits, each of dimension $d$ with Hilbert space $\Hilbert_d$, forming a system with Hilbert space $\Hilbert$ of dimension $D = d^N$. 

\begin{df} Let $\Pauli(d)$ denote the set of $d^2$ generalized Pauli operators in dimension $d$ \cite{GottesmanEtAl2001} \[X^a Z^b: a, b = 0, 1, \dots, d-1\] with the generalized Pauli operators acting on the computational basis as \[ X: \ket{j} \mapsto \ket{j+1 \bmod d}, Z: \ket{j} \mapsto \omega^j \ket{j}\] with $\omega = e^{2 \pi i/d}$. Note that some authors call these the \emph{Heisenberg-Weyl operators}.
\end{df}

It directly follows that $ZX = \omega XZ$. Further, we have that \[X^a: \ket{j} \mapsto \ket{j+a \bmod d}, Z^a: \ket{j} \mapsto \omega^{aj} \ket{j}\] and \[ \left(X^a\right)\dag: \ket{j} \mapsto \ket{j-a \bmod d}, \left(Z^b)\right)\dag: \ket{j} \mapsto \omega^{-aj} \ket{j}.\] We have the commutation relation \[X^a Z^b = \omega^{ab} Z^b X^a, Z^b X^a = \omega^{-ab} X^a Z^b\] for $a, b \in \Z_d$, which implies \[X^{a_1} Z^{b_1} X^{a_2} Z^{b_2} = \omega^{a_1 b_2 - a_2 b_1} X^{a_2} Z^{b_2} X^{a_1} Z^{b_1}\] for all $a_1, a_2, b_1, b_2 \in \Z_d$. We also note that the set $\Pauli(d) = \{X^a Z^b\,|\,a, b = 0, 1, \dots, d-1\}$ is a basis for $L(\Hilbert_d)$. This can be seen using the Hilbert-Schmidt inner product, which yields 
\bess
\tr (X^a Z^b)\dag X^{a'} Z^{b'} & = & \tr (Z^b)\dag (X^a)\dag X^{a'} Z^{b'} = \tr (Z^b)\dag X^{a'-a} Z^{b'} \\
& = &  \tr  X^{a'-a} Z^{b'}(Z^b)\dag = \tr X^{a'-a} Z^{b'-b} \\
& = & \sum_{j=0}^{d-1} \bra{j} X^{a'-a} Z^{b'-b} \ket{j} = \sum_{j=0}^{d-1} \omega^{(b'-b)j} \bra{j} X^{a'-a} \ket{j} \\
& = & \delta_{a',a} \sum_{j=0}^{d-1} \omega^{(b'-b)j} = d \delta_{a',a} \delta_{b',b}.
\eess
We can turn $\Pauli(d)$ into an orthonormal basis if we normalize by $\f{1}{\sqrt{d}}$. As we want $\Pauli(d)$ to be a set of unitary operator, we skip this normalization. 

Note that in the case of qubits, we have $d=2$ and the Pauli operators $\Pauli(2)$ are sometimes written as 
\bess
X^0 Z^0 & = & \Id, \\
X^1 Z^0 & = & \sigma_x, \\
X^0 Z^1 & = & \sigma_y, \text{ and } \\
X^1 Z^1 & = & \sigma_z.
\eess

Taking all possible tensor products of $N$ generalized Pauli operators yields the tensor-product Pauli operators which we will denote by \[ \Pauli(d,N) = \left\{\f{1}{\sqrt{D}} X^{a_1} Z^{b_1} \otimes \dots \otimes X^{a_N} Z^{b_N}\,|\,a_i, b_i \in \Z_d\right\}\] but we will use the short-hand notation \[X^{a_1} Z^{b_1} \otimes \dots \otimes X^{a_N} Z^{b_N} = X^{\v{a}} Z^{\v{b}}\] for $\v{a}, \v{b} \in \Z_d^N$. $\Pauli(d,N)$ is a basis for $L(\Hilbert)$ consisting of $d^{2N} = D^2$ elements. When $d$ and $N$ are clear from the context, we will write $\Pauli(D)$ instead of $\Pauli(d,N)$.

The commutation relation of these tensor-product Paulis can be deduced from the commutation relation of the generalized Pauli operators. Using the short-hand vector notation, we see that \[X^{\v{a}_1} Z^{\v{a}_1} X^{\v{a}_2} Z^{\v{b}_2} = \omega^{\v{a}_1 \v{b}_2 - \v{a}_2 \v{b}_1} X^{\v{a}_2} Z^{\v{b}_2} X^{\v{a}_1} Z^{\v{a}_1}\] for $\v{a}_1, \v{a}_2, \v{b}_1, \v{b}_2 \in \Z_d^N$. We can further simplify this expression by considering vectors $\v{x} = (\v{x}_a, \v{x}_b) \in \Z_d^{2N}$ together with the symplectic inner product $(\v{x}, \v{y})_{Sp} = \v{x}_a \cdot \v{y}_b - \v{x}_b \cdot \v{y}_a$ where $\v{x}_a$ denotes the vector consisting of the first $N$ components of $\v{x}$ and $\v{u} \cdot \v{v}$ denotes the usual inner product. Observe that the symplectic inner product is linear and $(\v{x}, \v{y})_{Sp} = -(\v{y}, \v{x})_{Sp}$. Together with the notation \[P_{\v{x}} \equiv X^{\v{x}_a} Z^{\v{x}_b}\] we can write the commutation relation in the concise form \be \label{eqn:pauli comm rel} P_{\v{x}} P_{\v{y}} = \omega^{(\v{x}, \v{y})_{Sp}} P_{\v{y}} P_{\v{x}}.\ee We note that $\f{1}{\sqrt{D}} P_{\v{x}}$ is normalized, but we need the property that $P_{\v{x}}$ is unitary, so we skip the normalization.

Sometimes, we will identify the elements of $\Pauli(d,N)$ with integers $j = 1, 2, \dots, D^2$. Ignoring global phases that are introduced by the commutation relation, we can treat $\Pauli(d,N)$ as the group of tensor-product Pauli operators. From the commutation relations, we have that 
\[P_{\v{x}} P_{\v{y}} = \omega^{-\v{y}_a \v{x}_b} P_{\v{x+y}}.\] Let $\Pauli'(d,N)$ by $\Pauli(d,N)$ and define the equivalence relation $P \equiv Q$ if and only if there is $\alpha \in \C$ such that $P = \alpha Q$. We can identify $\Pauli(d,N) = \Pauli'(d,N)/\equiv$ using \[P_{\v{x}} P_{\v{y}} \equiv P_{\v{x+y}}.\] The identity element is given by $P_{\v{o}} = \Id^{\otimes N}$. We will denote $\Pauli(d,N)$ with multiplication defined by ignoring phases as the \emph{Pauli group} with dimension $(d,N)$. This equivalence relation essentially ignores global phases caused by the commutation relation. This approach is reasonable if we consider conjugation by tensor-product Pauli operators, which will be one of the main tools used in this section.

\begin{df} Let $\Lambda$ be a completely-positive superoperator on $\Hilbert$. Define the \emph{Pauli-twirled} superoperator \[\Lambda_{P} = \f{1}{D^2} \sum_{j=1}^{D^2} \hat{P_j} \hat{\Lambda} \hat{P_j}\dag \] where \[\Lambda_{P}(\rho) = \f{1}{D^2} \sum_{j=1}^{D^2} P_j \Lambda( P_j\dag \rho P_j) P_j\dag\] for any $\rho \in L(\Hilbert)$. 

A \emph{Pauli superoperator} is a superoperator $\Lambda$ such that \[\Lambda(\rho) = \sum_{j=1}^{D^2} \alpha_j P_j \rho P_j\dag\] for all linear operators $\rho$, where $P_j \in \Pauli(D)$.
\end{df}

\begin{df} \label{def:Clifford group} The Clifford group $\Clifford(D)$ is the normalizer of the tensor-product Pauli group $\Pauli(D)$ under conjugation, i.e.\@ \[\Clifford (D) = \{V \in U(D)\,|\, V \Pauli V\dag \subseteq \Pauli\}.\] 
\end{df}

The Clifford group plays an important role in quantum error correction \cite{Gottesman1997} and has been used before to show a similar twirling result that already shows that 

\section{The Clifford Group is a Unitary $2$-Design}

\subsection{The Previous Result}

In 2001, \cite{DiVincenzoLeungTerhal2001} introduced the notion of a ``Clifford Twirl'' and showed a condition that is equivalent to a unitary $2$-design. We will present that part of their result and show that it is equivalent to a unitary $2$-design for $U(D)$ where $D = 2^N$.

\begin{theorem} For all states $\rho \in \Hilbert \otimes \Hilbert$, 
\be \label{eqn:clifford state twirl is unitary state twirl} \f{1}{|\Clifford(D)|} \sum_{C \in \Clifford(2^N)} (C \otimes C) \rho (C\dag \otimes C\dag) = \int_{U(D)} (U \otimes U) \rho (U\dag \otimes U\dag) dU.\ee
\end{theorem}
\begin{proof}
The proof is given in Section A.1 of \cite{DiVincenzoLeungTerhal2001}.
\end{proof}

It follows easily that $\Clifford(2^N)$ is a unitary $2$-design as the following corollary shows.
\begin{cor} \label{cor:Clifford is unitary 2-design using Debbies method} $\Clifford(2^N)$ is a unitary $2$-design.
\end{cor}
\begin{proof}
States are Hermitian matrices of trace $1$. First, we extend \Ref{eqn:clifford state twirl is unitary state twirl} to all Hermitian matrices using $\rho' = \rho + \f{1 - \tr \rho}{D} \Id$ is Hermitian with trace $1$. By the linearity of the sum and the integral, we only need to consider $\f{1 - \tr \rho}{D} \Id$. As $\Clifford(D)$ and $U(D)$ are unitary operators, \Ref{eqn:clifford state twirl is unitary state twirl} also holds for $\f{1 - \tr \rho}{D} \Id$.

We extend \Ref{eqn:clifford state twirl is unitary state twirl} to all linear operators $\rho \in \Hilbert \otimes \Hilbert$ using the fact that there is a Hermitian basis for $\Hilbert \otimes \Hilbert$.

By choosing $\rho$ appropriately, we can show the $2$-design condition \Ref{eqn:unitary 2-design condition} for all monomials of homogeneous degree $(2,2)$. By linearity, the result follows for all homogeneous polynomials of degree $(2,2)$.
\end{proof}

\subsection{A Different Proof}

Inspired by discussions with Daniel Gottesmann and \cite{Chau2005}, we will give a different proof that the Clifford group is a unitary $2$-design. The argument starts by showing that ``twirling'' a completely-positive superoperator by tensor-product Pauli operators gives a completely-positive superoperator with only tensor-product Pauli operators as operation elements. After that, the Clifford group symmetrizes their weights to give a unitarily invariant superoperator. In an argument slightly more complicated than Corollary \ref{cor:Clifford is unitary 2-design using Debbies method}, we deduce that this implies a unitary $2$-design.

\begin{lemma} \label{lem:sum over pauli characters}
Define $\chi_{\v{j}}(P_{\v{x}}) = \omega^{(\v{x}, \v{j})_{Sp}}$. Then $\chi_{\v{j}}$ is a character of $\Pauli$ for any $\v{j}$. Furthermore, for all $\v{j} \in \Z_d^{2N}, \v{j} \neq \v{o}$, \be \label{eqn:char sum} \sum_{\v{x}} \chi_{\v{j}}(P_{\v{x}}) = 0.\ee
\end{lemma}
\begin{proof}
Observe that 
\bess
\chi_{\v{j}}(P_{\v{x}}) \chi_{\v{j}} (P_{\v{y}}) & = & \omega^{(\v{x}, \v{j})_{Sp}}  \omega^{(\v{y}, \v{j})_{Sp}} \\
& = & \omega^{(\v{x+y}, \v{j})_{Sp}} = \chi_{\v{j}}(P_{\v{x+y}}) = \chi_{\v{j}}(P_{\v{x}} P_{\v{y}}).
\eess
As long as $\v{j} \neq \v{o}$, $\chi_{\v{j}}$ is a nontrivial character of $\Pauli$ and \Ref{eqn:char sum} is the well-known character sum formula. See \cite[Ch.~5]{LidlNiederreiter1994} for a proof.
\end{proof}

\begin{lemma} \label{lem:pauli twirling} Twirling a completely-positive superoperator $\Lambda$ with the tensor-product Pauli group $\Pauli$ yields a Pauli superoperator $\Lambda_{P}$.
\end{lemma}
\begin{proof}
We use that the tensor-product Pauli operators $P_{\v{j}} \in \Pauli(D)$ form an orthonormal basis for $\Hilbert$, thus we can write the superoperator \[\Lambda(\rho) = \sum_{k=1}^{\leq D^2} A_k \rho A_k\dag\] as \[\Lambda(\rho) = \sum_k \sum_{\v{r} \in \Z_d^{2N}} \alpha_{k,\v{r}} P_{\v{r}} \rho \sum_{\v{s} \in \Z_d^{2N}} \overline{\alpha_{k,\v{s}}} P_{\v{s}}\dag, \] where \[A_k = \sum_{\v{r} \in \Z_d^{2N}} \alpha_{k,\v{r}} P_{\v{r}}.\]

The Pauli-twirled superoperator can now be simplified to
\bess
\Lambda_P(\rho) & = & \f{1}{D^2} \sum_{\v{j} \in \Z_d^{2N}} P_{\v{j}}\dag \sum_k \sum_{\v{r},\v{s} \in \Z_d^{2N}} \alpha_{k,\v{r}} P_{\v{r}} P_{\v{j}} \rho P_{\v{j}}\dag \overline{\alpha_{k,\v{s}}} P_{\v{s}}\dag P_{\v{j}} \\
& = & \f{1}{D^2} \sum_{\v{j} \in \Z_d^{2N}} \sum_k \alpha_{k,\v{r}} \overline{\alpha_{k,\v{s}}} 
\sum_{\v{r},\v{s} \in \Z_d^{2N}} P_{\v{j}}\dag P_{\v{r}} P_{\v{j}} \rho P_{\v{j}}\dag P_{\v{s}}\dag P_{\v{j}} \\
& = & \f{1}{D^2} \sum_{\v{r},\v{s} \in \Z_d^{2N}} \sum_k \alpha_{k,\v{r}} \overline{\alpha_{k,\v{s}}} 
\sum_{\v{j} \in \Z_d^{2N}} \omega^{(\v{r}, -\v{j})_{Sp} + (\v{j}, -\v{s})_{Sp}} P_{\v{r}} \rho P_{\v{s}}\dag \\
& = & \f{1}{D^2} \sum_{\v{r},\v{s} \in \Z_d^{2N}} \sum_k \alpha_{k,\v{r}} \overline{\alpha_{k,\v{s}}} 
\sum_{\v{j} \in \Z_d^{2N}} \omega^{(\v{j}, \v{r} - \v{s})_{Sp}} P_{\v{r}} \rho P_{\v{s}}\dag
\eess
where we used the commutation relation \Ref{eqn:pauli comm rel}.

From Lemma \ref{lem:sum over pauli characters}, we have \[\sum_{\v{j} \in \Z_d^{2N}} \omega^{(\v{j}, \v{r} - \v{s})_{Sp}} = D^2 \delta_{r,s}.\] Thus we can simplify the expression of the Pauli-twirled superoperator to
\bess
\Lambda_P(\rho) & = & \f{1}{D^2} \sum_{\v{r},\v{s} \in \Z_d^{2N}} \sum_k \alpha_{k,\v{r}} \overline{\alpha_{k,\v{s}}} D^2 \delta_{r,s} P_{\v{r}} \rho P_{\v{s}}\dag \\
& = & \sum_{\v{r} \in \Z_d^{2N}} \sum_k \alpha_{k,\v{r}} \overline{\alpha_{k,\v{r}}} P_{\v{r}} \rho P_{\v{r}}\dag \\
& = & \sum_{\v{r} \in \Z_d^{2N}} \left( \sum_k | \alpha_{k,\v{r}}|^2\right) P_{\v{r}} \rho P_{\v{r}}\dag.
\eess
This shows that $\Lambda_P$ is indeed a Pauli superoperator with real coefficients \[\beta_{\v{r}} = \left( \sum_k | \alpha_{k,\v{r}}|^2\right) \geq 0.\] 
\end{proof}

The following theorem shows how a Pauli superoperator can be twirled into a unitarily invariant superoperator using the Clifford group. As $\Pauli(D)$ is by definition a normal subgroup of $\Clifford(D)$, it suffices to consider $\Clifford(D)/\Pauli(D)$ which is called the ``symplectic group'' $\SL(D)$ in \cite{Chau2005}. This name arises from the fact that the Clifford group needs to preserve the commutation relationships between the tensor-product Pauli. That, in turn, means that it needs to preserve the symplectic inner product as it specifies the commutation relation \Ref{eqn:pauli comm rel}.

\begin{theorem}\cite{Chau2005} \label{thm:symplectic twirling} Twirling a Pauli superoperator $\Lambda_{P}$ by the symplectic group $\SL(D)$ turns it into a unitarily invariant superoperator $\Lambda_{U}$.
\end{theorem}
\begin{proof}
Using the argument from \cite{DiVincenzoLeungTerhal2001}, we note that the symplectic group will map each non-identity Pauli $P_j$, $j > 1$, equally often to $w_d^l \Pauli_j$ for all $l = 1, 2, \dots, d$ as it is the coset group of the normalizer of the Pauli group $\Pauli(D)$. We need the identity that for all $\rho \in L(\Hilbert)$,
\bess
\sum_{j=1}^{D^2} P_j \rho P_j\dag & = & \sum_{j=1}^{D^2} \sum_{l=1}^{D^2} P_j \tr (\rho\dag P_l) P_l P_j\dag \\
& = & \sum_{j=1}^{D^2} \sum_{l=1}^{D^2} \omega_d^{(j,l)_{Sp}} P_j P_j\dag \tr (\rho\dag P_l) P_l \\
& = & \sum_{l=1}^{D^2} \tr (\rho\dag P_l) P_l \sum_{j=1}^{D^2} \omega_d^{(j,l)_{Sp}} \\
& = & D^2 \tr (\rho\dag \Id ) \Id = D^2 \tr \rho \Id
\eess
where we used Lemma \ref{lem:sum over pauli characters} in the same way we did in the proof of Lemma \ref{lem:pauli twirling}  and that $\Pauli(D)$ forms a basis for $L(\Hilbert)$.

Now a calculation shows that for all $\rho \in L(\Hilbert)$,
\bess
\Lambda_U & = & 
\f{1}{|\SL(D)|} \sum_{C \in \SL(D)} C\dag \Lambda_P(C \rho C\dag) C \\
& = & \f{1}{|\SL(D)|} \sum_{C \in \SL(D)} \sum_{j=1}^{D^2} \alpha_j C\dag P_j C \rho C\dag P_j\dag C \\
& = & \f{1}{D^2-1} \sum_{j=2}^{D^2} \left(\sum_{l=1}^{D^2} \alpha_l \right) P_j \rho P_j\dag \\
& = & \left(\alpha_1 - \f{1}{D^2-1} \left(\sum_{l=2}^{D^2} \alpha_l \right) \right) \rho + \f{1}{D^2-1} \left(\sum_{l=2}^{D^2} \alpha_l \right) \sum_{j=1}^{D^2} P_j \rho P_j\dag \\
& = & \left(\alpha_1 - \f{1}{D^2-1} \left(\sum_{l=2}^{D^2} \alpha_l \right) \right) \rho + \f{D^2}{D^2-1} \left(\sum_{l=2}^{D^2} \alpha_l \right) \tr \rho \Id.
\eess
This structure of $\Lambda_U$ shows that it is unitarily invariant.
\end{proof}

\begin{lemma} \label{lem:all unitarily invariant twirlings are the same} Let $\mu$ be a probability measure on $U(D)$ and let $\Lambda$ be a superoperator. Define the $\mu$-twirled superoperator \[\Lambda_{\mu} = \int_{U(D)} \hat{V} \hat{\Lambda} \hat{V}\dag d \mu(V).\] If $\Lambda_{\mu}$ is unitarily invariant, then $\Lambda_{\mu} = \Lambda_T$ where $\Lambda_T$ is the Haar-twirled superoperator (see Definition \ref{def:haar twirling}).
\end{lemma}
\begin{proof} Using the unitary invariance of $\Lambda_{\mu}$ and the normalization of the Haar measure, it follows that
\bess
\left(\Lambda_{\mu}\right)_T & = & \int_{U(d)} \hat{U} \hat{\Lambda}_{\mu} \hat{U}\dag dU \\
 & = & \int_{U(D)} \Lambda_{\mu} dU = \Lambda_{\mu}.
\eess
Using the unitary invariance of $\Lambda_T$ from Lemma \ref{lem:unitarily inv superop} and the normalization of the probability measure $\mu$, we have
\bess
\left(\Lambda_T\right)_{\mu} & = & \int_{U(d)} \hat{V} \hat{\Lambda}_T \hat{V}\dag d \mu(V) \\
 & = & \int_{U(D)} \Lambda_T d \mu(V) = \Lambda_T.
\eess
The linearity of the integral ensures that we can change the order of integration. Together with the unitary invariance of the Haar measure, we see that the order of twirling does not matter.
\bess
\left(\Lambda_{\mu}\right)_T & = & \int_{U(D)} \hat{U} \Lambda_{\mu} \hat{U}\dag dU \\
& = & \int_{U(D)} \hat{U} \int_{U(D)} \hat{V} \hat{\Lambda} \hat{V}\dag d \mu(V) \hat{U}\dag dU \\
& = & \int_{U(D)} \int_{U(D)} \hat{U} \hat{V} \hat{\Lambda} \hat{V}\dag \hat{U}\dag  d \mu(V) dU \\
& = & \int_{U(D)} \int_{U(D)} \hat{V} \hat{U'} \hat{\Lambda} \hat{U'}\dag \hat{V}\dag d \mu(V) dU' \\
& = & \int_{U(D)} \hat{V} \int_{U(D)} \hat{U'} \hat{\Lambda} \hat{U'}\dag dU' \hat{V}\dag d \mu(V)  \\
& = & \int_{U(D)} \hat{V} \Lambda_T \hat{V}\dag d \mu(V)  \\
& = & \left(\Lambda_T\right)_{\mu}
\eess
where we used the change of variables $U = V U' V\dag$ for the fourth line. Therefore $\Lambda_{\mu} = \Lambda_T$.
\end{proof}

\begin{df} Define \[\Sum_D(\rho, \Lambda) = \f{1}{|\Clifford(D)|} \sum_{C \in \Clifford(D)}  C\dag \Lambda(C \rho C\dag) C\] and \[\Int_D(\rho, \Lambda) = \int_{U(D)} U\dag \Lambda(U \rho U\dag) U dU.\] 
\end{df}

Observe that both $\Sum_D(\rho, \Lambda)$ and $\Int_D(\rho, \Lambda)$ are linear functions on $L(\Hilbert)$ for fixed $\Lambda$.

\begin{theorem} \label{thm:Haar avg = pauli + clifford avg} Twirling a superoperator $\Lambda$ by $\Clifford(D)$ is the same as Haar-twirling. Formally, for all linear operators $\rho$, we have that
\be \label{eqn:Haar avg = pauli + clifford avg} \Sum_D(\rho, \Lambda) = \Int_D(\rho, \Lambda) = p \rho + q \f{\tr \rho}{D} \Id.\ee
\end{theorem}
\begin{proof}
Lemma \ref{lem:pauli twirling} and Theorem \ref{thm:symplectic twirling} in conjunction with Lemma \ref{lem:repn-of-superop} show that \[ \Sum_D(\rho, \Lambda) = p \rho + q \f{\tr \rho}{D} \Id\] for some parameter $0 \leq p \leq 1$. Corollary \ref{cor:repn-of-superop nicer} shows that \[\Int_D(\rho, \Lambda) = p' \rho + q' \tr \rho \f{1}{D} \Id\] for some constants $p', q'$. Lemma \ref{lem:all unitarily invariant twirlings are the same} shows that $p = p', q = q'$.
\end{proof}

\begin{cor} For any $M, N \in L(\Hilbert)$, 
\be \label{eqn:Haar avg = pauli + clifford avg gen} \f{1}{|\Clifford(D)|} \sum_{C \in \Clifford(D)} C\dag M C N C\dag M\dag C = \int_{U(D)} U\dag M U N U\dag M\dag U dU.\ee
\end{cor}
\begin{proof}
We consider the superoperator $\Lambda(\rho) = M \rho M\dag$. \Ref{eqn:Haar avg = pauli + clifford avg gen} follows directly from \Ref{eqn:Haar avg = pauli + clifford avg} by looking at $\Lambda(N)$.
\end{proof}

\begin{lemma} For any Hermitian $M, N, O \in L(\Hilbert)$, 
\be \label{eqn:Haar avg = pauli + clifford avg more gen} \f{1}{|\Clifford(D)|} \sum_{C \in \Clifford(D)} C\dag M C N C\dag O C = \int_{U(D)} U\dag M U N U\dag M O dU.\ee
\end{lemma}
\begin{proof}
Fix an arbitrary Hermitian $N \in L(\Hilbert)$. Define the operators \[\Sum'(M, O) = \f{1}{|\Clifford(D)|} \sum_{C \in \Clifford(D)} C\dag M C N C\dag O C \] and \[\Int'(M, O) = \int_{U(D)} U\dag M U N U\dag O U dU.\] Then \Ref{eqn:Haar avg = pauli + clifford avg gen} reads \be \label{eqn: Haar = Clifford-Pauli-avg simplified} \Sum'(M, M) = \Int'(M, M)\ee where we used that $M, N, O$ are Hermitian operators. Furthermore, we can see that \be \label{eqn:sum', int' hermitian} \Sum'(M, O) = \Sum'(O, M)\dag \text{ and } \Int'(M, O) = \Int'(O, M)\dag\ee as $M$, $N$ and $O$ are Hermitian.

We can extend \Ref{eqn:Haar avg = pauli + clifford avg gen} to work with two different operators by considering  $M_1 = M + O$ and $M_2 = M + i O$. From \Ref{eqn: Haar = Clifford-Pauli-avg simplified}, we get $\Sum'(M_j, M_j) = \Int'(M_j, M_j)$, $j = 1, 2$. By the bilinearity of both $\Sum'$ and $\Int'$, we can expand both sides for $j = 1, 2$ and subtract \Ref{eqn: Haar = Clifford-Pauli-avg simplified}. Using \Ref{eqn:sum', int' hermitian}, we end up with
\bes
\label{eqn:sum'(m,o)} \Sum'(M, O) + \Sum'(M, O)\dag & = & \Int'(M, O) + \Int'(M, O)\dag, \\
\label{eqn:sum'(m,io)} i \Sum'(M, O) - i \Sum'(M, O)\dag & = & i \Int'(M, O) - i \Int'(M, O)\dag.
\ees
Observe that $i \Ref{eqn:sum'(m,o)} + \Ref{eqn:sum'(m,io)}$ yields
\[ 2i \Sum'(M, O) = 2i \Int'(M, O) \]
and \Ref{eqn:Haar avg = pauli + clifford avg more gen} follows.
\end{proof}

\begin{lemma} \Ref{eqn:Haar avg = pauli + clifford avg more gen} holds for any Hermitian $N$ and all $M, O \in L(\Hilbert)$.
\end{lemma}
\begin{proof}
$\Sum'(M, O)$ and $\Int'(M, O)$ from the previous lemma are bilinear forms on $L(\Hilbert)$. Thus 
the construction in the proof of Corollary \ref{cor:main} applies and \Ref{eqn:Haar avg = pauli + clifford avg more gen} holds for any linear operators $M, O$ and all Hermitian $N$.
\end{proof}

\begin{lemma} \Ref{eqn:Haar avg = pauli + clifford avg more gen} holds for all $M, N, O \in L(\Hilbert)$.
\end{lemma}
\begin{proof}
In a last step, we extend $N$ in \Ref{eqn:Haar avg = pauli + clifford avg more gen} from Hermitian to any linear operator. We fix $M, O \in L(\Hilbert)$ and define \[\Sum''(N) = \f{1}{|\Clifford(D)|} \sum_{C \in \Clifford(D)} C\dag M C N C\dag O C\] and \[\Int''(N) = \int_{U(D)} U\dag M U N U\dag O U dU.\]

It is immediate that $\Sum''(N)$ and $\Int''(N)$ are linear in $N$ for all $N \in L(\Hilbert)$. $\Sum''(N) = \Int''(N)$ only holds for Hermitian $N$, but can be extended to all $N \in L(\N)$ by linearity and the fact that we can express the canonical basis $\left\{\ket{k}\bra{l}\right\}_{k,l=1}^{D^2}$ for $L(\Hilbert)$ as linear combinations of Hermitian operators: \[\ket{k}\bra{l} = \ahalf \left(\ket{k}\bra{l} + \ket{l}\bra{k}\right) + \f{i}{2} i \left(\ket{k}\bra{l} - \ket{l}\bra{k}\right),\] where $\ket{k}\bra{l} + \ket{l}\bra{k}$ and $i \left(\ket{k}\bra{l} - \ket{l}\bra{k}\right)$ are Hermitian.
\end{proof}

\begin{cor} \label{cor:clifford is 2-design} $\Clifford(D) = SL(D) \circ \Pauli(D)$ is a unitary $2$-design.
\end{cor}

This concludes that twirling by the Clifford group yields a unitary $2$-design. However, it is not clear how to uniformly randomly sample from the Clifford group and only a randomized algorithm is known so far for the case of qubits, i.e.\@ $D = 2^N$ \cite{DiVincenzoLeungTerhal2001}. This algorithm uses $O(N^8)$ classical steps and produces a circuit of size $O(N^2)$, where the distribution is close to the uniform distribution over $\Clifford(D)$ in the $\littlel_1$-norm. We will see in the second section how this approximation shows up in the $2$-design condition.

\subsection{Efficient Approximate Construction}

In this section, we will prove that a subset of the Clifford group $\Clifford(D)$ already gives an approximate $2$-design in the induced superoperator norm. Our construction also only works for qubits, thus we also assume $D = 2^N$ here.

\begin{theorem} \label{thm:approximate symplectic twirling} For any $\epsilon > 0$, twirling a Pauli superoperator $\Lambda_{P}$ by a subset $SL_{\epsilon}(D) \subseteq SL(D)$ of the symplectic group turns it into a superoperator $\Lambda_{\epsilon}$ that such that \[\| \Lambda_{\epsilon} - \Lambda_U \|_{\diamond} \leq B(\Lambda) \left(\epsilon_0 + \epsilon\right)\] for $\epsilon_0 = \f{1}{2^N-2^{-N}}$ and $\Lambda_U$ the unitarily invariant channel from \ref{thm:symplectic twirling}. The norm is the induced operator norm from $L(\Hilbert)$ and the parameter $B(\Lambda)$ will be determined later.

The circuits in $SL_{\epsilon}(D)$ consist of \[O\left(N \log \f{1}{\epsilon}\right)\] single and two-qubit gates in depth \[O\left(\log N \log \f{1}{\epsilon}\right)\] and the constructions needs \[O\left(N \log \f{1}{\epsilon}\right)\] random bits. The subset is of size \[ | SL_{\epsilon}(D) | = 2^{O\left( N \log \f{1}{\epsilon} \right)}.\]
\end{theorem}
\begin{proof}

The task is to find a subset $\Clifford_{\epsilon}$ that uniformizes the tensor-product Paulis with high probability, i.e.\@ that maps a non-identity tensor-product Pauli to any tensor-product Pauli with almost equal probability.

We can choose using suitable phase factors for the Pauli operators as they are irrevelevant in the Kraus operator-sum representation $\sum_P P \rho P\dag$ for they will cancel out. Therefore, a typical Pauli will look like the following:
\[ \sigma_x \otimes \sigma_z \otimes \sigma_y \otimes \Id \otimes \sigma_z \otimes \Id \otimes \Id \otimes \sigma_y \otimes \sigma_z\]

\paragraph{(a) Basic Building Blocks}

In general, we consider a tensor product of Paulis that is not equivalent to the identity, thus at least one component is not $\Id$. We can use the cyclic shift generator $T = HP$, where \[H = \ket{+}\bra{+} - \ket{-}\bra{-}\] is the Hadamard gate and \[P = \ket{0}\bra{0} + i \ket{1}\bra{1}\] is the phase gate. Ignoring global phases, we see that 
\bess
T \sigma_x T\dag & = & \sigma_y, \\
T \sigma_y T\dag & = & \sigma_z, \text{ and } \\
T \sigma_z T\dag & = & \sigma_x
\eess
using the convenient $\sigma_j$ notation for a single-qubit Pauli. We thus have that $T$, $T^2$, and $T^3 = \Id$ generate permutations of a single-qubit Pauli, which will be used as a building block later in the construction.

\paragraph{How to Twirl Two Single-Qubit Paulis}
Now notice that we can conjugate pairs of Pauli operators in the tensor product by a $CNOT$ to create or annihilate identities. To see this, realize that the action of a $CNOT$ is 
\[ CNOT (X^{a_1} Z^{b_1} \otimes X^{a_2} Z^{b_2}) CNOT\dag = X^{a_1} Z^{b_1 - b_2} \otimes X^{a_2 - a_1} Z^{b_2}\] where the minus in the exponent is chosen to stay consistent with the general qudit case. Hence we create identities if $a_1 = a_2 = 1$ and $b_2 = 0$, with some back-action that will modify the Pauli on the control qubit. We will take care of that later and note that we will use either the $X^a Z^b$ or $\sigma_{a,b}$ notation, or even shorter $\sigma_i$ where $i \in \Z_4 = Z^2 \times Z^2$.

\paragraph{(b) Step 1: How to Generate a $\sigma_x$ or $\sigma_y$ with Constant Probability}

We want to use this construction to generate a tensor-product Pauli where a specific component has an $X$ or $Y$ Pauli with constant probability. We can reduce this to the much simpler problem of a binary string $x \in \{0,1\}^N$ that is guaranteed to have at least one $1$ and we can change a pair of positions by a controlled-NOT operation in the following way:
\begin{center}
	\begin{tabular}{c|c}
	$x$ & $CNOT(x)$ \\ \hline
	$00$ & $00$ \\ 
	$01$ & $01$ \\ 
	$10$ & $11$ \\ 
	$11$ & $10$
	\end{tabular}
\end{center}
This is the abstraction of conjugating a tensor-product Pauli by $CNOT$ gates if we identify $X^a Z^b$ with $a$. 

Now we can make use of the well-known fact that for $x \in \{0,1\}^N, x \neq 0^N$, \[P_{b \in \{0,1\}^N}(b \cdot x = 1) = \ahalf,\] where $b \cdot x = \sum_{i=1}^N b_i x_i \bmod 2$. We restrict $b \neq 0^N$ and observe that $0^N \cdot x = 0$ for all $x \in \{0,1\}^N$. Hence \[P_{b \in \{0,1\}^N, b\neq 0^N}(b \cdot x = 1) > \ahalf\] and we define \[p_{\text{success}} = \f{1}{2}.\] Pick a non-empty subset $B \subseteq \{1, 2, \dots, N\}$ uniformly at random and apply the $CNOT$ conjugation to the first bit position in $B$ from all the other positions in $B$. Then the first bit in $B$ will be $1$ with probability greater than $1/2$.

Going back to the tensor-product Paulis, this result implies that picking a random target qubit and conjugating with $CNOT$ gates from a random subset of the remaining qubits as controls will guarantee that the target qubit has an $X$ or $Y$ Pauli.

\paragraph{(c) Step 2: How to Generate an Almost Uniform Distribution over $\Pauli$}

We will now pick the target qubit from step 1 as our \emph{control qubit}. Then, we apply a single-qubit $T^{s_i}$ on each other qubit, for independently and randomly chosen $s_i \in \{0,1,2\}$. This will uniformize all non-identity Paulis on the target qubits. After that, we will independently twirl with a $CNOT$ gate on each of the other qubits as target and controlled on the control qubit, with probability $3/4$ each. We assume that the control qubit has an $X$ or $Y$ Pauli, which is guaranteed to happen with probability greater than $1/2$ by the previous step.

Consider a target qubit $t$. This either has the $\Id$ operator or $X$, $Y$, or $Z$ with probability $1/3$ each. Observe the effects of a $CNOT$-twirl that is applied with probability $3/4$.
\begin{center}
	\begin{tabular}{c|c|l}
		Target & Result of $CNOT$-Twirl & Probability \\ \hline
		$\Id$ & $\Id$ & 1/4 \\
		$\Id$ & $X$ & 3/4 \\
		$X$ 	& $X$ & 1/4 \\
		$X$ 	& $\Id$ & 3/4 \\
		$Y$ 	& $Y$ & 1/4 \\
		$Y$ 	& $Z$ & 3/4 \\
		$Z$ 	& $Z$ & 1/4 \\
		$Z$ 	& $Y$ & 3/4
	\end{tabular}
\end{center}
We see that if the Pauli on $t$ was $\Id$, it will be twirled to a non-identity Pauli with probability $3/4$. If the Pauli was not $\Id$, we see that it is twirled to an identity with probability $1/4$ and stays a non-identity with probability $3/4$. Once again, we insert a single-qubit twirl $T^{s_i}$ on each target qubit for independently and randomly chosen $s_i \in \{0,1,2\}$ to ensure that the non-identity Paulis on each target qubit will have probability $1/3$ each. The circuit generated so far is shown in Figure \ref{fig:circuit random cnots on all targets}. Note that by the back-action of the $CNOT$ twirl, it might change the control qubit Pauli from $X$ to $Y$ or vice-versa.

\begin{figure}[htbp]
	\centerline{
	\Qcircuit @C=1em @R=.7em
	{
		& \push{\f{3}{4}} & \push{\f{3}{4}} & & \push{\f{3}{4}} & \push{\f{3}{4}} \\
		& \targ \qw & \qw & \qdots \qw & \qw & \qw & \gate{T^{s_1}} \qw & \qw\\
		& \qw & \targ \qw & \qdots \qw & \qw & \qw & \gate{T^{s_2}} \qw & \qw\\
		& \ctrl{-2} \qw & \ctrl{-1} & \qdots \qw & \ctrl{2} \qw & \ctrl{3} \qw & \qw & \qw\\
		& & & \qddots & & \\
		& \qw & \qw & \qdots \qw & \targ \qw & \qw & \gate{T^{s_{N-1}}} \qw & \qw\\
		& \qw & \qw & \qdots \qw & \qw & \targ \qw & \gate{T^{s_N}} \qw & \qw
	} }
	\caption[$T$-Twirl and $CNOT$ Gates from a Randomly Chosen Control]{$T$-Twirl and $CNOT$ Gates with Probability $3/4$ each from a Randomly Chosen Control}
	\label{fig:circuit random cnots on all targets}
\end{figure}
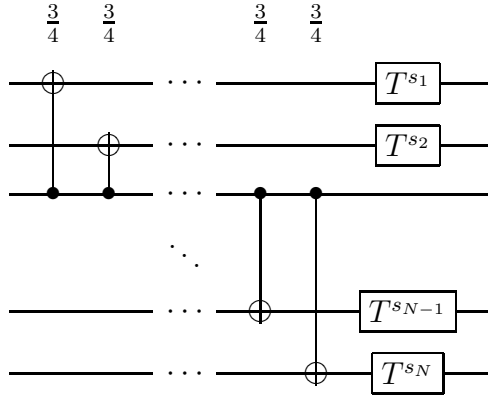

The next step is to randomize the control qubit. We note that all other qubits already have a uniformly chosen tensor-product Pauli on them, with probability $1/4^{n-1}$ each. 

\subparagraph{Observation:} If we apply any permutation of Paulis to these $N-1$ target qubits, it will not change their distribution for it is already uniform. 

We randomize the control qubit Pauli by a $P$-twirl with probability $1/2$. Note that $P$ permutes $X$ and $Y$. As the control qubit Pauli starts out in an unknown distribution of $X$ and $Y$ caused by the back-action of the previous step, the random $H$ twirl will set it to $X$ or $Y$ with probability $1/2$ each. Now we apply a random $CNOT$ with the former control qubit (which we will call the ``first'' qubit from now an) as target and controlled by each other qubit. Each $CNOT$ is applied with probability $1/2$, and we see that the back-action that might modify some of the now-control qubits does not affect the distribution over the Paulis on these qubits. The observation shows that the uniformity is not changed by a permutation caused by possible back-actions.

This procedure will sucessfully randomize the first qubit as it will change it from $X$ or $Y$ with probability $1/2$ each to $Z$ or $\Id$ with probability $1/2$ each, where a change occurs with probability $1/2$ if at least one of the other qubits has an $X$ or $Y$ Pauli. After that, it will be one of the four Paulis with probability $1/4$ each and other permutations by later $CNOT$ twirls will not change this uniform distribution due to our observation. The complete circuit of step 2 up to here is illustrated in Figure \ref{fig:step 2}.

\begin{figure}[htbp]
	\centerline{
	\Qcircuit @C=.7em @R=.7em
	{
		& & \push{\f{1}{2}} & \push{\f{1}{2}} & & \push{\f{1}{2}} & \push{\f{1}{2}} & & \push{\f{3}{4}} & \push{\f{3}{4}} & & \push{\f{3}{4}} & \push{\f{3}{4}} \\
		& \qw & \ctrl{2} \qw & \qw & \qdots \qw & \qw & \qw & \gate{T^{s_1}} \qw & \targ \qw & \qw & \qdots \qw & \qw & \qw & \gate{T^{t_1}} \qw & \qw\\
		& \qw & \qw & \ctrl{1} \qw & \qdots \qw & \qw & \qw & \gate{T^{s_2}} \qw & \qw & \targ \qw & \qdots \qw & \qw & \qw & \gate{T^{t_2}} \qw & \qw\\
		& \gate{T^{r_2}} \qw & \targ \qw & \targ & \qdots \qw & \targ \qw & \targ \qw & \gate{P^{r_1}} \qw & \ctrl{-2} \qw & \ctrl{-1} & \qdots \qw & \ctrl{2} \qw & \ctrl{3} \qw & \gate{T^{t_3}} \qw & \qw\\
		& & & & \qddots & & & & & & \qddots & & \\
		& \qw & \qw & \qw & \qdots \qw & \ctrl{-2} \qw & \qw & \gate{T^{s_{N-1}}} \qw & \qw & \qw & \qdots \qw & \targ \qw & \qw & \gate{T^{t_{N-1}}} \qw &  \qw \\
		& \qw & \qw & \qw & \qdots \qw & \qw & \ctrl{-3} \gategroup{1}{3}{7}{7}{.7em}{--} \qw & \gate{T^{s_N}} \qw & \qw & \qw & \qdots \qw & \qw & \targ \gategroup{1}{9}{7}{13}{.7em}{--} \qw & \gate{T^{t_N}} \qw &  \qw
	} }
	\caption[Step 2 of the Pauli Uniformization Process]{Step 2 of the Pauli Uniformization Process, Highlighting the Random $CNOT$ Parts.}
	\label{fig:step 2}
\end{figure}

However, the distribution obtained so far will not have any weight on the tensor-product Paulis that are $\Id$ on the first qubit and $\Id$ or $\Z$ on the other qubits, as this case prevents any change to the first qubit and it remains $X$ or $Y$ with probability $1/2$ each. Adding a $T^j$-twirl will at least randomize between $X$, $Y$, or $Z$. Thus the only non-reachable tensor-product Paulis are those with $\Id$ on the first qubit and $\Id$ and $Z$ on the other qubits. The sample circuit for $N=5$ in Figure \ref{fig:sample twirl circuit} illustrates the complete twirl.

\begin{figure}[htbp]
	\centerline{
	\Qcircuit @C=1em @R=.7em 
	{
& \push{\rule{0em}{1em}} \qw & \qw & \qw & \qw & \targ \qw & \qw & \qw & \gate{\Id} \qw & \qw & \qw & \ctrl{1} \qw & \gate{\sigma_x} \qw & \qw & \\
& \gate{T} \qw & \targ \qw & \targ \qw & \gate{H} \qw & \ctrl{-1} \qw & \ctrl{2} \qw & \ctrl{3} \qw & \qw & \targ \qw & \targ \qw & \targ \qw & \gate{\sigma_y} \qw & \qw & \\
& \qw & \qw & \ctrl{-1} \qw & \qw & \qw & \qw & \qw & \gate{\Id} \qw & \qw & \qw & \qw & \gate{\Id} \qw & \qw & \\
& \qw & \qw & \qw & \qw & \qw & \targ \qw & \qw & \gate{T^2} \qw & \qw & \ctrl{-2} \qw & \qw & \gate{\Id} \qw & \qw & \\
& \push{\rule{0em}{1em}} \qw & \ctrl{-3} \qw & \qw & \qw &\qw & \qw & \targ \qw & \gate{T} \qw & \ctrl{-3} \qw & \qw & \qw & \gate{\sigma_z} \qw & \qw \gategroup{1}{2}{5}{12}{2em}{--} &
	} }
	\caption{A Sample Circuit $C \in \Clifford_{\epsilon}$ for $N=5$.}
	\label{fig:sample twirl circuit}
\end{figure}

We will now show that this part of the random procedure will generate an almost uniform distribution over all possible tensor-product Paulis except the all-identity Pauli. For a precise estimation, we will consider the $\littlel_1$-distance between the uniform probability distribution 
\[u(\v{x}) = \begin{cases} \f{1}{4^N-1} & \v{x} \neq \v{o} \\ 0 & \v{x} = \v{o} \end{cases}\]
on all Paulis but the identity and the distribution $q(\v{x})$ obtained by this process. In case the tensor-product Pauli we started with had an $X$ or $Y$ at the randomly chosen control, the process will produce an almost uniform distribution $q$, which can be seen in Figure \ref{fig:l1distance}. 

\begin{figure}[htbp]
	\centering
		\psfrag{q}{$\f{1}{4^N-2^{N-1}}$}
		\psfrag{u}{$\f{1}{4^N-1}$}
		\psfrag{4n-2n}{$4^N - 2^{N-1}$}
		\psfrag{2n-1-1}{$2^{N-1} - 1$}
		\psfrag{Idfirst}[][][.75]{$\Id \otimes \left(\sigma_z, \Id\right)^{\otimes (N-1)}$} 
		\psfrag{Id}[][][.75]{$\Id^{\otimes N}$} 
		\includegraphics[width=0.80\textwidth]{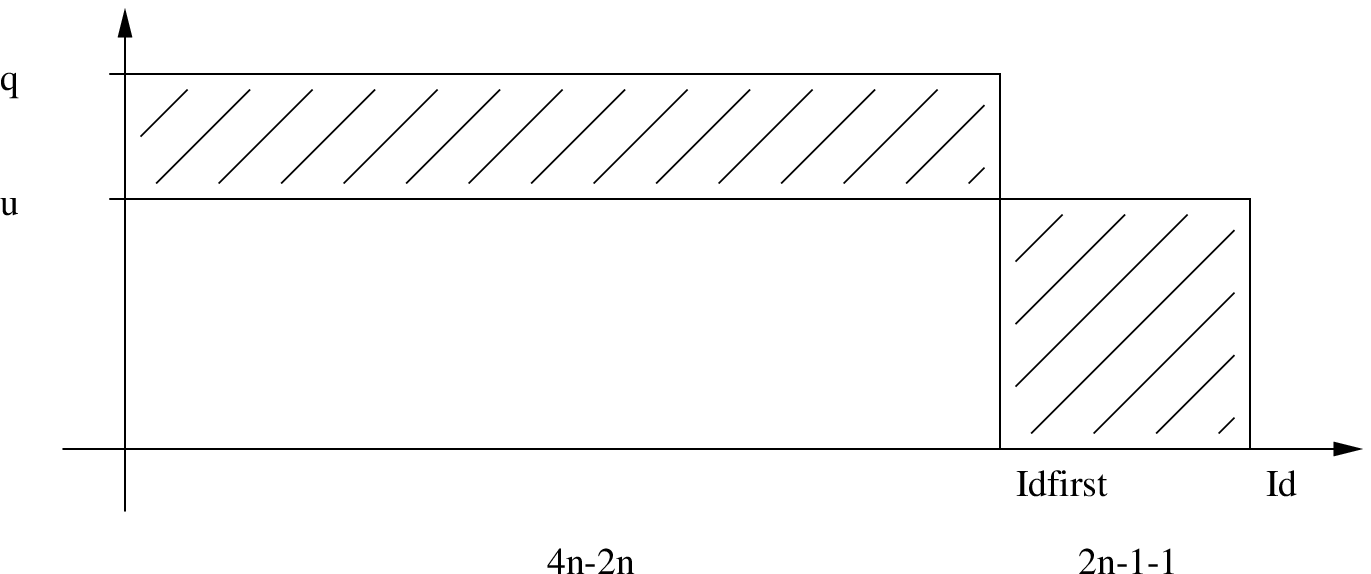}
	\caption{The $\littlel_1$-distance between $u$ and $q$.}
	\label{fig:l1distance}
\end{figure}

Precisely, the $l_1$-distance between $u$ and $q$ is given by 
\bess
\|u - q\|_{\littlel_1} & = & \sum_{\v{x} \in \Z_4^{N}} | u(\v{x}) - q(\v{x}) | \\
& = & \left( \f{1}{4^N - 2^N} - \f{1}{4^N-1} \right) \left( 4^N - 2^{N-1} \right) + \underbrace{\f{1}{4^N-1} \left(2^{N-1} - 1 \right)}_{\Id \otimes \stackrel{\sigma_z}{\Id} \otimes \dotsb \stackrel{\sigma_z}{\Id} \neq \Id \otimes \Id^{\otimes (N-1)}} \\
& = & 1 - \f{4^N - 2^{N-1}}{4^N-1} + \f{2^{N-1} - 1}{4^N-1} \\
& = & \f{2(2^{N-1}-1)}{4^N-1} \leq \f{2^N}{4^N-1} = \f{1}{2^N-2^{-N}} = \epsilon_0,
\eess
which is exponentially small in the number of qubits.

\paragraph{(d) Further Uniformizing the Distribution}
In the bad case where our choice of control will pick a qubit with an $\Id$ or $Z$, we use the fact that our random circuit is a probability distribution over permutations over the tensor-product Paulis. Denote $p$ the initial distribution over the tensor-product Paulis after the the random chain step. We have that for any permutation $\pi$ acting on a probability distribution $p$ that $\|\pi \circ p - u\| = \|p - u\|$ for any distance $\| \cdot \|$ as permutations will only permute the probabilities in the distribution, which has no effect on any norm. With a probability distribution $r$ over permutations $\pi_i$, we see that
\bess
\|\sum_i r(\pi_i) \pi_i \circ p - u\| 
& = & \|\sum_i r(\pi_i) (\pi_i \circ p - u)\| \leq \sum_i r(\pi_i) \|\pi_i \circ p - u\| \\
& = & \sum_i r(\pi_i) \|p - u\| = \|p - u\|
\eess
for any distance $\|\cdot\|$ using the triangle inequality and that $\sum_i r(\pi_i) = 1$. This is the convex-linearity of the distance.

This argument shows that in the case where we do not have $X$ or $Y$ on the control qubit, the process will not increase the distance to $u$. Denote $d_0 = \|p - u\|$, and by the convex-linearity of the distance we see that 
\[ d_1 \leq p_{\text{success}} \|q - u\| + (1 - p_{\text{success}}) \|p - u\| = p_{\text{success}} \|q - u\| + (1 - p_{\text{success}}) d_0.\] 
We can define the following recursion for the decrease in distance after steps 1 and 2 have been applied to an initial distribution $p'$. Denote $d_k$ the distance to $u$ after step $k$. The $\littlel_1$-distance for any $k \geq 0$ is given by the recursion
\bess \label{eqn:distance decrease recursion}
d_{k+1} & \leq & \sum_{\v{x} \in \Z_4^N} P(\text{Step 2 works for } \v{x}) \|q - u\|_1 + P(\text{Step 2 does not work for } \v{x}) d_k \\ 
\nn & \stackrel{d_k \geq \|q - u\|_1}{\leq} & \leq p_{\text{success}} \|q - u\|_1 + (1 - p_{\text{success}}) d_k \\
\nn & = & p_{\text{success}} \epsilon_0 + (1 - p_{\text{success}}) d_k 
\eess
as $p_{\text{success}}$ is a lower bound for $P(\text{Step 2 works for } \v{x})$. If $d_k < \|q - u\|_1$, we cannot use that lower bound but have to step back to \[d_{k+1} \leq p_{\text{success}} \|q - u\|_1 + d_k < (p_{\text{success}} + 1) \|q - u\|_1 < 2 \|q - u\|_1.\] Thus are analysis can only guarantee a bound twice as high, after which we are in the regime of \Ref{eqn:distance decrease recursion} again. 

We can solve \Ref{eqn:distance decrease recursion} analytically and see that 
\bess d_{k} 
& \leq & p_{\text{success}} \epsilon_0 \sum_{i=0}^{k-1} \left(1 -  p_{\text{success}} \right)^i + \left(1 -  p_{\text{success}} \right)^k d_0 \\
& = & p_{\text{success}} \epsilon_0 \f{1 - \left(1 -  p_{\text{success}} \right)^k}{1 - (1 - p_{\text{success}}} \\
& = & \epsilon_0 \left(1 - \left(1 -  p_{\text{success}} \right)^k\right) + \left(1 -  p_{\text{success}} \right)^k d_0 \\
& = & \epsilon_0 + \left(1 - p_{\text{success}} \right)^k \left( \epsilon_0 + d_0 \right) \\
& \leq & \epsilon_0 + \left(1 - p_{\text{success}} \right)^k \left( \epsilon_0 + 1 \right).
\eess

In order to be $\epsilon$-close to $\epsilon_0$, we need 
\[\left(1 - p_{\text{success}} \right)^k \left( \epsilon_0 + 1 \right) \leq \epsilon\] which implies
\[k \geq \f{\log (\epsilon_0 + 1) + \log \f{1}{\epsilon}}{\log \f{1}{1 - p_{\text{success}}}}\] and thus
\[k = O\left(\log \f{1}{\epsilon}\right).\]

\paragraph{(e) Optimizing the Circuit Complexity}

The random subset chosen in step 1 requires exactly $N$ random bits and at most $N-1$ $CNOT$ gates in depth at most $N - 1$. This can be optimized using techniques employed by parallel prefix adders. Suppose we were to to map \[\ket{x_1} \ket{x_2} \dotso \ket{x_N} \mapsto \ket{x_1} \ket{x_1 \oplus x_2} \ket{x_1 \oplus x_2 \oplus x_3} \dotso \ket{x_1 \oplus x_2 \oplus \dotsb \oplus x_N}.\]
Using a parallel prefix computation circuit from classical computation \cite{LadnerFischer1980}, we can decrease the depth to $\lceil\log N\rceil$ using at most $4N$ $CNOT$ gates. Figure \ref{fig:parallel prefix adder} shows what the parallel prefix adder looks like for $N = 16$.

\begin{figure}[htbp]
  \includegraphics{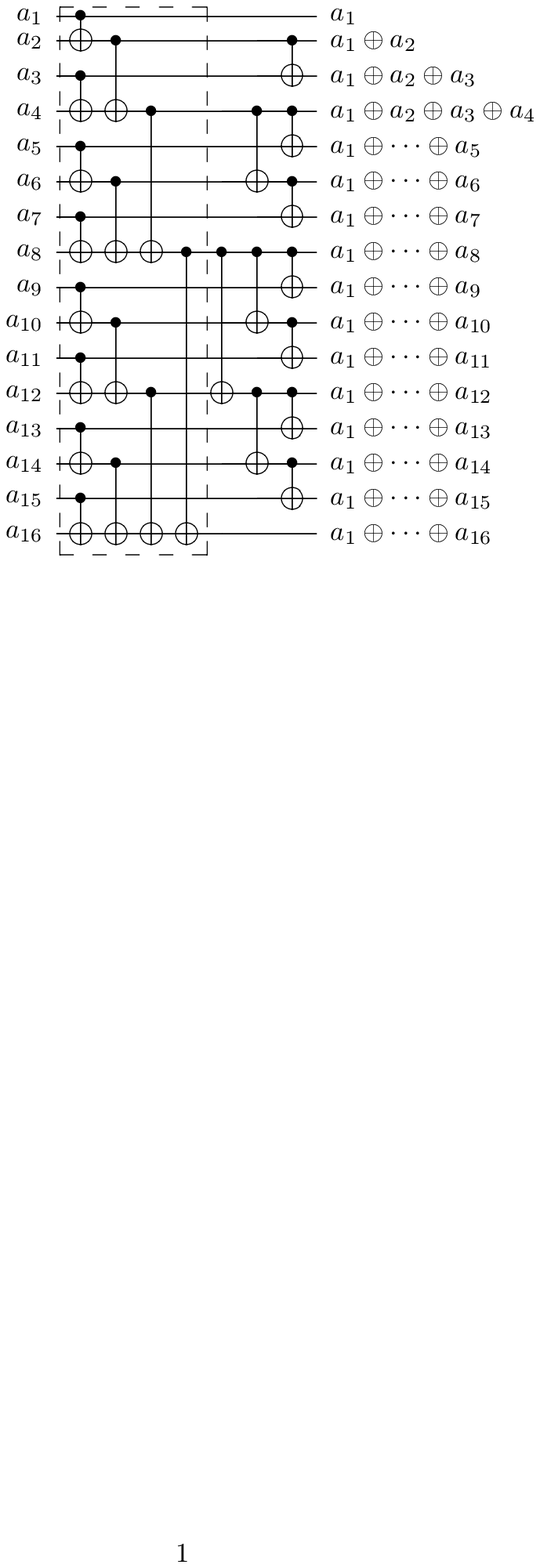}
	\caption{A Parallel Prefix Adder for $16$ Qubits.}
	\label{fig:parallel prefix adder}
\end{figure}

For our purposes, we only need to compute the parity of at most $N$ qubits and do not need the partial sums. This is similar to a parallel prefix circuit, except that we do not need the prefixes. Thus we generate the $CNOT$ circuit first using the appropriately chosen subset. We will end up with a circuit of $CNOT$ gates from $r$ qubits onto one qubit, which we will call the ``last qubit''. Then, we transform this circuit into a parallel prefix circuit, but we only consider gates that affect the last qubit and ignore the other gates. We are left with a circuit $C$ that is half of the circuit from Figure \ref{fig:parallel prefix adder}, highlighted using a dashed rectangle. Finally, we need to uncompute the intermediate results on all but the last qubit. This can be accomplished by applying the $CNOT$ gates in $C$ that do not involve the original control again in their reverse order. This yields an equivalent circuit of depth $O(\log N)$ and $O(N)$ $CNOT$ gates. 

The circuit for step 2 uses up to $2(N-1)$ $CNOT$ gates, $N$ single-qubit gates $\Id$, $T$, or $T^2$, and single $H$ gate to uniformize $X$ and $Y$. It has depth $2N$ and uses $2$ random bits per $CNOT$ for the first part of $CNOT$ gates to get probability $1/4$ and it uses a single random bit per $CNOT$ in the second part. The $X$-$Y$-uniformization costs a single random bit, and the $N$ $T^i$ twirls cost $\f{\log N}{\log 3}$ random bits. This gives a total of $O(N)$ gates in depth $O(N)$ using $O(N \log N)$ random bits.

\begin{figure}[htbp]
	\centerline{
	\Qcircuit @C=.7em @R=.7em
	{
	  & \gate{H} \qw & \ctrl{2} \qw & \gate{H} \qw & \qw & & & \targ \qw 		 & \qw \\
	  & 						 &     					&     				 &     & \push{\rule{.3em}{0em}\equiv\rule{.3em}{0em}} & & & \\
	  & \gate{H} \qw & \targ \qw    & \gate{H} \qw & \qw & & & \ctrl{-2} \qw & \qw 
	} }
	\caption{Hadamard Conjugation Flips Targets and Control of a $CNOT$ gate.}
	\label{fig:cnot h conjugation}
\end{figure}
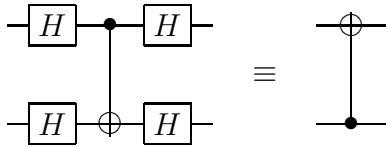

We can optimize the depth for this case as well. Once the circuit has been established, we can transform both $CNOT$ parts. For the first part, we use that conjugating a $CNOT$ with $H \otimes H$ swaps control and target as seen in Figure \ref{fig:cnot h conjugation}. Applying a Hadamard to all qubits before and after the first $CNOT$ part swaps the controls and targets of all $CNOT$ gates, using $H H = \Id$ between individual two $CNOT$ gates. This conjugation by Hadamard gates is illustrated in Figure \ref{fig:hadamard conjugation of random cnot circuit}.

\begin{figure}[htbp]
	\centerline{
	\Qcircuit @C=.7em @R=.7em
	{
		& \push{\f{3}{4}} & \push{\f{3}{4}} & & \push{\f{3}{4}} & \push{\f{3}{4}} & & & & & & \push{\f{3}{4}} & \push{\f{3}{4}} & & \push{\f{3}{4}} & \push{\f{3}{4}} \\
		& \targ \qw & \qw & \qdots \qw & \qw & \qw & \gate{T^{s_1}} \qw & \qw & & & \multigate{5}{H^{\otimes N}} \qw \qw & \ctrl{2} \qw & \qw & \qdots \qw & \qw & \qw & \multigate{5}{H^{\otimes N}} \qw & \gate{T^{s_1}} \qw & \qw\\
		& \qw & \targ \qw & \qdots \qw & \qw & \qw & \gate{T^{s_2}} \qw & \qw & & & \ghost{H^{\otimes N}} \qw & \qw & \ctrl{1} \qw & \qdots \qw & \qw & \qw & \ghost{H^{\otimes N}} \qw & \gate{T^{s_2}} \qw & \qw\\
		& \ctrl{-2} \qw & \ctrl{-1} & \qdots \qw & \ctrl{2} \qw & \ctrl{3} \qw & \gate{T^{s_3}} \qw & \qw & & & \ghost{H^{\otimes N}} \qw & \targ \qw & \targ & \qdots \qw & \targ \qw & \targ \qw & \ghost{H^{\otimes N}} \qw & \gate{T^{s_3}} \qw & \qw\\
		& & & \qddots & & & \qvdots & & \push{\rule{.3em}{0em}\equiv\rule{.3em}{0em}} & & & & & \qddots & & & &  \qvdots \\
		& \qw & \qw & \qdots \qw & \targ \qw & \qw & \gate{T^{s_{N-1}}} \qw & \qw & & & \ghost{H^{\otimes N}} \qw & \qw & \qw & \qdots \qw & \ctrl{-2} \qw & \qw & \ghost{H^{\otimes N}} \qw & \gate{T^{s_{N-1}}} \qw & \qw\\
		& \qw & \qw & \qdots \qw & \qw & \targ \qw & \gate{T^{s_N}} \qw & \qw & & & \ghost{H^{\otimes N}} \qw & \qw & \qw & \qdots \qw & \qw & \ctrl{-3} \qw & \ghost{H^{\otimes N}} \qw & \gate{T^{s_N}} \qw & \qw
	} }
	\caption{Hadamard Conjugation Flips Targets and Control in Circuit \ref{fig:circuit random cnots on all targets}.}
	\label{fig:hadamard conjugation of random cnot circuit}
\end{figure}
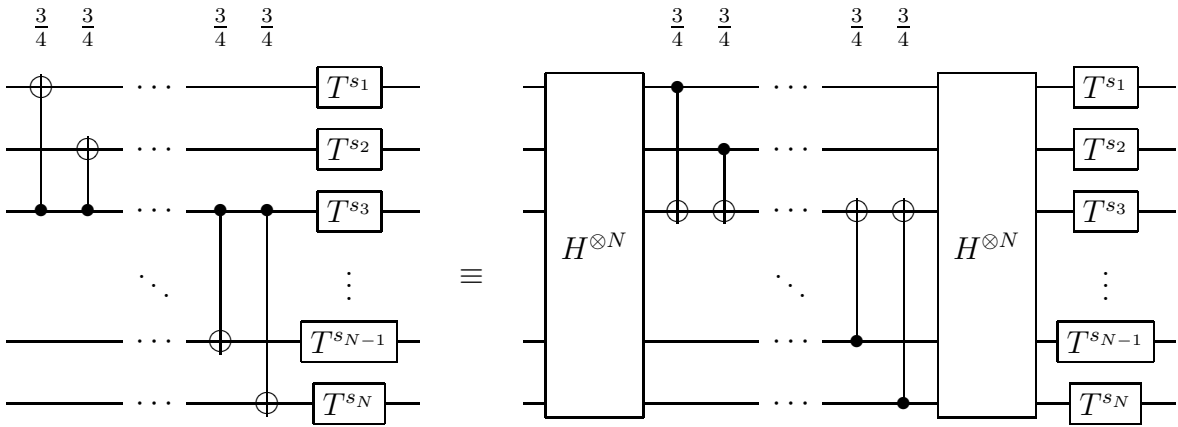

Using the same construction as in the optimization of step 1, we can also reduce the depth to $O(\log N)$ and $O(N)$ $CNOT$ gates. Counting the necessary $2N$ Hadamard gates, we end up with $O(N)$ gates. Note that the second part of $CNOT$ gates (see Figure \ref{fig:step 2}) can be directly transformed using this method, without conjugating by Hadamard gates as they already have the correction orientation.

These optimizations thus reduce the depth of step 1 and 2 to $O(\log N)$ and retain the circuit complexity of $O(N)$ $H$, $T$, $T^2$ and $CNOT$ gates. We need to repeat both steps $O\left(\log \f{1}{\epsilon}\right)$ times and the total complexity follows.

\paragraph{(5) Error Bound}

To bound the error, we consider $\Lambda_U$ that we would end up with had we perfectly symmetrized the Paulis. We assume the initial Pauli channel \[\Lambda(\rho) = \sum_{k=1}^{D^2} \alpha_j P_j \rho P_j\dag\] and denote the probability distribution of our conjugation process on $P_j$ by $\beta_{j,k}$ for $k = 2, 3, \dots, D^2$. 

Recall that \[\|\Lambda\|_{\diamond} = \sup_{\|\rho\|_1 = 1} \big\| (\Lambda \otimes \Id) (\rho)\bigr\|_1\] and observe that our conjugation process will not change the all-identity tensor-product Pauli. Thus 
\bess \left\| \Lambda_{\epsilon} - \Lambda_U \right\|_{\diamond} & = & \sup_{\|\rho\|_1 = 1} \left\| \sum_{j=2}^{D^2} \alpha_j \sum_{k=2}^{D^2} \beta_{j,k} (P_k \otimes \Id) \rho (P_k \otimes \Id)\dag - \sum_{j=2}^{D^2} \alpha_j \sum_{k=2}^{D^2} u(k) (P_k \otimes \Id) \rho (P_k \otimes \Id)\dag \right\|_1 \\
& = & \sup_{\|\rho\|_1 = 1} \left\|\sum_{j=2}^{D^2} \alpha_j \sum_{k=2}^{D^2} \left( \beta_{j,k} - u(k)\right) (P_k \otimes \Id) \rho (P_k \otimes \Id)\dag \right\|_1 \\
& \leq & \sup_{\|\rho\|_1 = 1} \sum_{j=2}^{D^2} |\alpha_j| \sum_{k=2}^{D^2} \left|\beta_{j,k} - u(k)\right| \left\| (P_k \otimes \Id) \rho (P_k \otimes \Id)\dag \right\|_1 \\
& = & \sup_{\|\rho\|_1 = 1} \sum_{j=2}^{D^2} |\alpha_j| \sum_{k=2}^{D^2} \left|\beta_{j,k} - u(k)\right| \left\|\rho \right\|_1 \\
& = & \sup_{\|\rho\|_1 = 1} \sum_{j=2}^{D^2} |\alpha_j| \left\|\rho \right\|_1 \sum_{k=2}^{D^2} \left|\beta_{j,k} - u(k)\right| \\
& \leq & \left(\epsilon_0 + \epsilon \right) \sum_{j=2}^{D^2} |\alpha_j|.
\eess
where we used that $\left\| (P_k \otimes \Id) \rho (P_k \otimes \Id)\dag \right\|_1 = \| \rho \|_1$ for $P_k \otimes \Id$ is unitary and $\|\cdot\|_1$ is unitarily invariant.
\end{proof}

We need to take the error bound into account to further derive the unitary $2$-design condition. As the input for Theorem \ref{thm:approximate symplectic twirling} is the Pauli-twirled superoperator from Lemma \ref{lem:pauli twirling}. Taking the completely-positive before the Pauli twirl to be \[\Lambda(\rho) = \sum_k A_k \rho A_k\dag,\] we see that $\alpha_j = \f{1}{D^2} \sum_k |\alpha_{k,j}|^2$, where $\sum \alpha_{k,j} P_j = A_k$ such that \[\alpha_{k,j} = \f{1}{D} \tr A_k\dag P_j = \f{1}{\sqrt{D}} \left(A_k, \f{1}{\sqrt{D}} P_j\right).\] Using the fact that $\big\{\f{1}{\sqrt{D}} P_j\,|\, j = 1, 2, \dots, D^2\bigr\}$ is an orthonormal basis, we conclude
\bess
\sum_{j=2}^{D^2} |\alpha_j| & = & \f{1}{D^2} \sum_{j=2}^{D^2} \sum_k |\alpha_{k,j}|^2 \\
& = & \f{1}{D^3} \left( \sum_k  \sum_{j=1}^{D^2} \left| \left(A_k, \f{1}{\sqrt{D}} P_j\right) \right|^2 - \sum_k \left| \left(A_k, \f{1}{\sqrt{D}} \Id\right)\right|^2 \right) \\
& = & \f{1}{D^3} \left( \sum_k \tr \left(A_k\dag A_k \right)- \f{1}{D} \sum_k \left|\tr A_k\dag \Id\right|^2 \right) \\
& = & \f{1}{D^4} \left(D \tr \left(\sum_k A_k A_k\dag\right) - \sum_k \left|\tr A_k\right|^2 \right) \\
& = & \f{1}{D^4} \left(D \tr \Lambda(\Id) - \tr \hat{\Lambda} \right)
\eess
using the formulas for $\tr \Lambda(\Id)$ and $\tr \hat{\Lambda}$ from Theorem \ref{thm:Haar avg explicit} in conjunction with the calculations in the proof of Corollary \ref{cor:avg-fidelity-explicit}.

This yields the bound \[\| \Lambda_{\epsilon} - \Lambda_U \|_{\diamond} \leq \f{D \tr \Lambda(\Id) - \tr \hat{\Lambda}}{D^4} \left(\epsilon_0 + \epsilon\right).\] We have that \[ B(\Lambda) = \f{D \tr \Lambda(\Id) - \tr \hat{\Lambda}}{D^4}. \]

\section{Discussion}

We introduced the notion of a $2$-design and showed that such an object was already used in the quantum information theory literature for bipartite state twirling \cite{DiVincenzoLeungTerhal2001}. However, it did not seem to be known that quantum operations can be twirled using the same object. 

We note that the private quantum channel result \cite{AmbainisEtAl2000} shows that the tensor-product Paulis $\Pauli(D)$ satisfy \[\f{1}{D^2} \sum_{j = 1}^{D^2} P_j \rho P_j\dag = \int_{U(D)} U \rho U\dag dU\] or all states $\rho$. It can be extended to all linear operators $\rho$ by linearity and the fact that the Hermitian operators form a basis for $L(\Hilbert)$, using the same arguments as in Section \ref{sec:motivation 2designs}. Hence we can see that this condition is equivalent to \[\f{1}{D^2} \sum_{j = 1}^{D^2} p(P_j) = \int_{U(D)} p(U) dU\] for all homogeneous polynomials of degree $(1,1)$, which we will call a \emph{unitary $1$-design}.

In the abstract formulation of Definition \ref{def:unitary 2-design}, it might turn out to be useful in a broader context where Haar-randomization can be reduced to randomization over a fairly small set of quantum gates that have efficient circuit decompositions. Such application beyong twirling are yet to be found or identified.

\clearpage
\thispagestyle{empty}
\cleardoublepage

\chapter{Conclusion and Future Research}
\label{ch:conclusion}
\markright{Conclusion and Open Problems}

\section{Conclusion}

We explored ways to efficiently estimate the average fidelity of a quantum channel or an implementation of a quantum algorithm $U$. It turned out that we re-discovered the previously known result that a complete set of mutually-unbiased bases gives a $2$-design for quantum states. This condition was shown to be equivalent to give an unbiased estimate of the average fidelity. Our contribution was an explicit circuit construction using $O(N^2)$ gates in depth $O(N)$ and $O(N)$ random bits.

Then, the notion of a $2$-design for quantum states was generalized to $2$-designs for unitary operators. Although the term ``$2$-design'' did not seem to have appeared before in the literature, the concept was implicitly used as early as at least 1996. It was independently proven \cite{DiVincenzoLeungTerhal2001, Chau2005} that the Clifford group is a unitary $2$-design by showing its use for quantum operation twirling and state twirling. An approximately uniform sampling algorithm over the Clifford group was proposed as well. Our contribution is the unified view of these different approaches as unitary $2$-design. Also, we showed that a subset of the Clifford group suffices to be exponentially close to a $2$-design for both applications. 

We have also seen that we can define the notion of a unitary $1$-design, which was already implicitly shown to exist \cite{AmbainisEtAl2000} in the context of a private quantum channel. Interestingly, the Pauli group was this unitary $1$-design.

\section{Directions for Future Research and Open Problems}

\subsection{Definition of Unitary $t$-Designs}

For future research, there are several areas to proceed in. First and foremost, it seems apparent to extend the notion of unitary designs to $t$-designs for arbitrary $t$. A possible application can be noise estimation scenarios where the time evolution of the average fidelity is of interest, as suggested in \cite{EmersonEtAl2005_Noise}. Imagine we model the evolution of our quantum systems from time $t_0$ to $t_1$ by the quantum operation $\E_1$, and from $t_1$ to $t_2$ by the quantum operation $\E_2$.

\begin{figure}[htbp]
	\centerline{
	\Qcircuit @C=1em @R=.7em
	{
	  \lstick{\rho} & \gate{U} \qw & \gate{\E_1} \qw & \gate{U\dag} \qw & \gate{\E_2} \qw & \gate{U} \qw & 
	} }
	\caption{Twirling two Successive Quantum Operations}
	\label{fig:two twirls}
\end{figure}
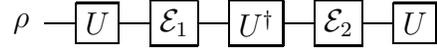

Suppose we twirl both operations with the same unitary as shown in Figure \ref{fig:two twirls}, so that we end up with the operation $\E'$ that maps
\[\E'(\rho) = \int_{U(D)} U \E_2(U\dag \E_1(U \rho U\dag) U) U\dag.\] The integral contains three occurances of each $U$ and $U\dag$, what suggests that estimating this integral requires a unitary $3$-design.

Estimating the fidelity decay with even finer temporal resolution seems to require higher-order unitary designs, so that the quest for unitary $t$-designs can be motivated from this experimental point of view.

\subsection{Proof Idea for Unitary $t$-Designs}

We will present a proof technique that might be useful in generalizing $2$-designs to $t$-designs for $t > 2$. It is based on representation theory and similar in spirit to the decomposition lemma of a unitarily invariant superoperator (Lemma \ref{lem:repn-of-superop}). This technique might be useful in determining subgroups of $U(d)$ that could serve as $2$-designs other than those already discovered. We will state where these ideas need to be extended and why they do not work yet.

Let $\Hilbert$ denote a Hilbert space of dimension $d$, and let $G \leq U(d)$ be a subgroup of the unitary group $U(d)$ such that an invariant measure exists on $G$.

\begin{df} \label{def:G-invariance} We will use the representation $\hat{U}: U(d) \rightarrow L(L(\Hilbert))$ from Definition \ref{def:unitarily inv}, which was defined as \[\hat{U} \rho = U \rho U\dag\] for all $U \in U(d)$. Note that this is also a representation of $G$. 

Furthermore, we will call a linear operator $\rho \in L(\Hilbert)$ \emph{$G$-invariant} if $\hat{V} \rho \hat{V}\dag = \rho$ for all $V \in G$.
\end{df}

Recall the definition of $U(d)$-invariance (Definition \ref{def:unitarily inv}), which becomes a special case of Definition \ref{def:G-invariance} if $G = U(d)$. We extend Definition \ref{def:haar twirling} to $G$-twirling in the obvious way.

\begin{df} \label{def:G-twirling} Let $\Lambda$ be a superoperator. Define the \emph{$G$-twirled superoperator} as \[\Lambda_G (\rho) = \int_G \hat{V} \hat{\Lambda} \hat{V}\dag dV \rho = \int_G V\dag \Lambda( V \rho V\dag) V dV.\]
\end{df}

\begin{lemma} The $G$-twirled superoperator $\Lambda_G$ is $G$-invariant.
\end{lemma}
\begin{proof} The proof is similar to the proof of Lemma \ref{lem:unitarily inv superop}. Let $U \in G$ and $\rho \in L(\Hilbert)$.
\bess
\hat{U} \hat{\Lambda_G} \hat{U}\dag \rho 
& = & U\dag \int_G V\dag \Lambda( V U \rho U \dag V\dag) V dV U \\
& = & \int_G (VU)\dag \Lambda( (VU) \rho (VU)\dag) (VU) dV \\
& = & \int_G (V')\dag \Lambda( V' \rho (V')\dag) V' dV \\
& = & \Lambda_G(\rho),
\eess
where we used the $G$-invariance of the measure $dV$ on $G$ and the substitution $V' = VU$.
\end{proof}

The following lemma is the critical point of this technique and needs to be shown.

\begin{conj} \label{conj:irreducible reps for G and U} If the irreducible representations of $\hat{U}$ of $U(d)$ are also irreducible for $G$, then $\Lambda_G$ is $U(d)$-invariant.
\end{conj}

\paragraph{Proof idea}

As the irreducible representations for $U(d)$ are also irreducible for $G$, Schur's Lemma (Fact \ref{fact:schurs lemma}) implies that $\Lambda_G$ will act as identity on the same subspaces as $\Lambda_T$. 

In the case of $t = 2$, these irreducible subspaces are known from Lemma \ref{lem:repn-of-superop} as the traceless linear operators and multiples of the identity, possibly with different coefficients than $\Lambda_T$ would. For this decomposition, we easily see that $\Lambda_G$ is unitarily invariant by calculating
\bess
\hat{U} \hat{\Lambda_G} \hat{U}\dag \rho & = & U\dag p \left(U \rho U\dag - \tr (U \rho U\dag) \f{\Id}{d}\right) U + U\dag q \tr (U \rho U\dag) \f{\Id}{d} U \\
& = & p\left( \rho - \tr \rho \f{\Id}{d}\right) + q \tr \rho \f{\Id}{d} = \Lambda_G(\rho).
\eess
Thus $\Lambda_G$ is $U(d)$-invariant.

However, to extend this to general $t$, we need to be able to show that $\Lambda_G$ is unitarily invariant either without knowing the explicit decomposition into the irreducible representations or by making use of this explicit decomposition.

If we had this lemma, we could show the following Corollary.

\begin{cor} \label{cor:G-twirl = U(d)-twirl} For any superoperator $\Lambda$, \[ \int_G \hat{V} \hat{\Lambda} \hat{V}\dag dV = \int_{U(d)} \hat{U} \hat{\Lambda} \hat{V}.\]
\end{cor}
\begin{proof} The invariant measure $dV$ on $G$ can be trivially extended to a probability measure $\mu$ on $U(d)$ by letting $\int_E d\mu = \int_{E \cup G} dV$. Lemma \ref{lem:all unitarily invariant twirlings are the same} implies that $\Lambda_{\mu} = \Lambda_G = \Lambda_T$ and the statement follows.
\end{proof}

The corollary that $G$ is a unitary $t$-design could be proven the following way: First, we could use Conjecture \ref{conj:irreducible reps for G and U} and thus we need to show that the irreducible subspaces of $\hat{U}$ for $U(d)$ remain irreducible for $G$. 

In order to prove that the Clifford group is a $2$-design, we could use \ref{lem:repn-of-superop} and show that the space of traceless Hermitian operators is irreducible under the Clifford group. However, we were not able to show that. Maybe the Clifford would turn out to be a $t$-design for $t > 2$. This should be subjecr of future research.

It also seems to be worthwhile looking into the random circuit construction (see Section \ref{sec:random circuits}) again and figure out how unitary $t$-designs might be derived using this approach.

\subsection{Find a Better Approximate Pauli Uniformization}

So far, the Pauli uniformization has an absolute lower bound of $\epsilon_0 \approx 1/2^N$ from Theorem \ref{thm:approximate symplectic twirling}. Maybe one could improve the analysis to get an arbitrarily small upper bound on $\epsilon_0$, or one might choose a slightly larger subset of the Clifford group which facilitates that.

\subsection{Extend the Approximate Pauli Uniformization}

The construction in Theorem \ref{thm:approximate symplectic twirling} only works for qubits. In order to make it work for qudits as well, we need to find an analogy to the generator of the single-Pauli twirl $T = HP$. Assuming $d > 3$ a prime, we could use the special phase gate \[P^d_{r} \ket{x} = \omega_d^{r x^2/2} \ket{x}\] for $r \in \F_d^*$ we can conjugate
\[P_{d,r} X^a Z^b (P^d_r)\dag = X^a Z^{b - a r^{-1}}\]
and hence we can uniformize the $Z$ component provided $a \neq 0$.

In order to uniformize the $X$ component, we could make use of the Quantum Fourier Transform modulo $d$, which is given by
\[ F_d \ket{x} = \sum_{y \in \F_d} \omega_d^{xy} \ket{y}. \] It acts on the Paulis by conjugation as 
\[ F_d X^a Z^b F_d\dag =  X^{-b} Z^a,\] which implies
\[ F_d^3 X^a Z^b (F_d^3)\dag =  X^b Z^a.\] This allows us to uniformize the $X$ component by conjugating with $F_d^3 P_{d,r} F_d^3$ as long as the $X$ component is non-zero.

The problem is the case where either the $X$ or the $Z$ component is zero, as those will not be reached by one of these two randomization steps. However, it seems conceivable that an alternating chain of conjugation by $F_d^3 P_{d,r} F_d^3$ and $P_{d,r}$ could create an almost uniform distribution over all non-identity Paulis. Maybe one could even find a generator of a cyclic shift that is analog to $T$ in the qubit case.

The next step is to find an analogy for the $CNOT$ operation. Observe that the obvious generalization of the $CNOT$ is $CPLUS_d$, which is defined by $CPLUS_d \ket{x}\ket{y} = \ket{x}\ket{y + 1}$ where addition is modulo $d$. It conjugates 
\[ CPLUS_d (X^{a_1} Z^{b_1} \otimes  X^{a_2} Z^{b_2}) CPLUS_d\dag = X^{a_1} Z^{b_1 + b_2} \otimes  X^{a_2 - a_1} Z^{b_2}\]
and thus seems a reasonable candidate for further investigation.

If that would turn out not to be a good choice, one could try the generalization \[CSUM^{\alpha, \beta}_d \ket{x}\ket{y} \ket{x}\ket{y} = \ket{\alpha x + \beta y}\ket{\beta a - \alpha_y},\] which is reversible if $\alpha, \beta \in \F_d$ and $(\alpha, \beta) \neq (0,0)$. It can be shown that
\[CSUM^{\alpha, \beta})_d (X^{a_1} Z^{b_1} \otimes  X^{a_2} Z^{b_2}) (CSUM^{\alpha, \beta}_d))\dag \\ 
= X^{\f{\alpha a_1 + \beta a_2}{\alpha^2 + \beta^2}} Z^{\alpha b_1 + \beta b_2} \otimes  X^{\f{\beta a_1 - \alpha a_2}{\alpha^2 + \beta^2}} Z^{\beta b_1 - \alpha b_2},\] 
which might be more suitable than $CPLUS_d$.

\appendix

\clearpage
\thispagestyle{empty}
\cleardoublepage

\chapter{Mathematical Background}
\label{ch:background}
\markright{Mathematical Background}


This appendix is intended to be a reference for the mathematical concepts and notations used throughout this thesis. See \cite{Ballentine1998} for an introduction to the basic concepts of linear algebra in both finite and infinite-dimensional settings that is streamlined to the description of quantum mechanics. For the finite-dimensional case of quantum computing and quantum information theory topics, \cite{NielsenChuang2000} is the most suitable reference to date.

\section{Vector Spaces}

\begin{df}
A \emph{vector space} over a field $\F$ is a set $V$ together with two binary operations
\begin{itemize}
	\item \emph{vector addition}: $V \times V \rightarrow V$, written $\v{u} + \v{v}$ with $\v{u}, \v{v} \in V$ and
	\item \emph{scalar multiplication}: $F \times V \rightarrow V$, denoted by $a \v{u}$ with $a \in \F, \v{u} \in V$
\end{itemize}
such that the following axioms hold:
\begin{enumerate}
	\item Associativity of vector addition: $\v{u} + (\v{v} + \v{w}) = (\v{u} + \v{v}) + \v{w}$ for all $\v{u}, \v{v}, \v{w} \in V$
	\item Commutativity of vector addition: $\v{u} + \v{v} = \v{v} + \v{u}$ for all $\v{u}, \v{v} \in V$
	\item Existence of an \emph{additive identity} $\v{o} \in V$ such that $\v{u} + \v{o} = \v{u}$ for all $\v{u} \in V$
	\item Existence of an \emph{inverse vector} $-\v{u}$ for all $\v{u} \in V$ such that $\v{u} + (-\v{u}) = \v{o}$
	\item Associativity of scalar multiplication: $a(b\v{u}) = (ab)\v{u}$ for all $a, b \in \F$ and $\v{u} \in V$
	\item $1 \v{u} = \v{u}$ for all $\v{u} \in V$
	\item Distributivity of scalar multiplication over vector addition: $a(\v{u} + \v{v}) = a\v{u} + a\v{v}$
	\item Distributivity of scalar multiplication over scalar addition: $(a+b)\v{u} = a\v{u} + b\v{u}$
\end{enumerate}
The elements of $\v{u} \in V$ are called \emph{vectors} and the elements of $\F$ are called \emph{scalars}.
\end{df}

\begin{df}
A \emph{real vector space} is a vector space over the real numbers. A \emph{complex vector space} is a vector space over the complex numbers.
\end{df}

\begin{df}
A \emph{subspace} $W$ of a vector space $V$ is a subset that is closed under vector addition and scalar multiplication. The intersection of all subspaces that contain a given set of vectors $S$ is called the \emph{span} of $S$. A set of vectors $S = \{\v{v_1}, \dots, \v{v_n}\} \subset V$ is called \emph{linearly independent} if \[a_1 \v{v_1} + \dots + a_n \v{v_n} = \v{0}\]  has only the trivial solution $a_1 = \dots = a_n = 0$. $S$ is called a \emph{basis} if the span of $S$ is $V$.
\end{df}

Every basis for a vector space $V$ has the same cardinality which is called the \emph{dimension} of $V$. All vector spaces over a given field $\F$ of the same dimension are isomorphic. 

Sometimes it is helpful to write a vector space as a sum of some of its subspaces. 
\begin{df} Let $V$ and $W$ be vector spaces over a field $K$. The \emph{direct sum} of $V$ and $W$ is the Cartesian product $V \times W$ endowed with the vector space operations
\begin{enumerate}
	\item $(\v{v}_1, v{w}_1) + (\v{v}_2, v{w}_2) = (\v{v}_1 + \v{v}_2, \v{w}_1 + \v{w}_2)$ for all $\v{v}_1, \v{v}_2 \in V$, $\v{w}_1, \v{w}_2 \in W$ and
	\item $a (\v{v}, v{w}) = (a \v{v}, a \v{w})$ for all $a \in K$, $\v{v} \in V$, $\v{w} \in W$.
\end{enumerate}
The resulting vector space is called the direct sum of $V$ and $W$ and written as $V \oplus W$.
\end{df}

\begin{df}
Let $V$ be a vector space over a subfield $\F \subseteq \C$ of the complex numbers. A \emph{norm} on $V$ is a function $|\cdot|: V \rightarrow \R$ such that the following  properties hold:
\begin{enumerate}
	\item Positivity: $|\v{v}| \geq 0$ for all $\v{v} \in V$
	\item Positive scalability: $|a\v{v}| = |a| |\v{v}|$ for all $a \in \F, \v{v} \in V$
	\item Triangle inequality: $|\v{u} + \v{v}| \leq |\v{u}| + |\v{v}|$
	\item Positive definiteness: $|\v{v}| = 0$ iff $\v{v} = \v{o}$
\end{enumerate}
\end{df}

\begin{df} A \emph{normed vector space} is a pair $(V, |\cdot|)$ such that $V$ is a vector space and $|\cdot|$ is a norm on $V$. A vector $\v{v} \in V$ is \emph{normalized} if $|\v{v}| = 1$.
\end{df}

\begin{df}
A function from a vector space $V$ to a vector space $W$ over the same field $\F$, $f: V \rightarrow W$, is called \emph{linear} if 
\begin{itemize}
	\item $f(\v{u} + \v{v}) = f(\v{u}) + f(\v{v})$ for all $\v{u}, \v{v} \in V$ and
	\item $f(a\v{u}) = a f(\v{u})$ for all $a \in \F, \v{u} \in V$.
\end{itemize}
A function $f: V \times W \rightarrow X$ for vector spaces $V, W, X$ over the same field $\F$ is \emph{bilinear} if
\begin{enumerate}
	\item $\v{v} \mapsto f(\v{v}, \v{w})$ is linear for every $\v{w} \in W$ and
	\item $\v{w} \mapsto f(\v{v}, \v{w})$ is linear for every $\v{v} \in V$.
\end{enumerate}
A function $f: V \times W \rightarrow X$ for vector spaces $V, W, X$ over a subfield $\F \subseteq \C$ of the complex numbers is \emph{sesquilinear} if it is bilinear except $f(a\v{v}, \v{w}) = \overline{a} f(\v{v}, \v{w})$.
\end{df}

\begin{df} Let $V, W$ be normed vector spaces with norms $|\cdot|_V, |\cdot|_W$. This norm induces a norm on the set of linear operators $L(V,W)$ from $V$ to $W$ defined as \[\|A\| = \max_{\v{x} \in V} \f{\|A \v{x}\|_W}{\|\v{x}\|_V}.\] We will call $\|\cdot\|: L(V, W) \rightarrow \R$ the \emph{induced operator norm} on $L(V,W)$.
\end{df}

\begin{df}
A complex inner product space is a vector space $V$ over $\C$ together with a map $( \cdot, \cdot ): V \times V \rightarrow \C$ such that
\begin{enumerate}
	\item $( \cdot, \cdot )$ is sesquilinear,
	\item $( \v{u}, \v{v} ) = \overline{( \v{v}, \v{u} )}$ for all $\v{u}, \v{v} \in V$,
	\item $( \v{u}, \v{v} ) \geq 0$ for all $\v{u}, \v{v} \in V$, and
	\item $( \v{v}, \v{v} ) = 0$ iff $\v{v} = \v{o}$ for all $\v{v} \in V$.
\end{enumerate}
\end{df}

\begin{df}
Let $V$ be a complex inner product space. Two vectors $\v{u}, \v{v} \in V$ are \emph{orthogonal} if $( \v{u}, \v{v} ) = 0$.
\end{df}


\begin{df} For every complex inner product space $V$, there is a norm $|\v{v}| = \sqrt{\langle \v{u}, \v{v} \rangle}$. $V$ is \emph{complete} with respect to that norm if every Cauchy sequence converges to an element of that space. A complete normed complex inner product space $\Hilbert$ is called a \emph{Hilbert space}. Note that in the mathematical literature, a distinction is made between complex and real Hilbert spaces, which are complete normed inner product spaces over the real and complex numbers, respectively. However, quantum computing literature understands a Hilbert space as defined above. We will use that definition throughout this thesis.
\end{df}

\begin{df}
An orthonormal basis for a Hilbert space $\Hilbert$ is a set $S \subset \Hilbert$ whose span is dense in $\Hilbert$ and whose elements are pairwise orthogonal and have norm one.
\end{df}

\begin{fact}
\begin{enumerate}
	\item Every finite-dimensional complex inner product space $\Hilbert$ is a Hilbert space.
	\item Every Hilbert space $\Hilbert$ has an orthonormal basis. Any two orthonormal bases of $\Hilbert$ have the same cardinality.
\end{enumerate}
\end{fact}

\begin{df}
Let $f$ be a sesquilinear function $f: V \times V \mapsto \C$ for a vector space $V$ over a subfield $\F \subseteq \C$ of the complex numbers. Given $\v{u}$, the map $\v{v} \mapsto f(\v{u}, \v{v})$ is called a \emph{linear functional} on $V$. The set of all linear functionals on $V$ forms a vector space under addition of functions and scalar multiplication. It is called the dual space of $V$ and denoted $V^*$.
\end{df}

\section{Dirac Notation}

Paul Dirac introduced a convenient notation for Hilbert spaces that has been widely accepted in quantum mechanics literature. This notation is sometimes referred to as ``bra-ket'' notation because the inner product of two vectors is denoted by a bracket $(\phi, \psi)$ or $\langle \phi, \psi \rangle$. The left part $\bra{\phi}$, is called ``bra'', and the right part $\ket{\psi}$ is called ``ket''. Let $\Hilbert$ be a $n$-dimensional Hilbert space. Most of the definitions also hold for infinite-dimensional Hilbert spaces as well, but they are not necessary for quantum computing.

\begin{df} Each vector in $\Hilbert$ is called ``ket'' and written as $\ket{\psi}$. $\psi$ denotes the vector and the bar and angle bracket denote that it is to be understood as the vector $\psi$, read ``ket psi''. 
\end{df}

\begin{fact}
For every $\ket{\psi} \in \Hilbert$, there is exactly one dual $\bra{\psi} \in \Hilbert^*$, read ``bra psi'', which is a continuous linear functional from $\Hilbert$ to $\C$: \[\bra{\psi} (\ket{\phi}) = \left( \ket{\psi}, \ket{\phi} \right) \text{ for all } \ket{\phi} \in \Hilbert.\] The converse is true as well as $\Hilbert$ and $\Hilbert^*$ are isometrically isomorphic.
\end{fact}

\begin{df}
A \emph{linear operator} on $\Hilbert$ is a linear function from $\Hilbert$ to $\Hilbert$. Operators act on kets from the left. Let $A$ be a linear operator on a Hilbert space $\Hilbert$, then $A\ket{\psi} = A(\ket{\psi})$. Operators can also act on bras from the right hand side, such that $\bra{\phi} A$ is understood as the operator that acts as $(\bra{\phi}A)(\ket{\psi} = \bra{\phi} \left( A\ket{\psi} \right) = \bra{\phi} A \ket{\psi}$. 
\end{df}

\begin{fact}
Let $A$ be a linear operator on $\Hilbert$. There is a unique linear operator $A\dag$ such that \[(\ket{\phi}, A \ket{\psi}) = (A\dag \ket{\phi}, \ket{\psi})\] for all $\ket{\phi}, \ket{\psi} \in \Hilbert$. If we define $\ket{\phi}$ as the linear operator that maps $\ket{\phi} (\ket{\psi}) = \bracket{\phi}{\psi}$, we have that $\ket{\phi}\dag = \bra{\phi}$. It follows that $(A\ket{\phi})\dag = \bra{\phi} A\dag$. We also note that $(AB)\dag = B\dag A\dag$ for $A, B \in L(\Hilbert)$.
\end{fact}

\begin{df}
Let $A$ be a linear operator on $\Hilbert$. $A$ is
\begin{itemize}
  \item \emph{invertible} if there is an operator $A^{-1}$ such that $A \circ A^{-1} = A^{-1} \circ A = \Id$ is the identity operator on $\Hilbert$.
	\item \emph{Hermitian} or \emph{self-adjoint} if $A = A\dag$.
	\item \emph{normal} if $A A\dag = A\dag A$.
	\item \emph{unitary} if $A A\dag = \Id$.
	\item \emph{positive} if $\bra{\psi} A \ket{\psi} \geq 0$ for all $\ket{\psi} \in \Hilbert$.
\end{itemize}
\end{df}

\begin{fact}
A unitary operator $U$ on $\Hilbert$ preserves inner products: \[(U \ket{\psi}, U\dag \ket{\phi}) = \bra{\psi} U U\dag \ket{\phi} = \bracket{\psi}{\phi}.\]
\end{fact}

\begin{fact}
Let $B = \{\ket{\psi_1}, \dots, \ket{\psi_n}\}$ be an orthonormal basis for $\Hilbert$ and let $A$ be a linear operator on $\Hilbert$. If we choose to represent vectors in $\Hilbert$ as column vectors with $n$ entries in $\C$, we can represent $A$ by the $n \times n$ matrix $(a_{i,j})$ with elements $a_{i,j} = \bra{\psi_i} A \ket{\psi_j}$.
\end{fact}

\begin{df}
Let $A \in L({\Hilbert})$, $\ket{\psi} \in \Hilbert$ and $\lambda \in \C$. $\ket{\psi}$ is called an \emph{eigenvector} of $A$ with eigenvalue $\lambda$ if \[A \ket{\psi} = \lambda \ket{\psi}.\]
\end{df}

\begin{fact}[Spectral Decomposition Theorem] Let $A$ be a normal linear operator on $\Hilbert$. Then there is an orthonormal basis $\{\ket{\psi_1}, \dots, \ket{\psi_n}\}$ of $\Hilbert$ and $\lambda_1, \dots, \lambda_n \in \C$ such that \[A = \sum_{i=1}^n \lambda_i \ket{\psi}\bra{\psi}.\]
\end{fact}

\begin{df}
We define the ``outer product'' of two vectors $\ket{\phi}$ and $\ket{\psi}$ as the operator \[\left(\ket{\phi}\bra{\psi}\right)(\ket{\chi}) = \ket{\phi} \bracket{\psi}{\chi} =  \bracket{\psi}{\chi} \ket{\phi}. \] This outer product notation is generally used to define projection operators. Given a normalized $\ket{\phi} \in \Hilbert$, we define the operator that projects onto the subspace spanned by $\ket{\phi}$ as $\ket{\phi}\bra{\phi}$.
\end{df}

\begin{fact}
The \emph{trace} is the unique linear function \[\tr: \Hilbert \mapsto \C\] such that
\begin{itemize}
	\item it is \emph{unitarily invariant}, $\tr A = \tr U A U\dag$ for all linear operators $A$ and unitary operators $U$, and
	\item $\tr \Id = n$.
\end{itemize}
Let $A$ be a linear operator on $\Hilbert$ and let $(a_{i,j})$ be a matrix representation of $A$ in some orthonormal basis. Then \[\tr(A) = \sum_{i=1}^n a_{i,i}.\] The trace function is well defined and does not depend on the specific representation of $A$. Furthermore, the following algebraic properties hold. Let $A, B \in \Hilbert$, then
\begin{itemize}
	\item $\tr(AB) = \tr(BA)$ (cyclic property)
	\item $\tr A = \tr A\dag$
\end{itemize}
\end{fact}

\begin{fact} The set of all linear operators on $\Hilbert$ forms an $n^2$ dimensional vector space and is denoted by $L({\Hilbert})$. $L({\Hilbert})$ is a Hilbert space with inner product $(A, B) = \tr A\dag B$. This inner product is called \emph{Hilbert-Schmidt} or \emph{trace} inner product.
\end{fact}

\begin{df}
Let $V$ and $W$ be Hilbert spaces of dimensions $m$ and $n$, respectively. Then the \emph{tensor product} of $V$ and $W$, written as $V \otimes W$, is an $mn$ dimensional complex vector space. $V \otimes W = V \times W$ and the vector addition and scalar multiplication satisfy the following restrictions:
\begin{enumerate}
	\item $a (\ket{\psi} \otimes \ket{\phi}) = (a\ket{\psi}) \otimes \ket{\phi} = \ket{\psi} \otimes (a\ket{\phi})$ for all $\ket{\psi} \in V$, $\ket{\phi} \in W$, $a \in \C$
	\item $(\ket{\psi} + \ket{\phi}) \otimes \ket{\chi} = \ket{\psi}\otimes\ket{\chi} + \ket{\phi}\otimes\ket{\chi}$ for all $\ket{\psi}, \ket{\phi} \in V, \ket{\chi} \in W$
	\item $\ket{\chi} \otimes (\ket{\psi} + \ket{\phi})  = \ket{\chi}\otimes\ket{\psi} + \ket{\chi}\otimes\ket{\phi}$ for all $\ket{\chi} \in V, \ket{\psi}, \ket{\phi} \in W$
\end{enumerate}
The tensor product of two vectors $\ket{\psi}\otimes\ket{\phi}$ is most often abbreviated as $\ket{\psi}\ket{\phi}$, $\ket{\psi, \phi}$, or even $\ket{\psi \phi}$.

Given linear operators $A$ on $V$ and $B$ on $W$, we can define the operator $A \otimes B$ by letting  \[(A \otimes B)(\ket{\psi} \otimes \ket{\phi}) = A\ket{\psi} \otimes B\ket{\phi}.\]
\end{df}

\begin{fact}
Let $V$ and $W$ be Hilbert spaces of dimensions $m$ and $n$. Then $V \otimes W$ is a Hilbert space of dimension $mn$ with inner product \[\left(\sum_i a_i \ket{\psi_i} \otimes \ket{\phi_i}, \sum_j b_i \ket{\psi_j'} \otimes \ket{\phi_j'}\right) = \sum_{i,j} a_i^* b_j \bracket{\psi_i}{\psi_j'}\bracket{\phi_i}{\phi_j'}.\]
\end{fact}

To make the discussion about tensor products a little more concrete, we will have a look at an example tensor product. Pick an orthonormal basis for Hilbert spaces $V$ and $W$ of dimensions $n$ and $m$, respectively. Then we can represent their elements as column vectors. Let $\ket{\psi} \in V, \ket{\phi} \in W$: \[
\ket{\psi} = \bv \psi_1 \\ \psi_2 \\ \vdots \\ \psi_n \ev, 
\ket{\phi} = \bv \phi_1 \\ \phi_2 \\ \vdots \\ \phi_m \ev .
\] 

The tensor product of $\ket{\psi}$ and $\ket{\phi}$ is given by the \emph{Kronecker product} if we think of these vectors as $1$-by-$n$ and $1$-by-$m$ matrices. Hence \[\ket{\psi} \otimes \ket{\phi} = \bv \psi_1 \phi_1 \\ \psi_1 \phi_2 \\ \vdots \\ \psi_1 \phi_m \\ \psi_2 \phi_1 \\ \vdots \\ \psi_n \phi_m.
\ev.\]

Linear operators on $V$ are represented by $n$-by-$n$ dimensional complex matrices in the usual way. Given two operators $A$ on $V$ and $B$ on $W$, \[A = \bv 
a_{1,1} & a_{1,2} & \dots & a_{1,n} \\
a_{2,1} & a_{2,2} & \dots & a_{2,n} \\
\vdots & \vdots & \ddots & \vdots \\
a_{n,1} & a_{n,2} & \dots & a_{n,n} 
\ev, 
B = \bv 
b_{1,1} & b_{1,2} & \dots & b_{1,m} \\
\vdots & \vdots & \ddots & \vdots \\
b_{m,1} & b_{m,2} & \dots & b_{m,m} 
\ev.\] 

The operator $A \otimes B$ that acts on $V \otimes W$ is now given by the Kronecker product of $A$ and $B$:
\[ A \otimes B = \bv
a_{1,1} B & a_{1,2} B & \dots & a_{1,n} B \\
a_{2,1} B & a_{2,2} B & \dots & a_{2,n} B \\
\vdots & \vdots & \ddots & \vdots \\
a_{n,1} B & a_{n,2} B & \dots & a_{n,n} B 
\ev\]
where $a_{i,j} B$ means that the submatrix $B$ with all entries multiplied by $a_{i,j}$ is to be inserted.

\section{The Bloch Sphere}
\label{sec:bloch sphere}

\begin{figure}[htbp]
	\centering
	 	\psfrag{x}{$x$}
	 	\psfrag{y}{$y$}
	 	\psfrag{z}{$z$}
	 	\psfrag{ket0}{$\ket{0}$}
	 	\psfrag{ket1}{$\ket{1}$}
		\includegraphics[width=0.50\textwidth]{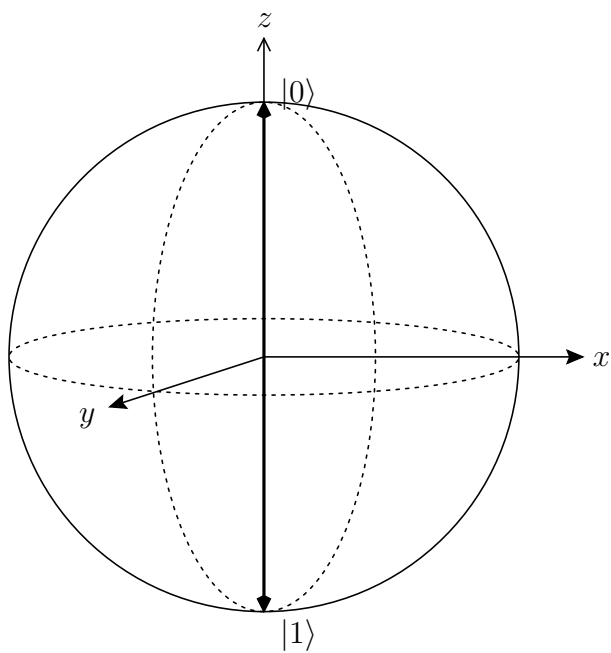}
	\caption{Bloch sphere representation of the computational basis states.}
	\label{fig:bloch sphere 1}
\end{figure}

We will make use of a nice geometrical interpretation of single qubit states. It is known that all the observable properties of a single qubit system can be described using the unit sphere. The state of a single qubit system is described by a unit vector $\ket{\psi}$ in a Hilbert space $\Hilbert_2$ of dimension $2$ with orthonormal basis $\{\ket{0}, \ket{1}\}$, \[\ket{\psi} = \alpha \ket{0} + \beta \ket{1}.\] We will call this basis the \emph{computational basis}. As $|\alpha|^2 + |\beta|^2 = 1$, we can rewrite \[\ket{\psi} = e^{i\gamma} \left(\cos \frac{\theta}{2}\ket{0} + e^{i\varphi}\sin\frac{\theta}{2}\ket{1}\right).\] The global phase factor has no observable properties, and hence this state is equivalent to \[\ket{\psi} = \cos \frac{\theta}{2}\ket{0} + e^{i\varphi}\sin\frac{\theta}{2}\ket{1}\] with two real parameters $\theta$ and $\varphi$. Now define 
\bess
x & = & \sin \theta \cos \varphi \\
y & = & \sin \theta \sin \varphi \\
z & = & \cos \theta
\eess
and we have a mapping from the set of pure quantum states to the unit sphere. Figure \ref{fig:bloch sphere 1} shows the Bloch sphere with the computational basis states $\ket{0}$ and $\ket{1}$.

The other two axes of the Bloch sphere correspond to the eigenbases of $X$, \[\left\{\frac{\ket{0}+\ket{1}}{\sqrt{2}}, \frac{\ket{0}-\ket{1}}{\sqrt{2}} \right\}\] and of $Y$, \[ \left\{ \frac{\ket{0}+i\ket{1}}{\sqrt{2}}, \frac{\ket{0}- i \ket{1}}{\sqrt{2}} \right\}.\] We will abbreviate the basis states and will use $\{\ket{+}, \ket{-}\}$ and $\{\ket{+i}, \ket{-i}\}$. Figure \ref{fig:bloch sphere 2} shows how these bases correspond to the three main axes of the Bloch sphere.


This representation can be used to describe single qubit unitary evolutions in a nice geometrical way. We will first introduce the Pauli matrices as they have a natural representation as rotations on the Bloch sphere.
\begin{df}
The Pauli operators are represented by the following matrices in the computational basis:
\bess
\sigma_x & = & \ket{+}\bra{+} - \ket{-}\bra{-} = \bv 0 & 1 \\ 1 & 0 \ev, \\
\sigma_y & = & \ket{+i}\bra{+i} - \ket{-i}\bra{-i} = \bv 0 & -i \\ i & 0 \ev, \\
\sigma_z & = & \ket{0}\bra{0} - \ket{1}\bra{1} = \bv 1 & 0 \\ 0 & -1 \ev. \\
\eess
\end{df}

\begin{figure}[htbp]
	\centering
	 	\psfrag{x}{$x$}
	 	\psfrag{y}{$y$}
	 	\psfrag{z}{$z$}
	 	\psfrag{ket0}{$\ket{0}$}
	 	\psfrag{ket1}{$\ket{1}$}
	 	\psfrag{ket+}{$\ket{+}$}
	 	\psfrag{ket-}{$\ket{-}$}
	 	\psfrag{ket+i}{$\ket{+i}$}
	 	\psfrag{ket-i}{$\ket{-i}$}
		\includegraphics[width=0.50\textwidth]{bloch2.eps}
	\caption[$\{\ket{0}, \ket{1}\}$, $\{\ket{+}, \ket{-}\}$ and $\{\ket{+i},\ket{-i}\}$ on the Bloch sphere.]{Bloch sphere representation of the $\{\ket{0}, \ket{1}\}$, $\{\ket{+}, \ket{-}\}$ and $\{\ket{+i},\ket{-i}\}$ bases.}
	\label{fig:bloch sphere 2}
\end{figure}

\begin{fact}
Any unitary operator $U$ on $\Hilbert_2$ can be decomposed as \[U = e^{i\alpha} e^{-i\theta \ahalf (n_x \sigma_x + n_y \sigma_y + n_z \sigma_z)}\] for real parameters $\alpha, \theta$ and a real unit vector $\hat{n} = (n_x, n_y, n_z)$. Acting on the Bloch sphere, $U$ is a rotation by $\theta$ about the $\hat{n}$ axis plus a global phase of $e^{i\alpha}$ that is not an observable property.
\end{fact}

We see that $\sigma_j$ corresponds to a rotation of $\pi$ about the $j$ axis.


\section{Density Operators}

\begin{df}
A \emph{density operator} is a positive operator $\rho \in L({\Hilbert})$ with $\tr \rho = 1$.
\end{df}

Density operators are used to describe ensembles of quantum states. If we are given a state and the promise that it is $\ket{\psi_i}$ with probability $p_i$, $i = 1, \dots, k$, we can incorporate our lack of knowledge about the state into a concise representation. This representation combines both the quantum mechanical concept of superpositions and the probability distribution over the set of states $\{\psi_i\,|\,i = 1, \dots, k\}$.

\begin{fact}
Every ensemble $\{p_i, \ket{\psi_i}\}_{i=1}^k$ has the associated density operator \[\rho = \sum_{i=1}^k p_i \ket{\psi_i}\bra{\psi_i}.\] Every density operator has an associated ensemble  $\{p_i, \ket{\psi_i}\}_{i=1}^k$.
\end{fact}

\begin{df} A \emph{pure state} is a single quantum state that is known exactly. The density operator of a pure state is of the form $\rho = \ket{\psi}\bra{\psi}$. A \emph{mixed state} is a state that is not pure.
\end{df}

\begin{fact} A state $\rho$ is pure if and only if $\tr (\rho^2) = 1$.
\end{fact}

We will state a useful fact that we will make use of later on.
\begin{fact}
The Pauli operators together with the identity form an orthonormal basis $\{\Id, \sigma_x, \sigma_y, \sigma_z\}$ for the space of linear operators $L(\Hilbert_2)$ on a $2$-dimensional Hilbert space.
\end{fact}

We can now reformulate the postulates of quantum mechanics in terms of density operators. We will mostly make use of this alternate notation in the remainder of this thesis. The formulation of the postulates has been taken from \cite{NielsenChuang2000}.

\paragraph{Postulate 1} To any isolated physical system is associated a Hilbert space, the \emph{state space} of the system. The system is completely described by its density operator acting on the state space. If a quantum system is in state $\rho_i$ with probability $p_i$, the density operator for the system is $\sum_i p_i \rho_i$.

\paragraph{Postulate 2} The evolution of a \emph{closed} quantum system from time $t_1$ to time $t_2$ is described by a unitary transformation $U$ that only depends on the times $t_1$ and $t_2$: \[\rho(t_2) = U \rho(t_1) U\dag\]

\paragraph{Postulate 3} A quantum measurement is described by a set of \emph{measurement operators} $\{M_m\}$, where $M_m$ is a measurement operator acting on the state space of the system. The index $m$ denotes the measurement outcome. The measurement operators satisfy the completeness relation \[\sum_m M_m\dag M_m = \Id.\] The probability of observing $m$ on a quantum system in state $\rho$ is \[p(m) = \tr (M_m\dag M_m \rho)\] and the state of the system after the measurement is \[\frac{M_m \rho M_m\dag}{p(m)}.\]

\paragraph{Postulate 4} The state space of a composite system is the tensor product of the state spaces of the component systems. If each component system $i$, $i = 1, \dots, n$ is prepared in the state $\rho_i$, then the joint state of the composite system is \[\bigotimes_{i=1}^n \rho_i.\]

Density operators are especially useful if we want to disregard some parts of a quantum system. There is an operation which is somehow inverse to the tensor product operation in the following way. Imagine we have a quantum system comprised of two subsystems, $A$ and $B$, with state spaces $\Hilbert_A$ and $\Hilbert_B$. The joint system has the state space $\Hilbert_A \otimes \Hilbert_B$. Let the system start in a state $\rho_0$, which we will write as $\rho_0^{AB}$ to denote that this is a state of the joint system. We let the joint system evolve to a state $\rho^{AB}$, but at some point we choose to ignore part $B$. If the joint state of the system is not a product state, i.e.\@ there are no density operators $\rho^A$ and $\rho^B$ such that $\rho^{AB} = \rho^A \otimes \rho^B$, we cannot just ignore system $B$. We must assume that $B$ will be modified later on behind our control, potentially being observed by an arbitrary measurement. It turns out that we can express our uncertainty about the future of system $B$ in a probability distribution over possible states of system $A$. The density operator notation allows us to end up with one density operator for system $A$ that covers both the state of system $A$ and our lack of knowledge about the future of system $B$. 
\begin{df} The \emph{reduced density operator} for system $A$ is defined as \[\rho^A = \tr_B \rho^{AB}.\] The \emph{partial trace} is defined for any product state $\sigma^A \otimes \sigma^B$ as \[\tr_B \sigma^A \otimes \sigma^B = \sigma^A \tr \sigma^B\] and extended to general density operators on $AB$ by linearity.
\end{df}

\section{Topology and Group Theory}

We assume basic familiarity with group theory and provide this section as a reference. We refer the interested reader to \cite{DummitFoote1991} and \cite{Willard1970,Munkres1975} for a more in-depth coverage. Most of the definitions were taken from \cite{Rudin1967}. We will first introduce the basic notions of topology and group theory and merge both of them to define topological groups later.

\subsection{Topology}

\begin{df} A \emph{topology} $\tau$ is a family of subsets of a set $S$ if
\begin{itemize}
	\item $S \in \tau$ and $\emptyset \in \tau$ and
	\item $\tau$ is closed under finite intersections and arbitrary unions.
\end{itemize}
A set $S$ with a topology $\tau$ is a \emph{topological space}, but most often $\tau$ is assumed from the context and $S$ itself is called the topological space. The elements $A \in \tau$ are defined as \emph{open sets}, their complements in $S$ are \emph{closed sets}. The elements of $S$ are sometimes referred to as \emph{points} in $S$.

The smallest closed set containing $A \subseteq S$ is the \emph{closure} of $A$, written as $\bar{A}$. The largest open set contained in $A$ is the \emph{interior} of $A$, denoted $\mathring{A}$. An \emph{interior point} of $A$ is an element $p \in \mathring{A}$. If $p$ is an interior point of $A$, then $A$ is a \emph{neighbourhood} of $p$.
\end{df}

\begin{df} Let $\tau$ by a topology on $S$, and let $T \subseteq S$ be a subset of $S$. Then $T$ becomes a topological space with topology $\tau' = \{X \cap T\,|\, X \in \tau\}$. $\tau'$ is called the \emph{subspace topology} induced by $T$.
\end{df}

\begin{df}
A topological space $S$ is called \emph{Hausdorff} if for every pair of distinct points $p_1, p_2 \in S$, there are disjoint neighbourhoods $N_1$ and $N_2$ of $p_1$ and $p_2$, respectively.
\end{df}

\begin{df}
A set $A \subseteq S$ is called \emph{compact} if each family of open sets whose union contains $A$ has a finite subfamily whose union contains $A$.
\end{df}

\begin{fact} Every closed subset of a compact space is compact. Every compact subset of a Hausdorff space is closed.
\end{fact}

\begin{df}
A function $f: X \rightarrow Y$ from a topological space $X$ to a topological space $Y$ is \emph{continuous} if $f^{-1}(E) = \{p \in X\,|\, f(p) \in E\}$ is open in $X$ for every open set $E \subseteq Y$. If $f(E)$ is open in $Y$ for every open set $E$ in $X$, then $f$ is called an \emph{open map}. If $f$ is one-to-one, $f(X) = Y$, and both $f$ and $f^{-1}$ are continuous, then $f$ is called a \emph{homeomorphism} of $X$ onto $Y$.
\end{df}

\begin{fact}
Let $X$ and $Y$ be topological spaces. If $K \subseteq X$ is compact and $f$ is continuous, then $f(K)$ is compact. 
\end{fact}

\begin{df}
Let $S$ be a topological space. Denote by $C(S)$ the set of all bounded continuous complex-valued functions on $S$. The \emph{support} $\supp f$ of a complex function $f$ in $S$ is the closure of $\{p \in S\,|\, f(p) \neq 0\}$. The set of all functions $f \in C(S)$ with compact support is denoted by $C_c(S)$. 

Let $f \in C(S)$. If for any $\epsilon > 0$, there is a compact set $K$ in $S$ such that $|f(p)| < \epsilon$ holds for all $p \in S \backslash K$, then $f$ \emph{vanishes at infinity}. The set of all $f \in C(S)$ that vanish at infinity is denoted $C_0(S)$.
\end{df}

\begin{fact} Let $S$ be a compact space. Then $C(S) = C_c(S) = C_0(S)$.
\end{fact}


\begin{df}
Let $S_1, S_2, \dots, S_n$ be topological spaces. Then $S = S_1 \times S_2 \times \dots \times S_n$ can be given the following \emph{product topology}. For a choice of indices $i_1, i_2, \dots, i_k$ and open sets $V_{i_j} \subseteq S_{i_j}$, $1 \leq j \leq k$, define $V = \{(p_1, \dots, p_n) \in S\,|\, p_{i_j} \in V_{i_j}, 1 \leq j \leq k\}$ and define a subset $E$ of $S$ as open iff it is the union of such sets $V$.
\end{df}

\begin{fact} Let $S_1, S_2, \dots, S_n$ be Hausdorff spaces. Then $S = S_1 \times \dots \times S_n$ is Hausdorff as well. If the $S_i$ are compact, then $S$ is compact.
\end{fact}

\subsection{Topological Groups}

\begin{df} A group is a pair $(G, \star)$ of a set $G$ and a binary operation $\star: G \times G \rightarrow G$ such that
\begin{itemize}
	\item for all $a, b,c \in G: a \star (b \star c) = (a \star b) \star c$ (associativity),
	\item there is an element $e \in G$ such that $a \star e = e \star a = a$ for all $a \in G$ (identity element), and
	\item for all $a \in G$, there is an element $a^{-1} \in G$ such that $a \star a^{-1} = a^{-1} \star a = e$ (inverse element).
\end{itemize}
We will usually omit the operation $\star$ and will write $ab$ to denote $a \star b$.
\end{df}

\begin{df} The \emph{left translate} of a subset $S \subseteq G$ by an element $a \in G$ is the set $a \star S = aS = \{a \star s\,|\, s \in S\}$. The \emph{right translate} of $S$ by $a$ is $Sa = \{s \star a\,|\, s \in S\}$.
\end{df}

\begin{df}
A \emph{homomorphism} $\phi: G \rightarrow H$ from a group $(G, \circ)$ to a group $(H, \star)$ is a mapping that satisfies \[\phi(x \circ y) = \phi(x) \star \phi(y)\] for all $x, y \in G$.
\end{df}

\begin{fact} The set of all complex numbers of absolute value $1$ forms a group under multiplication. With the usual topology taken from the complex numbers it forms the compact group $\T$.
\end{fact}

\begin{df} A character of a group $G$ is a homomorphism $\chi: G \mapsto \T$ into the multiplicative group of complex numbers $\alpha \in \C$ such that $|\alpha|=1$. We will call a character trivial if $\chi(g) = 1$ for all $g \in G$.
\end{df}

\bfact Let $\chi$ be a character of a finite group $G$. Then $\chi(g)^{|G|} = \chi(1_G) = 1$. Thus the values of $\chi$ are $|G|$-th roots of unities.
\efact

\begin{df}
A \emph{topological group} is a group $G$ that is a topological Hausdorff space with a topology $\tau$ such that the map $(x, y) \mapsto x y^{-1}: G \times G \rightarrow G$ is continuous. If the whole group $G$ is compact, we will call it a compact group.
\end{df}

In the following, let $G$ be a topological group. 

\begin{fact}
The translation map $t_x(y) = x y$ and the inversion $x \mapsto x^{-1}$ are homeomorphisms of $G$ onto itself. If $A$ is an open set of $G$ and $B \subseteq G$, then $AB$ is open. If $A$ and $B$ are compact, $AB$ is compact.
\end{fact}

\begin{df} Let $f$ be a complex-valued function on $G$. Denote by $\widetilde{f}$ the function \[\widetilde{f}(g) = \overline{f(g^{-1})}\] for all $g \in G$.
\end{df}

\begin{fact}
The set of all invertible linear operators on an $n$-dimensional Hilbert space $\Hilbert$ forms a group under multiplication, the \emph{general linear group} $GL(n)$. It becomes a topological space with the topology of element-wise convergence in $\C$ of the coordinate functions.  The subset of all unitary operators forms a group under multiplication. Given the subspace topology, it becomes as compact group which we will denote by $U(n)$. 
\end{fact}

\section{Haar Measure and Integration on Compact Groups}


In this section we will present the notion of an integral over a compact group $G$. We will introduce the basic concepts of measure theory first and show what is typically understood as an integral over $G$. We refer the reader to and make use of the notation given in \cite{Rudin1967} and \cite{Edwards1972} for a concise presentation of measure and integration theory for compact and locally compact abelian groups. See \cite{Rudin1973} for an introduction to functional analysis.

For the introduction to measure theory, let $X$ be a compact Hausdorff space. Let $\Borel$ be the smallest family of subsets of $X$ such that 
\begin{itemize}
	\item $\Borel$ contains all closed subsets of $X$,
	\item it is closed under finite unions, and
	\item it is closed under complementation.
\end{itemize}
The elements of $\Borel$ are called \emph{the Borel sets} of $X$.

\begin{df}
A \emph{measure} on $X$ is a set function $\mu: \Borel \rightarrow \C$ such that
\begin{enumerate}
	\item $\mu$ is \emph{countably additive}, i.e. \[\mu\left( \bigcup_i E_i \right) = \sum_i \mu(E_i)\] for a countable family of pairwise disjoint Borel sets $E_i \in \Borel$, and
	\item $\mu(E)$ is finite for all $E \in \Borel$.
\end{enumerate}
\end{df}

\begin{df} Let $\mu$ be a measure on $X$. $\mu$ is positive if $\mu$ is real-valued and $\mu(E) \geq 0$ for all Borel sets $E$.
\end{df}

\begin{df} A measure $\mu$ is a \emph{probability measure} if $\mu$ is positive and $\mu(G) = 1$.
\end{df}

\begin{df} Let $\mu$ be a measure on $X$. We define the \emph{total variation} of $\mu$ by \[|\mu|(E) = \sup \sum_i |\mu(E_i)|\] where the supremum is taken over all finite collections of disjoint Borel sets $E_i$ whose union is $E$. The total variation measures the largest variation of $\mu$ over all possible subdivisions of a Borel set $E$ and is used to define a norm on an arbitrary complex-valued measure. 
\end{df}

\begin{fact} For every measure $\mu$, $|\mu|$ is a measure as well. If $\mu$ is a positive measure, then $\mu = |\mu|$.
\end{fact}

\begin{df} A measure $\mu$ on $X$ is called \emph{regular} if \[|\mu|(E) = \sup |\mu|(K) = \inf |\mu|(V)\] where $K$ ranges over all compact subsets of $E$, and $V$ ranges over all open supersets of $E$. Let $\|\mu\| = |\mu|(X)$. Define \[M(X) = \{\mu\,|\, \mu \text{ measure on } X, \|\mu\| \text{ finite }\}.\]
\end{df}

Let $G$ denote a compact group.

\begin{df} The \emph{left translation operator} for $a \in G$ is a function $L_a: C(G) \rightarrow C(G)$ that maps $f$ to its left translate \[L_a f: x \mapsto f(a^{-1}x).\] Analogously, the \emph{right translation operator} $R_a$ is given by \[R_a f: x \mapsto f(x a^{-1}).\]
\end{df}

\begin{df}
A measure $\mu$ is \emph{left translationally invariant} if $\mu(aS) = \mu(S)$ for all $S \in \Borel$, $a \in G$. $\mu$ is \emph{right translationally invariant} if $\mu(Sa) = \mu(S)$ for all $S \in \Borel$, $a \in G$.
\end{df}

\begin{fact}[Existence of the Haar Measure]
There is a unique left and right translationally invariant measure on $G$ such that $m(G) = 1$. This measure $m$ is called the \emph{Haar Measure} on $G$.
\end{fact}

Using the theory of \emph{Lebesgue integration}, it is possible to define the notion of integrating over the compact group $G$ with respect to some measure $\mu$. We also assume the definition of an integrable function. See \cite{Rudin1973} for a detailed introduction to Lebesgue integration.

\begin{df}
Let $\mu$ be a measure on $G$ and $f$ an integrable complex-valued function on $G$. Denote the set of all integrable functions on $G$ as $\Integrable(G)$. We denote the \emph{Lebesgue integral} as \[\int_G f d\mu\] or \[\int_G f(g) d\mu(g)\] if the variable of integration is not clear from the context. If $\mu$ is the Haar measure $m$ and the group $G$ is clear from the context, we will also write $\int f(g)dg$. Note that every $f \in C(G)$ is integrable.
\end{df}

\begin{df} A measure $\mu \in M(G)$ is called \emph{discrete} if $\mu(G) = \mu(H)$ for some countable subset $H$ of $G$. We call $\mu$ \emph{continuous} if $\mu(E) = 0$ for every countable set $E$. $\mu \in M(G)$ is \emph{absolutely continuous} if $\mu(E) = 0$ whenever $m(E) = 0$ for all Borel sets $E$.
\end{df}

\section{Fubini-Study Measure}


It will be important to integrate over the set of all pure quantum states. We have seen that the set of all pure quantum states of a single qubit system can be identified by a real three-dimensional unit sphere. In that case, integration over the set of all pure states is equivalent to integration over the real unit sphere. Although there is no clear geometrical picture of  the state space of multi-qubit systems, we can still define an invariant measure and thus integration over the set of pure states of an $N$ qubit system.

\begin{fact} \label{def:fubini-study} There is a unitarily invariant measure on the set of all pure quantum states of a Hilbert space $\Hilbert$. This measure is typically referred to as the \emph{Fubini-Study} measure and written as \[ \int_{\text{F-S}} f(\ket{\psi}) d\ket{\psi}.\]
\end{fact}

Note that the Fubini-Study measure is also the unitarily invariant uniform measure on $CS^{d-1}$. See \cite[Ch.~11]{VilenkinKlimyk2} for a more rigorous introduction of the invariant measure on $CS^{d-1}$, which is denoted as $P_C^{d-1}$ in there.

\section{Representation Theory}
\label{sec:rep theory}

We will introduce some of the fundamental concepts of representation theory in this section. We assume some basic familiarity with the topic and refer to \cite{Edwards1972} for a brief introduction to representation theory. \cite{Boerner1967,Boerner1970} provides a complete but lengthy approach including proofs for all results. \cite{FultonHarris1991} presents a more modern approach to representation theory and especially the representation theory of the general linear group $GL(n)$ and its subgroups, especially $U(n)$. \cite{Hill2000} derives the representation for quite a number of elementary groups.

Let $G$ be a compact group.

\begin{df}
A \emph{representation} of $G$ is a homomorphism \[U: G \mapsto GL(n)\] of $G$ into the general linear group of invertible linear operators on an $n$-dimensional complex vector space $\Hilbert_U$. $\Hilbert_U$ is the \emph{representation space} of $U$, and $\dim \Hilbert_U$ is called the \emph{dimension} of the representation $U$. 

\begin{df} Let $V$ be a representation of $G$. The action of $G$ on $\Hilbert_V$ is defined as \[g\v{v} = V(g) \v{v}.\]
\end{df}

We will require $\Hilbert_U$ to be a finite-dimensional Hilbert space with basis $\{\ket{\psi_1}, \ket{\psi_2}, \dots, \ket{\psi_n}\}$ and $U$ to map to the group of unitary operators on $\Hilbert_U$. Furthermore, we will equip $\Hilbert_U$ with its usual topology as a complex Euclidean space. We require that the coordinate functions \[g \mapsto \bra{\psi_i} U(g) \ket{\psi_j}\] are continuous for all $1 \leq i, j \leq n$.
\end{df}

Our restrictive definition is justified by the fact that all continuous representations are unitarily equivalent and that every finite-dimensional measureable representation is continuous \cite{Edwards1972}.

\begin{df} A representation $U$ is called \emph{irreducible} if there is are no subspaces of $\Hilbert_U$ other than the trivial ones, $\{\v{0}\}$ and $\Hilbert_U$, which are invariant under $U(g)$ for all $g \in G$. Otherwise, the representation is called \emph{reducible}.
\end{df} 

\begin{df} Let $V$, $W$ be representations of $G$. A $G$-homomorphism from $\Hilbert_V$ to $\Hilbert_W$ is a linear map that respects the group action, i.e. \[\forall g \in G, \v{v} \in \Hilbert_V: \varphi(g \v{v}) = g(\varphi(\v{v})).\] Writing the group action explicitly, this becomes \[\forall g \in G, \v{v} \in \Hilbert_V: \varphi(V(g) \v{v}) = W(g) \varphi(\v{v}).\]
\end{df}

\begin{fact}[Schur's lemma] \label{fact:schurs lemma} Let $V$ and $W$ be irreducible representations of a group $G$ and let $\varphi: \Hilbert_V \rightarrow \Hilbert_W$ be a $G$-homomorphism. 
\begin{enumerate}
	\item Then $\varphi = 0$ or $\varphi = \lambda \Id$ for some $\lambda \in \C$.
	\item If $\Hilbert_V = \Hilbert_W$, $\varphi$ is an isomorphism.
\end{enumerate}
\end{fact}

\begin{fact} Every reducible representation $U$ on $\Hilbert_U$ can be decomposed into a finite direct sum of irreducible representations acting on invariant subspaces of $\Hilbert_U$: \[U(g) = U_1(g) \oplus U_2(g) \oplus \dots \oplus U_k(g)\] where $U_i$ is an irreducible representation of $G$ and $U_i(g)$ is a unitary operator on a subspace $\Hilbert_i$ such that $\bigoplus_{i=1}^k \Hilbert_i = \Hilbert_U$.
\end{fact}

\begin{df} Two representations $U$ and $V$ of $G$ are equivalent if there is an isomorphism $A$ from $\Hilbert_U$ onto $\Hilbert_V$ such that \[A U(g) = V(g) A\] for all $g \in G$. If $A$ is unitary, then $U$ and $V$ are called unitarily equivalent.
\end{df}

Let $U$ be a representation of $G$.

\begin{df} We will call the matrix elements $U_{i,j}(g)$ the \emph{coordinate functions} of the representation $U$.
\end{df}

\begin{df}
The \emph{character} of a representation $U$ of $G$ is a function $\chi_U \in C(G)$ defined as \[\chi_U(g) = \overline{\tr U(g)}.\] We will usually write $\chi$ if the representation used is clear from the context.
\end{df}

\begin{fact}
Every character $\chi$ of $G$ is continuous.
\end{fact}

\begin{fact}
Let $\chi$ be a character of $G$. Then \[\chi(g^{-1}) = \overline{\chi(g)}\] for all $g \in G$.
\end{fact}

As an example, we will consider the representations of the compact group $U(d)$ for some $d \in N$. We refer to \cite[Ch.~6]{VilenkinKlimyk1} for a treatment of $SU(2)$ and $GL(2)$. See \cite[Ch.~11]{VilenkinKlimyk2} for various analytical expressions of the irreducible representations of $U(d)$ for general $d$. As the actual matrices of the irreducible represenations of $U(d)$ are rather complicated, we will skip them here and present only the necessary formulas for the dimensions of its irreducible representations. 

\bfact \label{fact:dim irreps u(d)}
The irreducible representations $D^s$ of $U(d)$ are labelled by two integers \cite{VilenkinKlimyk2} $s = (k,l)$, $k, l \in \N$ and their dimension is given by \[d_{(k,l)} = \f{k + l + d - 1}{d-1} \binom{k + d - 2}{k}\binom{l + d - 2}{l}.\]
\efact

\section{Fourier Analysis}
\label{sec:fourier analysis}


The material in this section follows the notation set in \cite{Edwards1972}. We refer to \cite{Rudin1973} for the details of Banach space theory. See \cite{Barut1980} for the approximation of Fourier series. However, we note slight differences in the placement of complex conjugates and global dimensionality factors between \cite{Edwards1972} and \cite{Barut1980}. In this section, we assume $G$ a compact group and $m$ the Haar measure on $G$.

\subsection{Banach Spaces}

The concept of a metric is the generalization of the concept of distance in a Euclidean space. It is generalized to arbitrary sets in the following way.

\begin{df} A \emph{metric} on a set $X$ is a function $d: X \times X \rightarrow \R$ such that
\begin{enumerate}
	\item $d(x,y) \geq 0$ for all $x, y \in X$ (non-negativity),
	\item $d(x,y) = 0$ if and only if $x = y$ (definiteness),
	\item $d(x,y) = d(y,x)$ for all $x, y \in X$ (symmetry), and
	\item $d(x,z) \leq d(x,y) + d(y,z)$ for all $x,y,z \in X$ (triangle inequality).
\end{enumerate}
The pair $(X, d)$ is called a \emph{metric space}.
\end{df}

\begin{fact} A normed vector space $(V, \|\cdot\|)$ is a metric space with respect to the metric $d(\v{u},\v{v}) = \|\v{u} - \v{v}\|$. $V$ can be given the usual topology induced by its norm to turn $V$ into a \emph{topological vector space}.
\end{fact}

\begin{df} A \emph{Banach space} is a complete normed complex vector space.
\end{df}

\begin{fact} Every Hilbert space is a Banach space.
\end{fact}

\begin{fact} Let $p$ be a positive real number. Then the function $\|\cdot\|_p: \Integrable(G) \mapsto \R$ defined as \[\|f\|_p = \sqrt[p]{\int_G |f(g)|^p dm(g)}\] is a norm on $\Integrable(G)$.
\end{fact}

\begin{df} A function $f \in C(G)$ is \emph{zero almost everywhere} if $m\left(\{g \in G\,|\, f(g) = 0\} \right) = 0$. Two functions $f, g \in C(G)$ are \emph{equal almost everywhere} if $f-g$ is zero almost everywhere.
\end{df}

\begin{fact} $C(G)$ is a Banach space with addition of functions $f, g$ defined in the usual way as $(f+g)(x) = f(x) + g(x)$. $L^p(G) = \{f \in \Integrable(G)\,|\,\|f\|_p \text{ is finite} \}$, with two functions identified if they are equal almost everywhere, is a Banach space.  

$L^2(G)$ is a Hilbert space with inner product \[(f, g) = \int_G f(x) \overline{g(x)} dm(x),\] where $\overline{g}$ denotes the usual complex conjugate.

For $1 \leq q \leq p$, $L^p(G) \subseteq L^q(G)$. Furthermore, $C(G) \subseteq L_p(G)$ for any $p \geq 1$.
\end{fact}

It is important to notice that functions in $L^p(G)$ are not defined point-wise, as $m(\{x\}) = 0$ for all $x$. The definition only makes sense if we are interested in the way they are integrated against certain measures.

\subsection{Fourier Analysis}

\begin{df} Let $\hat{G}$ be the set of all representations of $G$ where equivalent representations are identified and one representative of each equivalence class is chosen for $\hat{G}$. Thus $\hat{G} = \{D^s\}$ is the set of pairwise inequivalent, irreducible unitary representations of $G$. We will order the irreducible representations $D^s$ by their increasing dimensionality $d_s$. Note that the trivial representation $D^s(g) = 1$ for all $g \in G$ has dimension $d_0 = 1$.
\end{df}

\begin{df} Let $f \in L^1(G)$ and $U \in \hat{G}$. The \emph{Fourier transform} of $f$ is defined for each representation $D^s$ as \[\hat{f}(D^s) = \int_G f(g) D^s(g) dg.\] Note that $\hat{f}(D^s) \in L({\Hilbert_U})$. As $\hat{f}(D^s)$ is a $d_s \times d_s$ complex matrix, it will not be useful to treat $\hat{f}$ as a function.
\end{df}

\begin{df} The \emph{convolution} of two functions $f, g \in \Integrable(G)$ is defined as \[(f \star g)(x) = \int f(y) g(y^{-1} x) dm(y).\] Convolution is an associative operation.
\end{df}

\begin{fact} Let $f, g \in \Integrable(G)$, $D^s \in \hat{G}$. Then \[\widehat{f \star g}(D^s) = \hat{f}(D^s) \hat{g}(D^s).\]
\end{fact}

We can also define the Fourier transform for measures, which represent a more general class of functions than those in $L^1(G)$. This transformation is typically referred to as the \emph{Fourier-Stieltjes Transform}. We will first introduce a correspondence between measures and functions in $L^1$.

\begin{fact} If $f \in L^1(G)$, then the measure $\mu(E) = \int_E f dm$ is in $M(G)$ and absolutely continuous. For every absolutely continuous measure $\mu \in M(G)$, there is a function $f \in L^1(G)$ such that $\mu(E) = \int_E f dm$ for all Borel sets $E$. Furthermore, $\|\mu\| = \|f\|_1$.
\end{fact}

\begin{df} The \emph{Fourier-Stieltjes transform} of a measure $\mu \in M(G)$ is given by \[\hat{\mu}(D^s) = \int_G D^s(g) d \mu(g).\]
\end{df}

\begin{df} Associate with every Borel set $E$ of $G$ the set $E^2 = \{(x, y) \in G\times G\,|\, xy \in E\}$. Then $E^2$ is a Borel set of $G^2$. The \emph{convolution} of two measures $\mu, \lambda \in M(G)$ is defined as \[(\mu \star \lambda)(E) = (\mu \times \lambda)(E^2),\] where $\mu \times \lambda$ is the product measure on the product space $G^2$.
\end{df}

\begin{fact} For $\mu, \lambda \in M(G)$, we have $\mu \star \lambda \in M(G)$. Convolution is associative and commutative. Finally, $\| \mu \star \lambda\| \leq \|\mu\| \|\lambda\|$.
\end{fact}

\begin{fact} Let $\mu, \lambda \in M(G)$, $D^s \in \hat{G}$. Then \[\widehat{\mu \star \lambda}(D^s) = \hat{\mu}(D^s) \hat{\lambda}(D^s).\]
\end{fact}


\begin{fact} \label{fact:ortho rel} The normalized coordinate functions $\sqrt{d_s} D^s_{i,j}: g \mapsto \sqrt{d_s} D^s_{i,j}(g)$ form a complete orthonormal set for the Hilbert space $L^2(G)$. The \emph{orthogonality relations} read \[\int_G D^s_{i,j}(g) \overline{D^{s'}_{m,n}(g)} dg = \frac{1}{d_s} \delta_{s,s'} \delta_{i,m} \delta_{j,n}.\] It follows that \[ \int_G \tr \left( D^s(g) (D^s(g))\dag \right) dg = \delta_{s,s'}.\] 
\end{fact}

\begin{fact} \label{fact:uniqueness thm} Let $f \in L^p(G)$ for $p \geq 1$ or $p = \infty$. If $\hat{f}(D^s) = 0$ for all $D^s \in \hat{G}$, then $f(g) = 0$ for almost all $g$. As a corollary, let $f, g \in L^p(G)$. If $\hat{f}(D^s) = \hat{g}(D^s)$ for all $D^s \in \hat{G}$, then $f = g$ almost everywhere.
\end{fact}

\begin{fact} \label{fact:parseval} The \emph{Parseval formula} is the following integral identity for $f \in L^2(G)$: \[\|f\|_2 = \sqrt{\int_G |f(g)|^2 dg} = \sqrt{\sum_{s \in \hat{G}} d_s \tr \hat{f}(D^s) \hat{f}(D^s)\dag}.\]
\end{fact}

\begin{fact} The \emph{Peter-Weyl-Theorem} states as a direct consequence that \[f(g) = \sum_{s \in \hat{G}} d_s \tr \hat{f}(D^s) D^s(g)\dag\] almost everywhere, the limit being the strong limit in $L^2(G)$ of its partial sums over finite $P \subseteq \hat{G}$.
\end{fact}

\begin{fact} \label{fact:riemann-lebesgue} The \emph{Riemann-Lebesgue Lemma} states that for any $f \in L^1(G)$, \[ \lim_{s \rightarrow \infty} \|\hat{f}(D^s)\| = 0.\]
\end{fact}

\begin{df} A function $\phi \in L^1(G)$ is called positive-definite if for all $f \in C(G)$, \[(f \star \phi \star \widetilde{f})(e) = \int_G \int_G \phi(h^{-1}g) \overline{f(g)} f(h) dg dh \geq 0.\]
\end{df}

\begin{df} $P(G)$ is the set of all continuous positive-definite functions on $G$.
\end{df}

\begin{fact} A function $\phi \in L^1(G)$ is positive-definite if and only if $\hat{\phi}(D^s)$ is positive self-adjoint for all $s$.
\end{fact}

\begin{df} \label{def:nice pd} A ``nice'' positive-definite function is a positive-definite function $\phi$ on $G$ such that
\begin{itemize}
\item $\phi$ is continuous (i.e.\@ $\phi \in P(G)$) or 
\item there is a number $m_{\phi}$ and a neighbourhood $N_{\phi}$ of $e \in G$ such that \[(f \star \phi \star \widetilde{f})(e) \leq m_{\phi} \|f\|_1^2\] for all $f \in C(G)$ whose support is contained in $N_{\phi}$.
\end{itemize}
\end{df}

\begin{fact}[\cite{Barut1980,Edwards1972}] Another version of the \emph{Peter-Weyl-Theorem} describes that the Fourier series of a nice positive-definite function $f$ converges uniformly for all (almost all if $f$ is not continuous) $g \in G$, \[f(g) = \lim_{S \rightarrow \infty} \sum_{s \leq S} d_s \tr \hat{f}(D^s) D^s(g)\dag\] where we ordered the irreducible representations $D^s$ by their increasing dimensionality $d_s$.
\end{fact}

\begin{fact}[\cite{Barut1980,Edwards1972}] \label{fact:peter weyl approx} A consequence is the \emph{Peter-Weyl Approximation Theorem}. For all nice positive-definite $f$ and all $\epsilon > 0$, there is a number $N_{\epsilon}$ such that 
\[\left| f(g) - \sum_{s=0}^{N_{\epsilon}} d_s \tr \hat{f}(D^s) D^s(g)\dag \right| < \epsilon\] for all (almost all if $f$ is not continuous) $g \in G$.
\end{fact}

\section{Fields and Rings}

Although we have used fields and assumed basic familiarity with them as mathematical objects, it it necessary to give the exact definition. It will be important to distinguish fields from rings to understand different constructions involved in this thesis. We refer to \cite{LidlNiederreiter1994} and \cite{McDonald1974} for a general treatment of finite fields and finite rings. \cite{Wan1997} and \cite{Wan2003} deal with Galois fields and Galois rings in particular.

\begin{df} A \emph{ring} $(R, +, \star)$ is a set $R$ together with two binary operations $+$ (addition) and $\star$ (multiplication) such that
\begin{itemize}
	\item $(R, +)$ is an abelian group,
	\item $a \star (b \star c) = (a \star b) \star c)$ for all $a, b, c \in R$ (multiplicative associativity), and
	\item $a \star (b + c)= (a \star b) + (a \star c)$ and $(a + b) \star c) = (a \star c) + (b \star c)$ for all $a, b, c \in R$ (distributivity).
\end{itemize}
We will typically denote the ring operations as addition and multiplication and we will understand that $a b$ means $a \star b$.
\end{df}

\begin{df} A \emph{field} $(\F, +, \star)$ is a set $\F$ together with two binary operation $+$ and $\star$ such that
\begin{itemize}
	\item $(\F, +, \star)$ is a ring where we denote the additive identity with $0$,
	\item $(\F \backslash \{0\}, \star) = \F^*$ is an abelian group with multiplicative identity $1 \neq 0$, and
	\item $ab = 0$ implies $a = 0$ or $b = 0$ for all $a, b \in \F$.
\end{itemize}
If $\F$ is finite, we will call it a \emph{finite field} or \emph{Galois field}.
\end{df}

\begin{df} A \emph{subring} of a ring is a subset $S \subseteq R$ such that $S$ is closed under $+$ and $\star$, and forms a ring with respect to these operations.
\end{df}

\begin{df} An \emph{ideal} of a ring $R$ is a subset $J \subseteq R$ such that $J$ is a subring of $R$ and $ar, ra \in J$ for all $a \in J, r \in R$.
\end{df}

\bfact Let $R$ be a commutative ring with multiplicative identity $1$. An ideal $J$ is \emph{principal} if there is an $a \in R$ such that $J = (a) = \{ra\,|\, r \in R\}$. We will call $J$ \emph{generated} by $a$.
\efact

An ideal $J$ partitions a ring $R$ into disjoint cosets $[a] = a + J = \{a + j\,|\, j \in J\}$. Elements $a$ and $b$ in the same coset or \emph{residue class} of $J$ are called \emph{congruent modulo $J$} and we will write $a \equiv b \mod J$. This is equivalent to $a - b \in J$. 

\bfact The set of residue classes of a ring $R$ modulo an ideal $J$ forms a ring if we define addition and multiplication of residue classes by letting 
\begin{itemize}
	\item $(a + J) + (b + J) = (a + b)+ J$ and
	\item $(a + J)(b + J) = (ab + J)$
\end{itemize}
for any $a, b \in R$. It is called the \emph{residue class ring} and denoted by $R/J$.
\efact

\begin{df} The \emph{characteristic} of a ring $R$ is the smallest positive integer $n \in \Z$ such that $nr = 0$ for all $r \in R$. If there is no such integer $n$, we say that $R$ has characteristic $0$.
\end{df}

\bfact Let $R$ be a commutative ring with prime characteristic $p$. Then \[(a + b)^{p^n} = a^{p^n} + b^{p^n}\] for all $a, b \in R$ and all $n \in N$.
\efact

\bfact Any finite field has prime characteristic. 
\efact

\bfact For any prime power $p^k$, all finite fields with $p^k$ elements are isomorphic and we write $\F_{p^k}$ or $GF(p^k)$ to denote the finite field with $p^k$ elements. All finite fields have a prime power number of elements.
\efact

\bfact For a ring $R$, the set of polynomials \[p(X) = \sum_{i=0}^n a_i X^i\] with $a_i \in R$ and $X$ a formal variable form a ring under usual addition and multiplication of polynomials, with $0$ the zero polynomial.
\efact

\begin{df}
The ring of polynomials over $R$ is called the \emph{polynomial ring} over $R$ and is denoted by $R[X]$. 
\end{df}

\begin{df}  A polynomial $p(X) = \sum_{i=1}^n a_i X^i$ is called \emph{monic} if $a_n = 1$.
\end{df}

\subsection{Galois Fields}
\label{sec:galois fields}

For any prime $p$, a usual method to construct the finite field $\F_p$ is to take the integers modulo $p$ which forms a field with $p$ elements. We will now describe how fields with a prime power number of elements $p^m$, $m \in N$, can be constructed. 

For matters of simplicity, we will use a simpler approach than the one presented in \cite{LidlNiederreiter1994}. The approach taken here is streamlined to facilitate later constructions and ease understanding for the purpose of applications of finite fields to this thesis. Furthermore this section should provide the reader with some intuition about the structure of finite fields.

\begin{df}
A polynomial $p(X) \in \F_p[X]$ is \emph{primitive} if there are no polynomials $r(X), s(X) \in \F_p[X]$ such that $p(X) = r(X) s(X)$ and $r(X), s(X) \neq p(X)$ and $r(X), s(X) \neq 1$. Intuitively, this is similar to the definition of a prime number and will serve an analogous purpose.
\end{df}

This definition implies that a primitive polynomial $p(X)$ is irreducible, as it cannot have any root $\xi$ for it would lead to a factorization $(X - \xi) | p(X)$. We will now use a monic primitive polynomial to define $GF(p^m)$ as a residue class ring of $\F_p[X]$ which will turn out to be a field.

\begin{theorem} 
Let $h(X) \in \F_p[X]$ be a monic primitive polynomial of degree $m \in N$. Then $\F_p[X] / (h(X))$ is a finite field with $p^m$ elements. We will denote this field by $GF(p^m)$. It is sometimes called an \emph{extension field}.
\end{theorem}
\begin{proof}
$h(X)$ is a monic polynomial of degree $m$, hence the remainders of polynomials in $\F_p[X]$ after division by $h(X)$ are polynomials of degree up to $m - 1$. If we pick the lowest-degree representative for the coset in $\F_p[X] / (h(X))$, we have that \[\F_p[X] / (h(X)) \cong \{a_0 + a_1 X + \dots + a_{m-1} X^{m-1} \,|\, a_0, a_1, \dots, a_{m-1} \in \F_p\}. \] The Extended Euclidean Algorithm shows that for any $f(X) \in GF(p^m)$, there is an inverse $f^{-1}(X)$ such that $f(X) f^{-1}(X) = 1$ and that there are no zero divisors.
\end{proof}

\bfact Any extension fields over $\F_p$ with monic, primitive polynomials $h_1(X), h_2(X)$ of degree $m$ are isomorphic. That justifies the label $GF(p^m)$ that is independent of the primitive polynomial that generates the extension field, which justified to speak of \emph{the} finite or Galois field with $p^m$ elements.
\efact

From the proof of the preceding theorem it is apparent that we can identify polynomials in $GF(p^m)$ as vectors with $m$ components. Usually, elements of the extension as well as elements of the base field $\F_p$ are denoted by latin letters. To avoid confusion, we will use greek letters for the extension field and latin letters for the base field in case we will have to mix both. We will use \[\v{a} = \bv a_0 \\ a_1 \\ \hdots \\ a_{m-1} \ev\] to denote the vector associated to $a$ when we consider the vector space $\F_p^m$.

\bfact $GF(p^m)$ is an $m$-dimensional vector space over $\F_p$, denoted by $\F_p^m$. We can also equip $\F_p^m$ with an inner product to turn it into an inner product space. We will conveniently use the standard inner product \[(\v{\alpha}, \v{\beta}) = \sum_{i=0}^{m-1} \alpha_i \beta_i.\]
\efact

As a basis, we can pick the polynomials $\{1, X, X^2, \dots, X^{m-1}\}$. Then, the vector representation of a polynomial $p(X) \in GF(p^m)$ is given by the column vector of its $m$ coefficients. It follows from the distributivity of multiplication and addition that multiplication is a linear function on $GF(p^m)$ as a vector space. 
\bfact For every $a \in GF(p^m)$, there is a matrix $M_a \in \F_p^{m \times m}$ such that $\v{ab} = M_a \v{b}$ for all $b \in GF(p^m)$.
\efact

There is a very important function from $GF(p^m)$ to $\F_p$, the trace mapping.
\begin{df} The \emph{trace} is a mapping $\tr_{GF(p^m)}: GF(p^m) \rightarrow \F_p$ such that \[\tr_{GF(p^m)} (\alpha) = \sum_{i=0}^{m-1} \alpha^{p^i}.\] If it is unambiguously clear from the context, we will write ``$\tr$'' instead of ``$\tr_{GF(p^m)}$''. Note that this function is referred to as \emph{absolute trace} as it maps to the prime field $\F_p$.
\end{df}
The trace is a very nice function with interesting properties.
\bfact \label{fact:field trace} The trace is a linear functional on $GF(p^m)$, i.e.\@
\begin{itemize}
	\item $\tr(\alpha + \beta) = \tr(\alpha) + \tr(\beta)$ for all $\alpha, \beta \in GF(p^m)$ and
	\item $\tr(c \alpha) = c \tr(\alpha)$ for all $\alpha \in GF(p^m), c \in \F_p$.
\end{itemize}
From the linearity of the trace it follows that there is a vector $\v{t}$ such that $\tr \alpha = (\v{t}, \v{\alpha})$. Furthermore
\begin{itemize}
	\item $\tr (a) = ma$ for all $a \in \F_p$ and
	\item $\tr (\alpha^p) = \alpha$ for all $\alpha \in GF(p^m)$.
\end{itemize}
\efact

\begin{fact} \label{fact:Gauss sums} Let $a \in GF(p^m)$. Then \[\sum_{x \in GF(p^m)} \left(e^{2 \pi i/p}\right)^{\tr ax} = \begin{cases} p^m & a = 0 \\ 0 & a \neq 0 \end{cases}. \]
\end{fact}

\subsection{Galois Rings}
\label{sec:galois rings}

The construction of Galois rings is quite similar to the construction of Galois fields. However, we will not consider the most general case of Galois rings but restrict ourselves to the case of rings over the base ring $\Z_4$. We refer to \cite{Wan1997} for a complete coverage of Galois rings over $\Z_4$ and \cite{Wan2003} for the more general case of a Galois ring over $\Z_n$ for arbitrary $n \in \N$. Note that we will write $\Z_2$ instead of $\F_2$ for easier reading. Furthermore, we will give a slightly stricter definition of a Galois ring than is usually adopted in the literature to focus on the specific results needed for later constructions.

\begin{df}
The map $\bar{}: \Z_4[X] \rightarrow Z_2[X]$ is defined for $f(X) = a_0 + a_1 X + \dots + a_n X^n$ as \[\bar{f}(X) = (a_0 \mod 2) + (a_1 \mod 2) X + \dots (a_n \mod 2) X^n.\]
\end{df}

\begin{df} We will call a polynomial $h(X) \in \Z_4[X]$ basic primitive if $\bar{h}(X)$ is a primitive polynomial in $\Z_2[X]$.
\end{df}

\begin{df} Let $h(X) \in \Z_4[X]$ be a monic, basic primitive polynomial of degree $m$. Then the residue class ring \[\Z_4[X] / (h(X)) \cong \{a_0 + a_1 X + \dots + a_{m-1} X^{m-1}\,|\, a_0, a_1, \dots, a_{m-1} \in \Z_4\}\] is called the \emph{Galois ring} and denoted by $GR(4^m)$. We say that $h(X)$ \emph{generates} the Galois Ring. We will call this polynomial representation of $GR(4^m)$ the \emph{additive representation}.
\end{df}

\bfact \label{fact:galos rings isomorphic} Any two Galois rings over $\Z_4$ with monic, basic primitive polynomials $h_1(X), h_2(X)$ of degree $m$ are isomorphic. That justifies the label $GR(4^m)$ which is independent of the primitive polynomial that generates the residue class ring.
\efact

\bfact $GR(4^m)$ has characteristic $4$.
\efact

There is a close connection between Galois fields and Galois rings that stems from the fact that the generating polynomial $h(X)$ is basic primitive and thus $\bar{h}(X)$ is primitive in $\Z_2$. Hence the Galois field with $2^m$ elements is contained in the Galois ring with $4^m$ elements.

\bfact $\overline{GR(4^m)} \cong GR(4^m) / (2) \cong GF(2^m).$
\efact

However, it is not true that the Galois ring is a trivial product of a Galois field and another simple structure. It turns out $h(X)$ ensures a richer structure that cannot easily be derived from $\bar{h}(X)$. Besides the additive representation, there is a second representation that gives more insight into the structure of $GR(4^m)$. This second representation is called the \emph{$2$-adic representation}.

\bfact The element $X \in GR(4^m)$ is of order $2^m - 1$ and is the root of a unique monic basic primitive polynomial $h(X)$ of degree $m$ that generates $GR(4^m)$.
\efact

The $2$-adic representation is facilitated by the powers of $X$.

\begin{df} \label{def:teichmueller set} The set $\Tau_m = \{0, 1, X, X^2, X^3, \dots, X^{2^m-2}\}$ is called the \emph{Teichm\"{u}ller set} of the Galois ring $GR(4^m)$.
\end{df}

Notice that $T \backslash \{0\}$ is a cyclic multiplicative group generated by $X$.

\bfact Let $\Tau_m$ be the Teichm\"{u}ller set of $GR(4^m)$. Then for any element $c \in GR(4^m)$, there are unique $a, b \in \Tau_m$ such that \[c = a + 2b.\]
\efact

Using the $2$-adic representation we can define a trace function for Galois rings.

\begin{df} The \emph{generalized Frobenius map} of $GR(4^m)$ is defined as \[f: GR(4^m) \rightarrow GR(4^m), c = a + 2b \mapsto a^2 + 2 b^2.\]
\end{df}

\bfact The generalized Frobenius map is a ring automorphism of $GR(4^m)$. The fixed elements of $f$ are the elements of $\Z_4$. Furthermore, $f$ is of order $m$.
\efact

\begin{df} The \emph{generalized trace} is a mapping $\tr_{GR(4^m)}: GR(4^m) \rightarrow \Z_4$ such that \[\tr_{GR(4^m)} (a + b) = \sum_{i=0}^{m-1} a^{2^i} + 2 b^{2^i}\] where $a + 2b \in GR(4^m)$ is the $2$-adic representation of an element of $GR(4^m)$. If it is unambiguously clear from the context, we will write ``$\tr$'' instead of ``$\tr_{GR(4^m)}$''. 
\end{df}
In analogy to the field trace, we also have the nice property that the trace is linear.
\bfact The trace is a linear functional from $GF(4^m)$ to $\Z_4$.
\efact


\bibliographystyle{alpha} 

\clearpage
\thispagestyle{empty}
\cleardoublepage

\markright{Bibliography}
\bibliography{thesis} 

\newcommand{\etalchar}[1]{$^{#1}$}
\begin{thebibliography}{AMTdW00}

\bibitem[Abr04]{Abrahamsson2004}
Bj\"{o}rn Abrahamsson.
\newblock {\em Architectures for Multiplication in Galois Rings}.
\newblock Master's thesis, University of Linköping, 2004.
\newblock http://www.ep.liu.se/exjobb/isy/2004/3549/.

\bibitem[All80]{Alltop1980}
W.~O. Alltop.
\newblock Complex sequences with low periodic correlations.
\newblock {\em IEEE Transactions on Information Theory}, 26:350--354, 1980.

\bibitem[AMTdW00]{AmbainisEtAl2000}
Andris Ambainis, Michele Mosca, Alain Tapp, and Ronald de~Wolf.
\newblock Private quantum channels.
\newblock In {\em Proceedings of the 41st Annual Symposium on Foundations of
  Computer Science (FOCS)}, page 547, Los Alamitos, CA, 2000. IEEE Press.
\newblock ArXiv.org Preprint quant-ph/0003101.

\bibitem[Arc05]{Archer2005}
Claude Archer.
\newblock There is no generalization of known formulas for mutually-unbiased
  bases.
\newblock {\em Journal of Mathematical Physics}, 46:022106, 2005.

\bibitem[Bal98]{Ballentine1998}
Leslie~E. Ballentine.
\newblock {\em Quantum Mechanics: A Modern Development}.
\newblock World Scientific, Singapore, 1998.

\bibitem[Bar02]{Barnum2002}
Howard Barnum.
\newblock Information-disturbance tradeoff in quantum measurement on the
  uniform ensemble and on the mutually unbiased bases.
\newblock {\em ArXiv.org Preprint quant-ph/0205155}, 2002.

\bibitem[BB84]{BB84}
Charles~H. Bennett and Gilles Brassard.
\newblock Quantum cryptography: Public key distribution and coin tossing.
\newblock In {\em Proceedings of IEEE International Conference on Computers
  Systems and Signal Processing, Bangalore India, December 1984}, pages
  175--179, 1984.
\newblock
  http://www.research.ibm.com/people/b/bennetc/bennettc198469790513.pdf.

\bibitem[BBF02]{BeauregardEtAl2003}
Stephane Beauregard, Gilles Brassard, and Jose~M. Fernandez.
\newblock Quantum arithmetic on galois fields.
\newblock {\em ArXiv.org Preprint quant-ph/0301163}, 2002.

\bibitem[BBRV02]{BandyopadhyayEtAl2002}
Somshubhro Bandyopadhyay, P.~Oscar Boykin, Vwani Roychowdhury, and Farrokh
  Vatan.
\newblock A new proof for the existence of mutually unbiased bases.
\newblock {\em Algorithmica}, 34(4):512--528, 2002.
\newblock ArXiv.org Preprint quant-ph/0103162.

\bibitem[BCH{\etalchar{+}}05]{BuhrmanEtAl2005}
Harry Buhrman, Matthias Christandl, Patrick Hayden, Hoi-Kwong Lo, and Stephanie
  Wehner.
\newblock On the (im)possibility of quantum string commitment.
\newblock {\em ArXiv.org Preprint quant-ph/0504078}, 2005.

\bibitem[BDSW96]{PoulinEtAl2004}
Charles~H. Bennett, David~P. DiVincenzo, John~A. Smolin, and William~K.
  Wootters.
\newblock Mixed state entanglement and quantum error correction.
\newblock {\em Physical Review~A: General Physics}, 54:3824--3851, 1996.
\newblock ArXiv.org Preprint quant-ph/9604024.

\bibitem[Ben82]{Benioff1982}
Paul Benioff.
\newblock Quantum mechanical models of turing machines that dissipate no
  energy.
\newblock {\em Physical Review Letters}, 48:1581--1585, 1982.

\bibitem[Ben98]{Benioff1998}
Paul Benioff.
\newblock Models of quantum turing machines.
\newblock {\em Fortschritte der Physik}, 46:423--442, 1998.
\newblock ArXiv.org Preprint quant-ph/9708054.

\bibitem[Boe67]{Boerner1967}
Hermann Boerner.
\newblock {\em Darstellungen von Gruppen: mit Ber\"{u}cksichtigung der
  Bed\"{u}rfnisse der modernen Physik}.
\newblock Springer-Verlag, Berlin, 1967.

\bibitem[Boe70]{Boerner1970}
Hermann Boerner.
\newblock {\em Representations of groups with special consideration for the
  needs of modern physics}.
\newblock North-Holland Publishers, Amsterdam, 1970.

\bibitem[BOS{\etalchar{+}}02]{BowdreyEtAl2002}
Mark~D. Bowdrey, Daniel K.~L. Oi, Anthony~J. Short, Konrad Banaszek, and
  Jonathan~A. Jones.
\newblock Fidelity of single qubit maps.
\newblock {\em Physics Letters~A}, 294:258--260, 2002.
\newblock arxiv quant-ph/0201106.

\bibitem[Car98]{Carlet1998}
Claude Carlet.
\newblock One-weight $\mathbb{Z}_4$-linear codes.
\newblock In Johannes Buchmann, Tom H{\o}holdt, Henning Stichtenroth, and
  Horacio Tapia-Recillas, editors, {\em Coding Theory, Cryptography and Related
  Areas}, Berlin, 1998. Springer Verlag.

\bibitem[CBK{\etalchar{+}}02]{CerfEtAl2002}
Nicolas~J. Cerf, Mohamed Bourennane, Anders Karlsson, , and Nicolas Gisin.
\newblock Security of quantum key distribution using $d$-level systems.
\newblock {\em Physical Review Letters}, 88:127902, 2002.
\newblock ArXiv.org Preprint quant-ph/0107130.

\bibitem[Cha02]{Chaturvedi2002}
S.~Chaturvedi.
\newblock Aspects of mutually unbiased bases in odd-prime-power dimensions.
\newblock {\em Physical Review~A: General Physics}, 65(044301), 2002.

\bibitem[Cha05]{Chau2005}
H.~F. Chau.
\newblock Unconditionally secure key distribution in higher dimensions by
  depolarization.
\newblock {\em IEEE Transactions on Information Theory}, 51:1451--1468, 2005.
\newblock ArXiv.org Preprint quant-ph/0405016.

\bibitem[dBCW02]{BeaudrapEtAl2002}
J.~Niel de~Beaudrap, Richard Cleve, and John Watrous.
\newblock Sharp quantum versus classical query complexity separations.
\newblock {\em Algorithmica}, 34:449--461, 2002.

\bibitem[Deu85]{Deutsch1985}
David Deutsch.
\newblock Quantum theory, the church-turing principle and the universal quantum
  computer.
\newblock {\em Proceedings of the Royal Society of London. Series A},
  400(1818):97--117, 1985.

\bibitem[DF91]{DummitFoote1991}
D.~S. Dummit and R.~M. Foote.
\newblock {\em Abstract Algebra}.
\newblock Prentice-Hall, Englewood Cliffs, NJ, 1991.

\bibitem[DLT02]{DiVincenzoLeungTerhal2001}
David~P. DiVincenzo, Debbie~W. Leung, and Barbara~M. Terhal.
\newblock Quantum data hiding.
\newblock {\em IEEE Transactions on Information Theory}, 48(3):580--599, 2002.
\newblock ArXiv.org Preprint quant-ph/0103098.

\bibitem[Dur05]{Durt2005}
Thomas Durt.
\newblock About mutually-unbiased bases in even and odd prime power dimensions.
\newblock {\em Journal of Physics~A: Mathematical and General}, 38:5267--5283,
  2005.

\bibitem[dW99]{deWolf1999}
Ronald de~Wolf.
\newblock Quantum computation and shor's factoring algorithm.
\newblock http://homepages.cwi.nl/~rdewolf/publ/qc/survey.ps, 1999.

\bibitem[EAZ05]{EmersonEtAl2005_Noise}
Joseph Emerson, Robert Alicki, and Karol \.{Z}yczkowski.
\newblock Scalable noise estimation with random unitary operators.
\newblock {\em ArXiv.org Preprint quant-ph/0503243}, 2005.

\bibitem[Edw72]{Edwards1972}
Robert~E. Edwards.
\newblock {\em Integration and Harmonic Analysis on Compact Groups}.
\newblock Cambridge University Press, Cambridge, UK, 1972.

\bibitem[ELL05]{EmersonEtAl2005_RandomUnitaries}
Joseph Emerson, Etera Livine, and Seth Lloyd.
\newblock Convergence conditions for random quantum circuits.
\newblock {\em ArXiv.org Preprint quant-ph/0503210}, 2005.

\bibitem[ES03]{ErdosSuranyi2003}
Paul Erd\H{o}s and J\'{a}nos Sur\'{a}nyi.
\newblock {\em Topics in the Theory of Numbers}.
\newblock Springer-Verla, New York, 2003.

\bibitem[fAOBR80]{Barut1980}
\fontencoding{T1}\selectfont Asim O.~Barut and Ryszard R\k{a}czka.
\newblock {\em Theory of Group Representations and Applications}.
\newblock Polish Scientific Publishers, Warszawa, 1980.

\bibitem[Fey82]{Feynman1982}
Richard~P. Feynman.
\newblock Simulating physics with computers.
\newblock {\em International Journal of Theoretical Physics}, 21:467, 1982.

\bibitem[FH91]{FultonHarris1991}
William Fulton and Joe Harris.
\newblock {\em Representation Theory: A First Course}.
\newblock Springer-Verlag, New York, 1991.

\bibitem[GKP01]{GottesmanEtAl2001}
Daniel Gottesman, Alexei Kitaev, and John Preskill.
\newblock Encoding a qubit in an oscillator.
\newblock {\em Physical Review~A: General Physics}, 64:012310, 2001.
\newblock ArXiv.org Preprint quant-ph/0008040.

\bibitem[Got97]{Gottesman1997}
Daniel Gottesman.
\newblock {\em Stabilizer Codes and Quantum Error Correction}.
\newblock Ph.d. thesis, Caltech, 1997.
\newblock ArXiv.org Preprint quant-ph/9705052.

\bibitem[Hav03]{Havel2003}
Timothy~F. Havel.
\newblock Procedures for converting among lindblad, kraus and matrix
  representations of quantum dynamical semigroups.
\newblock {\em Journal of Mathematical Physics}, 44(2):534--557, 2003.
\newblock ArXiv.org Preprint quant-ph/0201127.

\bibitem[HHH99]{Horodecki1999}
Michal Horodecki, Pawel Horodeck, and Ryszard Horodecki.
\newblock General teleportation channel, singlet fraction, and
  quasidistillation.
\newblock {\em Physical Review~A: General Physics}, 60(1888), 1999.
\newblock ArXiv.org Preprint quant-ph/9807091.

\bibitem[HHH05]{HayashiEtAl2005}
A.~Hayashi, M.~Horibe, and T.~Hashimoto.
\newblock Mean king's problem with mutually unbiased bases and orthogonal latin
  squares.
\newblock {\em Physical Review~A: General Physics}, 71:052331, 2005.
\newblock ArXiv.org Preprint quant-ph/0502092.

\bibitem[IV00]{Hill2000}
Victor E.~Hill IV.
\newblock {\em Groups and Characters}.
\newblock Chapman \& Hall/CRC, Boca Raton, FL, 2000.

\bibitem[Iva81]{Ivanovic1981}
I.~D. Ivanovi\'{c}.
\newblock Geometrical description of quantal state determination.
\newblock {\em Journal of Physics~A: Mathematical and General}, 14:3241--3245,
  1981.

\bibitem[KR04]{KlappeneckerRoetteler2004}
Andreas Klappenecker and Martin R\"{o}tteler.
\newblock Constructions of mutually unbiased bases.
\newblock In {\em Lecture Notes in Computer Science}, volume 2948, 2004.
\newblock ArXiv.org Preprint quant-ph/0309120.

\bibitem[KR05a]{KlappeneckerRoetteler2005}
Andreas Klappenecker and Martin R\"{o}tteler.
\newblock Mutually unbiased bases are complex projective $2$-designs.
\newblock {\em ArXiv.org Preprint quant-ph/0502031}, 2005.

\bibitem[KR05b]{KlappeneckerRoetteler2005_MK}
Andreas Klappenecker and Martin R\"{o}tteler.
\newblock New tales of the mean king.
\newblock {\em ArXiv.org Preprint quant-ph/0502138}, 2005.

\bibitem[KSV02]{Kitaev2002}
A.~Yu. Kitaev, A.~H. Shen, and M.~N. Vyalyi.
\newblock {\em Classical and Quantum Computation}.
\newblock AMS, Providence, RI, 2002.

\bibitem[LBZ02]{LawrenceEtAl2002}
Jay Lawrence, Caslav Brukner, and Anton Zeilinger.
\newblock Mutually unbiased binary observable sets on $n$ qubits.
\newblock {\em Physical Review~A: General Physics}, 65:032320, 2002.
\newblock ArXiv.org Preprint quant-ph/0104012.

\bibitem[LF80]{LadnerFischer1980}
Richard~E. Ladner and Michael~J. Fischer.
\newblock Parallel prefix computation.
\newblock {\em Journal of the ACM}, 27:831--838, 1980.

\bibitem[LN94]{LidlNiederreiter1994}
Rudolf Lidl and Harald Niederreiter.
\newblock {\em Introduction to finite fields and their applications}.
\newblock Cambridge University Press, Camridge, UK, 1994.

\bibitem[McD74]{McDonald1974}
Bernard~R. McDonald.
\newblock {\em Finite Rings with Identity}.
\newblock Marcel Dekker, New York, 1974.

\bibitem[Meg05]{Meglicki2005}
Zdis{\l}aw Meglicki.
\newblock Introduction to quantum computing (m743).
\newblock http://beige.ucs.indiana.edu/M743/M743.pdf, 2005.
\newblock Chapter 3.

\bibitem[Mos99]{Mosca1999}
Michele Mosca.
\newblock {\em Quantum Computer Algorithms}.
\newblock D. phil. thesis, University of Oxford, 1999.
\newblock http://www.cacr.math.uwaterloo.ca/~mmosca/moscathesis.ps.

\bibitem[Mun75]{Munkres1975}
J.~Munkres.
\newblock {\em Topolgy: a first course}.
\newblock Prentice-Hall, Englewood Cliffs, NJ, 1975.

\bibitem[NC00]{NielsenChuang2000}
Michael~A. Nielsen and Issac~L. Chuang.
\newblock {\em Quantum Computation and Quantum Information}.
\newblock Cambridge University Press, Cambdrige, UK, 2000.

\bibitem[Nie02]{Nielsen2002}
Michael~A. Nielsen.
\newblock A simple formula for the average gate fidelity of a quantum dynamical
  operation.
\newblock {\em Physics Letters~A}, 303:249--252, 2002.
\newblock ArXiv.org Preprint quant-ph/0205035.

\bibitem[Pap94]{Papadimitriou1994}
Christos~H. Papadimitriou.
\newblock {\em Computational Complexity}.
\newblock Addison-Wesley, Reading, MA, 1994.

\bibitem[PBKLO04]{BennettEtAl1996}
David Poulin, Robin Blume-Kohout, Raymond Laflamme, and Harold Ollivier.
\newblock Exponential speed-up with a single bit of quantum information:
  Testing the quantum butterfly effect.
\newblock {\em Physical Review Letters}, 92:177906, 2004.
\newblock ArXiv.org Preprint quant-ph/0310038.

\bibitem[PR05]{PlanatRosu2005}
Michel Planat and Haret Rosu.
\newblock Mutually unbiased phase states, phase uncertainties, and gauss sums.
\newblock {\em European Physical Journal D}, 36(1):133--139, 2005.
\newblock ArXiv.org Preprint quant-ph/0506128.

\bibitem[RBKSS05]{RomeroEtAl2005}
J.~L. Romero, G.~Bjork, A.~B. Klimov, and L.~L. Sanchez-Soto.
\newblock On the structure of the sets of mutually unbiased bases for $n$
  qubits.
\newblock {\em ArXiv.org Preprint quant-ph/0508129}, 2005.

\bibitem[Rud67]{Rudin1967}
Walter Rudin.
\newblock {\em Fourier Analysis on Groups}.
\newblock Interscience Publishers, New York, 1967.

\bibitem[Rud73]{Rudin1973}
Walter Rudin.
\newblock {\em Functional analysis}.
\newblock McGraw-Hill, New York, 1973.

\bibitem[Sch60]{Schwinger1960}
Julian Schwinger.
\newblock Unitary operator bases.
\newblock {\em Proceedings of the National Academy of Sciences of the USA},
  46:570--579, 1960.

\bibitem[Sho96]{Shor1994}
Peter~W. Shor.
\newblock Algorithms for quantum computation: discrete logarithms and
  factoring.
\newblock In {\em Proceedings of the 35 th Annual Symposium on Foundations of
  Computer Science (FOCS)}, Los Alamitos, CA, 1996. IEEE Press.
\newblock ArXiv.org Preprint quant-ph/9508027.

\bibitem[VK91]{VilenkinKlimyk1}
N.~Ja Vilenkin and A.~U. Klimyk.
\newblock {\em Representation of Lie Groups and Special Functions, Volume 1}.
\newblock Kluwer Academic Publishers, Netherlands, 1991.

\bibitem[VK93]{VilenkinKlimyk2}
N.~Ja Vilenkin and A.~U. Klimyk.
\newblock {\em Representation of Lie Groups and Special Functions, Volume 2}.
\newblock Kluwer Academic Publishers, Netherlands, 1993.

\bibitem[Wan97]{Wan1997}
Zhe-Xian Wan.
\newblock {\em Quaternary Codes}.
\newblock World Scientific, Singapore, 1997.

\bibitem[Wan03]{Wan2003}
Zhe-Xian Wan.
\newblock {\em Lectures on Finite Fields and Galois Rings}.
\newblock World Scientific, Singapore, 2003.

\bibitem[WB04]{WojcanBeth2004}
Pawel Wojcan and Thomas Beth.
\newblock New construction of mutually unbiased bases in square dimensions.
\newblock {\em ArXiv.org Preprint quant-ph/0407081}, 2004.

\bibitem[WF89]{WoottersFields1989}
William~K. Wootters and Brian~D. Fields.
\newblock Optimal state-determination by mutually unbiased measurements.
\newblock {\em Annals of Physics}, 191:363--381, 1989.

\bibitem[WHE{\etalchar{+}}04]{WeinsteinEtAl2004}
Yaakov~S. Weinstein, Timothy~F. Havel, Joseph Emerson, Nicolas Boulant, Marcos
  Saraceno, Seth Lloyd, and David~G. Cory.
\newblock Quantum process tomography of the quantum fourier transform.
\newblock {\em Journal of Chemical Physics}, 121(13):6117--6133, 2004.
\newblock ArXiv.org Preprint quant-ph/0406239.

\bibitem[Wil70]{Willard1970}
Stephen Willard.
\newblock {\em General Topology}.
\newblock Addison-Wesley, Reading, MA, 1970.

\bibitem[Zau99]{Zauner1999}
Gerhard Zauner.
\newblock {\em Quantendesigns - Grundzüge einer nichtkommutativen
  Designtheorie}.
\newblock Dissertation, University of Vienna, 1999.
\newblock http://www.mat.univie.ac.at/~neum/ms/zauner.ps.gz.

\end{thebibliography}

\end{document}